\documentclass[12pt]{article}

\usepackage{amsthm,amssymb,stmaryrd,wasysym}

\input amssym.def

\newtheorem{proposition}{Proposition}
\newtheorem*{theorem}{Theorem}
\newtheorem{lemma}{Lemma}
\newtheorem{corollary}{Corollary}

\theoremstyle{definition}
\newtheorem{remark}{Remark}

\newlength{\blength}
\renewcommand{\proof}[1]{\vspace{-.05cm}
\begin{list}{\bf Proof:}
{\listparindent=\parindent\parsep=0pt \labelwidth=-0.5cm
\labelsep=\parindent \addtolength{\labelsep}{-\blength}
\addtolength{\labelsep}{1.5cm}
\itemindent=-\blength
\addtolength{\itemindent}{\parindent} \leftmargin=1.0cm}
\item
#1~$\qedsymbol$\end{list}
\vspace{.0cm}}

\begin{document}

\thispagestyle{empty}

\begin{flushright}
{\tt arXiv:0807.4223 [hep-th]} 
\end{flushright}
\vskip 2cm

 \begin{centering}

{\large {\bfseries Comments on higher-spin symmetries}}

 \vspace{2cm}
Xavier Bekaert\\
\vspace{2mm}
{\small Laboratoire de Math\'ematiques et Physique Th\'eorique}\\
{\small Unit\'e Mixte de Recherche $6083$ du CNRS}\\
{\small F\'ed\'eration de Recherche $2964$ Denis Poisson}\\
{\small Universit\'e Fran\c{c}ois Rabelais, Parc de Grandmount}\\
{\small 37200 Tours, France} \\
\vspace{1mm}{\tt \footnotesize Xavier.Bekaert@lmpt.univ-tours.fr}

\vspace{1.2cm}

\end{centering}

\begin{abstract}
The unconstrained frame-like formulation of an infinite tower of completely symmetric tensor gauge fields is reviewed and examined  
in the limit where the cosmological constant goes to zero. 
By partially fixing the gauge and solving the torsion constraints, 
the form of the gauge transformations in the unconstrained metric-like formulation
are obtained till first order in a weak field expansion.
The algebra of the corresponding gauge symmetries is shown to be equivalent, at this order and modulo
(unphysical) gauge parameter redefinitions, to the Lie algebra of Hermitian differential operators on ${\mathbb R}^n$, 
the restriction of which to the spin-two sector is the Lie algebra of infinitesimal diffeomorphisms.
\end{abstract}

\vspace{1.5cm}


\vspace{.5cm}


\vspace{4.5cm}

\pagebreak

\pagenumbering{arabic}

\section{Introduction}

Although remarkable results have been found at the level of equations of motion by Vasiliev
when the cosmological constant is nonvanishing (see e.g. \cite{Vasiliev:2004qz,BCIV} 
for recent reviews), the old Fr\o nsdal programme of introducing consistent couplings among
higher-spin gauge fields \cite{Fronsdal:1978rb} is still far away
from completion at the level of the action.\footnote{For introductory reviews on higher-spin gauge theories, see \cite{Bouatta:2004kk} and references therein.}
Various strategies for constructing cubic vertices have been explored over the years, such as 
working in the light-cone gauge \cite{Bengtsson:1983pd} (see \cite{Metsaev} for latest results),
applying the Noether method \cite{Berends:1984rq} 
and its modern BRST reformulation \cite{Barnich:1993vg} (see the review \cite{Bekaert:2006us} as a tentative summary of the state-of-the-art) or
mimicing string field theory \cite{Bengtsson:1987jt} (see \cite{Buchbinder:2006eq} for a review including the recent developments). 
But the most successful approach still remains the frame-like formulation making use of spinorial oscillators \cite{Fradkin:1987ks}.
Despite this series of encouraging results, no consistent vertex has ever been constructed beyond cubic order.

Up to now, it proved to be extremely fruitful to compare higher-spin gauge theories with gravity when looking for inspiration. In a perturbative analysis of Einstein's theory around some fixed background, gravity appears as a non-Abelian gauge theory of spin-two particles where the geometric origin of the self-interactions is obscure. By analogy, the lack of a deeper understanding of higher-spin interactions can be traced back to the fact that the underlying geometry (if any!) remains elusive. This unsatisfactory situation may call for a comparative look on the development of both subjects.

\subsection{Higher-spins \textit{vs} spin-two}

On the mathematics side Cartan's moving frames are of course posterior to pure Riemannian geometry and, analogously, on the physics side general relativity was initially discovered in the metric form by Einstein and Hilbert, more than a decade before the introduction of vielbeins by Weyl. Free higher-spin gauge theories also first appeared in ``metric'' version \cite{Fronsdal:1978rb} but soon later they were presented in ``frame'' version \cite{Vframe80} whereas for the interactions between higher-spin gauge fields and gravity the story is quite different: the frame-like formulation appeared first and somehow remains the only one available (at all orders).
One of the virtues of the metric formulation of gravity with respect to the frame formulation is that it involves a minimal number of ingredients and so its geometrical interpretation is more direct. Not surprisingly, the first attempts of unraveling some geometry for free higher-spin theories were performed in the metric-like formulation \cite{deWit:1979pe}. But an advantage of the frame formulation of gravity is that it roughly ressembles to a Yang-Mills gauge theory for the isometry algebra of its maximally symmetric background. For higher-spins, the frame-like formulation of \cite{Vframe87} proved to be an effective starting point for writing interactions \cite{Fradkin:1987ks} through a generalisation of MacDowell-Mansouri action.\footnote{The frame-like version of the so-called triplet from \cite{Bengtsson:1987jt} has been recently constructed \cite{Sorokin:2008tf}. Note also the proposal \cite{Engquist:2007kz} in the spirit of the Chern-Simons gravity theories reviewed in \cite{Zanelli:2002qm}.} A drawback of the frame formulation of gravity is that its geometrical interpretation is subtler than its similarities with Yang-Mills theory would suggest. In fact the ``local translations'' are \textit{not} symmetries because they do not preserve the torsion constraint. This issue has an analogue for higher-spins: setting the torsion-like two-forms to zero is not consistent with the expected gauge transformations beyond the lowest order. This difficulty has been circumvented by Vasiliev at the level of field equations by his ``doubling of oscillators'' which leads to a perturbative reconstruction of consistent deformations of the initial torsion constraints, field equations, \textit{etc}.
Although impressive analyses have been carried out \cite{Sezgin:2000hr}, it seems technically out of reach to perform this reconstruction in closed form for the unfolded equations \cite{Vasiliev:1990en} till the order where one could get some insights on the (would-be) corresponding quartic vertices. 

The specific features of higher spins with respect to the spin two should not be hidden by their similarities, because the former have yet prevented a better understanding of their geometry. In particular, the trace conditions of \cite{Fronsdal:1978rb,Vframe80} on the gauge fields and parameters are somewhat unnatural from a geometrical perspective. A formulation of higher-spin gauge theories is nowadays referred to as ``($\,$un)$\,$constrained'' whether trace constraints are imposed ($\,$or not). Foregoing these algebraic constraints opens a wide window of possible geometrical interpretations. For instance, Dubois-Violette and Henneaux elegantly encoded the structure of the linear unconstrained metric-like theory in a generalised complex \cite{DuboisVH}. The possibility of relaxing the trace constraints at the level of the action was unraveled by Francia and Sagnotti in the metric-like formalism \cite{Francia:2002aa}.
Then, it was natural to look for removing the trace constraints in the frame-like formulation as well.
At the free level and in flat space-time, it is easy to check that the unconstrained analogue of the field equations of \cite{Vframe87}
are equivalent to the Bargmann-Wigner equations and their higher-dimensional analogues (as was briefly mentioned during \cite{Dubna}).
A decisive step was performed by Sagnotti, Sezgin and Sundell who imposed their ``strong $Sp\,(2\,,\mathbb R)$ condition'' in order to implement the ``off-shell'' (\textit{i.e.} unconstrained) higher-spin algebra in the unfolded formalism \cite{Sagnotti:2005}. A systematic and detailed analysis of the unconstrained frame-like formalism at the free level has recently been performed \cite{Engquist:2007yk} while the unconstrained metric-like formalism keeps being developped in a large number of directions (see \textit{e.g.} \cite{FMS} for some of them). 

\subsection{Non-Abelian symmetries at lowest order}\label{nonabsym}

These various considerations motivate a thorough examination of the non-Abelian higher-spin gauge symmetries in the metric-like formalism arising from the frame-like one \cite{Vasiliev:2004qz,Sagnotti:2005,Vasiliev:2005zu} hopefully looking for a simpler formulation.
For the spin two case, it is well known \cite{Fang:1978rc} that, even from an analysis at first order in the coupling constant, one may already recognise the structure of the diffeomorphism algebra and of the Lie derivative.
This paradigmatic example is closely followed and applied to the higher-spin case in the next sections, thereby leading to our main result summarised as follows:\vspace{3mm}

Let $\varphi_{\mu_1\ldots\mu_s}(x)$ be a tower of completely symmetric tensor gauge fields.
Consider the gauge transformations of the unconstrained frame-like formulation arising from the Minkowski off-shell higher-spin algebra.
By partially fixing the gauge and solving the torsion constraints, the gauge transformations of the corresponding unconstrained metric-like formulation are obtained and read, modulo perturbative redefinitions of the gauge fields and parameters:
\begin{equation}
\delta_\varepsilon\varphi\,=\,\{\,\varepsilon\,,\phi\,\}_{_M}\,+\,{\cal K}\,+\,{\cal O}(\varphi^2)\,,
\label{nonabeliangaugetransfos}
\end{equation}
where $\varepsilon$ is a smooth function in the position $x^\mu$ and a power series in the auxiliary variables $p_\nu$, as well as the function
\begin{equation}
\phi(x,p)\,\,=\,\frac12\,p^2\,+\,\varphi(x,p)\,,
\label{phi}
\end{equation}
with $p^2=\eta^{\mu\nu}p_\mu p_\nu$ and
\begin{equation}
\varphi(x,p)\,\,=\,\sum\limits_{s}\,\frac{1}{s\,!}\,\,\varphi^{\mu_1\ldots\mu_s}(x)\,p_{\mu_1}\ldots p_{\mu_s}\,.
\label{varphi}
\end{equation}
The term $\cal K$ denotes a linear function of the linearised curvature tensors and their derivatives,
while the term ${\cal O}(\varphi^2)$ is at least quadratic in the tensor gauge fields.\footnote{Both of these last two terms have not been computed explicitly and it may be that they simply are vanishing, or can be eliminated via suitable redefinitions.}
Moreover, the Lie bracket $\{f,g\}_{_M}$ of two functions $f(x,p)$ and $g(x,p)$ is defined as
\begin{equation}
\{\,f\,,\,g\,\}_{_M}(x,p)\,=\,\frac{2}{\lambda}\,\,f(x,p)\,\sin\Big[\,\frac{\lambda}{2}\,\,\Big(\frac{\overleftarrow{\partial}}{\partial x^{\mu}}\frac{\overrightarrow\partial}{\partial p_{\mu}}-\frac{\overleftarrow{\partial}}{\partial p_{\mu}}\frac{\overrightarrow\partial}{\partial x^{\mu}}\Big)\Big]\,
\,g(x,p)
\label{explicitbracket}
\end{equation}
where $\lambda$ is a constant with the dimension of a length and the arrows indicate on which factor each derivative acts. 
The commutator of two gauge transformations (\ref{nonabeliangaugetransfos}) reads
\begin{equation}
[\,\delta_{\varepsilon_1},\delta_{\varepsilon_2}]\varphi\,=\,\delta_{\{\,\varepsilon_1,\varepsilon_2\,\}_{_M}}\varphi\,+\,{\cal O}(\varphi)\,.
\label{commutatoralgebra}
\end{equation}
Therefore, at lowest order in the weak field expansion, the non-Abelian algebra of the non-linear gauge symmetries (\ref{nonabeliangaugetransfos}) is isomorphic to the real Lie algebra of Hermitian differential operators acting on the Hilbert space of square-integrable functions on ${\mathbb R}^n\,$.\vspace{3mm}

This very last result was already mentioned in the proceedings \cite{Bekaert:2007mi}. By analogy with gravity, the gauge fields $\varphi^{\mu_1\ldots\mu_s}$ are assumed to be dimensionless and the gauge parameters $\varepsilon^{\mu_1\ldots\mu_s}$ to have the dimension of a length. A simple dimensional analysis shows that the auxiliary variables $p^\mu$ must be dimensionless.
At lowest order in the weak field expansion $\varphi\,$, the gauge transformation (\ref{nonabeliangaugetransfos}) reproduces the celebrated symmetrised derivative of the gauge parameter unraveled by Fr\o nsdal in \cite{Fronsdal:1978rb},
\begin{equation}
\delta_\varepsilon\varphi\,=\,p_\mu\eta^{\mu\nu}\frac{\partial\,\varepsilon}{\partial x^\nu}\,+\,{\cal O}(\varphi)\,,
\label{abeliangaugetransfos}
\end{equation}
as it should. The term $\cal K\,$, built out of the curvature tensors investigated along several lines in \cite{deWit:1979pe,DuboisVH}, is strictly gauge invariant under (\ref{abeliangaugetransfos}) at lowest order, hence it does not play any role in the gauge algebra (\ref{commutatoralgebra}) at this order.
The $p^2$ term in (\ref{phi}) must be interpreted as the Minkowski background while (\ref{varphi}) is the perturbation. The Lie bracket (\ref{explicitbracket}) is nothing more than the Moyal bracket of real functions on the phase space. Notice that at lowest order in $\lambda$ (\textit{i.e.} in some low energy limit) this bracket is equal to the canonical Poisson bracket of classical observables. 
So the restriction of the bracket in (\ref{nonabeliangaugetransfos}) to the pure spin $s=2$ sector, with $\phi=\frac12 \,g^{\mu\nu}p_\mu p_\nu$ and $\varepsilon=\xi^\mu\,p_\mu\,$, reproduces the Lie derivative of the (inverse) metric,
\begin{equation}
\{\,\frac12\, g^{\mu\nu}p_\mu p_\nu\,,\,\xi^\rho\,p_\rho\,\}_{_M}\,=\,\frac12\,({\cal L}_\xi g^{\mu\nu})\,p_\mu p_\nu\,+{\cal O}(\lambda^2)\,,
\label{spintwononabeliangaugetransfos}
\end{equation}
and the last term in (\ref{spintwononabeliangaugetransfos}), which is of homogeneity degree two in $\lambda$ and zero in $p^\mu\,$, 
can be dropped consistently in such a restriction. More generally, the restriction to gauge parameters $\varepsilon=\xi^\mu\,p_\mu$ which are linear in the fibre, reproduces at lowest order in $\lambda$ the Lie derivative for all symmetric tensors,
\begin{equation}
\{\,\phi\,,\,\xi^\rho\,p_\rho\,\}_{_M}\,=\,{\cal L}_\xi \phi\,+{\cal O}(\lambda^2)\,.
\label{anyspinonabeliangaugetransfos}
\end{equation}
It is still somewhat consistent to drop the extra term of order at least two in $\lambda\,$, in the sense that it can be removed from the gauge transformation (\ref{nonabeliangaugetransfos}) at linearised order in $\varphi$ via a field redefinition.

The result presented above on the algebra of gauge symmetries is presumably not so surprising due to two well known properties: the Moyal product in the fibre should induce a star product on the cotangent bundle via the higher-spin equations of motion and the Moyal product is the somewhat unique star product on $T^*{\mathbb R}^n\,$. One might even say that our result is an expected corollary of some recent works \cite{Vasiliev:2005zu,Grigoriev:2006tt} making link between the unconstrained frame-like formulation and the Fedosov construction \cite{Fedosov}. The deep relation existing between Vasiliev's unfolded formulation and Fedosov's deformation quantisation was pointed out very early \cite{Vasiliev:1999ba} but has been clarified during these last years \cite{Vasiliev:2005zu,Grigoriev:2006tt}. Nevertheless, it should be emphasised that (to the author's knowledge) the final step of inducing the Moyal product on the cotangent bundle from the one on the fibre, only by making use of the torsion constraints (in fact, the unfolded formalism is never used here), has not been performed explicitly or analysed in details before. This analysis is the main goal of this paper.

\subsection{Plan of the paper}
 
The perturbative analysis of the gravity theory formulated along the lines of Cartan's view of geometry is briefly introduced in Section \ref{spintwo} in order to enlighten the subsequent discussion of its higher-spin generalisation.
The AdS/CFT and Minkowski higher-spin algebras are reviewed in Section \ref{Higherspinalgebras} with many details and emphasis on various interpretations they allow. The corresponding frame-like formulations are motivated and introduced in Section \ref{frameformul} by analogy with the example of gravity. These sections (\ref{spintwo}, \ref{Higherspinalgebras} and \ref{frameformul}) are intended to constitute a review of higher-spin theories underlying the interplay between the algebraic and geometrical perspectives, but notice that some new results are included in Section \ref{Higherspinalgebras}. 

The main result of the present paper concerns the metric-like non-Abelian gauge symmetries discussed in Section \ref{metricformul}. More precisely, this result has been already stated in the last subsection but its proof is presented in Subsection \ref{proof}, just after the Subsection \ref{mlikeAbelian} on the Abelian transformations. Suggestive algebraic and geometrical properties of the Moyal bracket and its relatives are discussed in Subsection \ref{remarks} together with a deformation of the Abelian gauge symmetries in the presence of a cosmological constant.
The section \ref{conclusion} is the conclusion. In order to be as self-contained as possible without weighing down the core of the text, a series of useful mathematical definitions, which might be less familiar to physicists, are reviewed in the appendices.\vspace{2mm}

\subsection{Notation}

Let $\cal A$ be an algebra with product $\star\,$. The commutator is denoted by $[\,\,\,\stackrel{\star}{,}\,\,\,]$ and this bracket acts as
$[\,a\,\stackrel{\star}{,}\,b\,]:=a\star b-b\star a$ where $a,b\in\cal A\,$.

The symmetric tensor product $\vee$ is defined by $A\vee B=A\otimes B+B\otimes A$ while the antisymmetric tensor product $\wedge$ is defined by $A\wedge B=A\otimes B-B\otimes A\,$. The wedge product of differential forms is implicit in the present paper in order to lighten the formulas. Curved (respectively, square) brackets over a set of indices denote complete (anti)symmetrisation over all this indices, with weight one, \textit{i.e.} $S^{(\mu_1\ldots \mu_r)}=S^{\mu_1\ldots \mu_r}$ and $A^{[\mu_1\ldots \mu_r]}=A^{\mu_1\ldots \mu_r}$ respectively for $S\in\vee^{\,r}({\mathbb R}^n)$ and $A\in\wedge^r({\mathbb R}^n)\,$.

Let $\mathbb K$ be a field. The commutative algebra denoted by ${\mathbb K}[X^a]$ (respectively, ${\mathbb K}[[X^a]]$) is spanned by the polynomials (respectively, the formal power series around the origin) in the variables $X^a$ with coefficients in $\mathbb K\,$. Non-commuting variables or operators, say $\textsc{X}^a\,$, are slanted while commuting variables or symbols of operators, like $X^a\,$, are in italic.

\section{Gravity example}\label{spintwo}

The subsection \ref{Cartan} is a review of Cartan's approach to Riemannian geometry and largely finds its inspiration in the textbooks \cite{Ortin} (for a physicist point of view) and \cite{Erlangen} (for a mathematician one). Though this material is standard, it is provided here with emphasis on the geometrical interpretation in order to prepare the ground for the frame-like formulation of higher-spin gauge fields.
The subsection \ref{perturbative} briefly reviews the perturbative approach to gravity as a non-Abelian spin-two gauge theory.
An inspiring survey of the early story of this approach is \cite{Fang:1978rc}. A comprehensive review is developed in the chapter 3 of \cite{Ortin}.

\subsection{Cartan \textit{versus} Riemann}\label{Cartan}

Let $\cal M$ be a manifold of dimension $n\,$.
An arbitrary basis in (co)tangent space is defined by a set of $n$ vectors (respectively, linear forms): the
(co)frame basis $e_a:=e^\mu_a\,\partial_\mu$ (respectively, $e^a:=e_\mu^a\,dx^\mu$) such that det($e$)$\,\neq 0\,$. Latin indices $a,b,...$ will denote ``tangent'' (anholonomic) indices while Greek indices $\mu,\nu,...$ will denote ``world'' (holonomic) indices. 
The world tensors transform under infinitesimal diffeomorphisms (via the Lie derivative) while the tangent tensors transform under infinitesimal local $GL(n)$ transformations (via the corresponding tensor representation of $\mathfrak{gl}(n)\,$). In more fancy terminology, one may say that the ``(co)frame bundle'' is defined as the principal $GL(n)$-bundle associated with the (co)tangent bundle.

\subsubsection{Moving frames}\label{movingfr}

Given an affine connection $\Gamma$, covariantising with respect to the world indices, and a Ehresmann connection $\omega$ for the structure group $GL(n)\,$, covariantising with respect to the tangent indices, one defines the corresponding total covariant derivative \DH\,, covariantising with respect to all the indices. For instance, acting on the coframe, it is given by
\begin{equation}
\mbox{\DH}^{}_\mu e_\nu^a\,=\,\partial^{}_\mu e_\nu^a\,-\,\Gamma_\mu{}^\rho{}_\nu\, e_\rho^a\,+\,\omega_\mu{}^a{}_b\,e_\nu^b\,.
\end{equation}
The (world \textit{vs} tangent) covariant derivatives are (respectively) denoted by $\nabla_\mu:=\partial_\mu+\Gamma_\mu$ and $D_\mu:=\partial_\mu+i\,\omega_\mu\,$.

The affine connection one-form $$\Gamma_\mu\,:=\,\Gamma_\mu{}^\rho{}_\nu\,\, \frac{\partial}{\partial x^\rho}\otimes dx^\nu\,,$$
takes values in the Lie algebra of endomorphisms on the tangent space $\mathfrak{gl}(T{\cal M})$ of basis $\frac{\partial}{\partial x^\rho}\otimes dx^\nu$ while the linear connection one-form $\omega_\mu :=\omega_\mu{}^a{}_b\textsc{M}_a{}^b$ takes values in the general linear Lie algebra $\mathfrak{gl}(n)$ of basis $\textsc{M}_a{}^b\,$.
The components of the curvature two-form $R:=(d+\Gamma)^2$ for the affine connection $\Gamma$ are given by the Riemann tensor
\begin{equation}
R_{\mu\nu}{}^\sigma{}_\rho\,
=\,2\,\Big(\,\partial_{[\mu}\Gamma_{\nu]}{}^\sigma{}_\rho\,+\,\Gamma_{[\mu|}{}^\sigma{}_\tau\,\Gamma_{|\nu]}{}^\tau{}_\rho\,\Big)\,,
\end{equation}
while the components of the curvature two-form $\cal R$ defined by $i\,{\cal R}=(d+i\,\omega)^2$ for the Ehresmann connection $\omega$ read
\begin{equation}
{\cal R}_{\mu\nu}{}^a{}_b\,
=\,2\,\Big(\,\partial_{[\mu}\omega_{\nu]}{}^a{}_b\,+\,\omega_{[\mu|}{}^a{}_c\,\omega_{|\nu]}{}^c{}_b\,\Big)\,.
\end{equation}
The torsion of the affine connection is the world tensor $T_{\mu\nu}{}^\rho:=2\,\Gamma_{[\mu}{}^\rho{}_{\nu]}\,$.

The ``first vielbein postulate''
\begin{equation}
\mbox{\DH}^{}_\mu e_\nu^a=0
\label{firstvp}
\end{equation}
allows to convert tangent into world indices inside the total covariant derivative and it implies the following relations between the connections\footnote{Actually in \cite{Hehl:1994ue}, a converse viewpoint was adopted: the condition (\ref{firstvp}) was argued not to be a ``postulate'' but as a mere statement of the inhomogeneous transformation law (\ref{relomviel}) relating the components of the same connection expressed either in the anholonomic or holonomic bases.}
\begin{equation}
\omega_{\mu}{}^a{}_b\,=\,e_b^\nu\Big(\Gamma_{\mu}{}^\rho{}_\nu\, e_\rho^a\,-\,\partial_\mu e_\nu^a\Big)\,,
\label{relomviel}
\end{equation}
and the curvatures
\begin{equation}
e^\sigma_a\,e_\rho^b\,{\cal R}_{\mu\nu}{}^a{}_b\,
=\,R_{\mu\nu}{}^\sigma{}_\rho\,.
\label{structure}
\end{equation}
Moreover, the first vielbein postulate gives an important relation between the torsion and the frame. Taking the antisymmetric part \DH$^{}_{[\mu} e_{\nu]}^a=0$ of the postulate, one obtains that the antisymmetric part of the tangent covariant derivative of the vielbein is equal to
\begin{equation}
D^{}_{[\mu} e_{\nu]}^a\,:=\,\partial^{}_{[\mu}e_{\nu]}^a\,+\,\omega_{[\mu|}{}^a{}_b\, e_{|\nu]}^b\,=\,T_{\mu\nu}{}^a\,:=\,T_{\mu\nu}{}^\rho\,e_\rho^a\,.
\label{antivielbpost}
\end{equation}
The relations (\ref{structure})-(\ref{antivielbpost}) are sometimes called Cartan's structure equations. They allow to provide an interpretation of the torsion and Riemann tensors in terms of the ``Cartan connection'' denoted by ${\cal A}$ and defined as the one-form
\begin{equation}
{\cal A}_\mu\,:=\,e_\mu^a\,\textsc{P}_a\,+\,\omega_\mu{}^b{}_c\,\textsc{M}_b{}^c\,,
\label{Connection}
\end{equation}
taking values in the Lie algebra $\mathfrak{igl}(n)={\mathbb R}^n\niplus \mathfrak{gl}(n)$ of the affine group,
spanned by the basis $\{\textsc{P}_a,\textsc{M}_b{}^c\}\,$. In other words, the Cartan connection (\ref{Connection}) may be seen as the pullback of a Ehresmann connection for a $IGL(n)={\mathbb R}^n\rtimes GL(n)$ principal bundle with base $\cal M\,$.
If the first vielbein postulate (\ref{firstvp}) is obeyed, then the restriction of the curvature two-form 
\begin{equation}
{\cal F}=d{\cal A}+{\cal A}^2
\label{curvtwof}
\end{equation}
of this principal $IGL(n)$-bundle is related to the torsion and Riemann tensors by 
\begin{equation}
{\cal F}\,=\,T^a\,\textsc{P}_a\,+\,R^b{}_c\,\textsc{M}_b{}^c\,.
\label{torsionRiemann}
\end{equation}

The ``Cartan covariant derivative'' is defined by
$${\cal D}\,:=\,d+i\,{\cal A}\,=\,D+i\,e^a \,\textsc{P}_a$$
The local $GL(n)$ transformations are adjoint transformations
\begin{equation}
{\cal D}\rightarrow \textsc{U}^{-1}{\cal D}\textsc{U}\,,
\label{adjointransfo}
\end{equation}
where $\textsc{U}=\exp(i\,\varepsilon^b{}_c\textsc{M}_b{}^c)$ is generated by the $\mathfrak{gl}(n)$ basis elements,
while the ``local translations'' are gauge transformations (\ref{adjointransfo}) where $\textsc{U}=\exp(i\,\varepsilon^a\textsc{P}_a)$ is generated by the translation generators $\textsc{P}_a\,$. Notice that, in general when $R^b{}_c\neq 0\,$, the torsion-free condition $T^a=0$ is \textit{not} preserved by local translations. Hence, if the torsion is set to zero, the local affine group of symmetries is in general broken to the local $GL(n)$ transformation subgroup. This is consistent with the fact that the torsion-free condition $T^a=0$ implies that a diffeomorphism of the coframe can be exchanged with the combined action of a local translation and a local $GL(n)$ transformation due to the equality
\begin{equation}
{\cal L}_\xi\, e^a_\mu\,:=\,\xi^\nu\partial_\nu e^a_\mu\,+\,\partial_\mu\xi^\nu e^a_\nu\,=\,\xi^\nu T_{\nu\mu}{}^a
\,+\,D_\mu(\xi^\nu e_\nu^a)\,-\,(\xi^\nu\omega_\nu{}^a{}_b)\, e^b_\mu\,,
\label{diffcofr}
\end{equation}
where (\ref{antivielbpost}) has been used. However, a diffeomorphism of the Ehresmann connection $\omega$ (and therefore also of the Cartan connection $\cal A$) \textit{cannot} be interpreted as the combined action of a local translation and a local $GL(n)$ transformation when the curvature does not vanish,
\begin{equation}
{\cal L}_\xi\, \omega_\mu{}^a{}_b\,:=\,\xi^\nu\partial_\nu \omega_\mu{}^a{}_b\,+\,\partial_\mu\xi^\nu \omega_\nu{}^a{}_b\,=\,\xi^\nu R_{\nu\mu}{}^a{}_b
\,+\,D_\mu(\xi^\nu \omega_\nu{}^a{}_b)\,.
\label{diffconn}
\end{equation}
There is no contradiction since the diffeomorphisms (\ref{diffcofr})-(\ref{diffconn}) preserve the torsion constraint while the local affine transformations do not.

\subsubsection{Klein \& Cartan view of geometry}

From a mathematical perspective, the previous definitions originate from Cartan's generalisation of the Erlangen programme (presented in details in the textbook \cite{Erlangen}).

The celebrated definition of a homogeneous geometry by Klein states that a ``geometry'' is a (transitive and effective) Lie group action on a (connected) manifold. In other words, the following group-theoretical data are required to speak about a homogeneous geometry: a symmetry Lie group $G$ and one of its (closed) subgroup $H\subseteq G\,$. The set of ``geometrical'' points is defined as the (connected) coset manifold $G/H$ on which $G$ acts and where $H$ is the stabiliser (or isotropy group) of any point. The study of such a ``geometry'' is essentially the study of the properties that are preserved by the group $G$ of symmetries. Notice that $G$ may be seen as a principal $H$-bundle with base $G/H\,$. The Maurer-Cartan form is precisely a one-form on this principal $H$-bundle $G\,$, taking values in $\mathfrak{g}\,$, which identifies each tangent space with the Lie algebra. 

For instance\footnote{If $G=IO(n)$ (respectively, $G=O(n+1)$ or $G=O(n-1,1)\,$) and $H=O(n)\,$, then one obtains the Euclidean (respectively, elliptic or hyperbolic) geometries on ${\mathbb R}^n$ (respectively, on $S^n$ or $H_n$).}, when $G=IGL(n)$ and $H=GL(n)\,$, one obtains the affine geometry on the affine space $IGL(n)/GL(n)\cong {\mathbb R}^n\,$.
Since any manifold $\cal M$ is locally homeomorphic to ${\mathbb R}^n\,$, the natural generalisation proposed by Cartan considers a principal $GL(n)$-bundle with base $\cal M\,$, equipped with a one-form ${\cal A}$ taking values in $\mathfrak{igl}(n)$ satisfying the vielbein postulates, as in the previous subsection. One makes contact with usual differential geometry via the frame bundle, which is thus endowed with the Ehresmann connection $\omega$ taking values in $\mathfrak{gl}(n)\,$. More generally, one considers a principal $H$-bundle with base $\cal M$ such that the tangent spaces of this principal bundle are isomorphic to $\mathfrak{g}\,$. In this context, the homogeneous geometry $G/H$ is called the ``model space.'' 
Let us assume that the subalgebra $\mathfrak{h}$ is reductive in the Lie algebra $\mathfrak{g}\,$, \textit{i.e.} the latter decomposes as $\mathfrak{g}=\mathfrak{h}\,\oplus\,\mathfrak{m}$ where each summand is an $\mathfrak{h}$-module for the adjoint representation: $[\mathfrak{h},\mathfrak{m}]\subseteq \mathfrak{m}\,$. A (reductive) ``Cartan connection'' is a $\mathfrak{g}$-valued one-form ${\cal A}$ defined as the sum of the $\mathfrak{m}$-valued solder (also called fundamental) form $e$ and an Ehresmann connection $\omega$ taking values in $\mathfrak{h}\,$. The homogeneous geometry $G/H$ corresponds, locally, to the particular case of a flat Cartan connection, which is essentially the Maurer-Cartan form on $G\,$. Thus, Cartan geometries are curved analogues of Klein geometries. An alternative definition of a Cartan connection is as the pullback of a Ehresman connection on a principal $G$-bundle onto a principal $H$-subbundle such that this pullback contains the solder form performing the identification between each tangent space and the Lie algebra $\mathfrak{g}\,$. Roughly, a Cartan connection is a prescription for attaching a copy of the model space $G/H$ to each point of $\cal M$ and thinking of that model space as being tangent to (and infinitesimally identical with) the manifold at the point of contact. 

In this language, the torsion constraint allows a geometrical interpretation. 
Consider an Ehresmann connection of a principal $G$-bundle with base $\cal M$.
By definition, it provides a correspondence between a curve in $\cal M$ and its horizontal lift in $G\,$. 
One obtains an induced connection on the associated bundle with fibre $G/H$ and structure group $G\,$.
This leads to a correspondence between paths in $\cal M$ and their horizontal lifts in $G/H\,$.
If the induced connection is accompanied with a reduction of the structure group from $G$ to $H\,$, then let $\cal A$ be the inherited connection
on the subbundle with fibre $G/H\,$. 
Yet another equivalent definition of a Cartan connection is that the pullback of $\cal A$ by the preferred section performs the isomorphism between the tangent spaces of $\cal M$ and the vertical spaces.
The geometry of the manifold is infinitesimally identical to that of the homogeneous geometry, but globally can be quite different, the Cartan connection supplies a way of connecting the infinitesimal model spaces within the manifold by means of parallel transport. The preferred section identifies the point of contact between the manifold $\cal M$ and the tangent space $\mathfrak{m}\cong\mathfrak{g}/\mathfrak{h}$ of the model space. 
The horizontal lift of curves is called ``development'' in the case of a Cartan connection.
Development corresponds to the intuitive idea of rolling (without slipping) the tangent copies of the model space along curves in the manifold.
If the curvature two-form $\cal F$ takes values in $\mathfrak{h}$ only, then the parallel transport (of some given point) defines a well defined (\textit{i.e.} path independent) correspondence between (end)points in $\cal M$ and (end)points in $G/H\,$, at least infinitesimally. Still, there is no reason that such a local section of the associated bundle with fibre $G/H$ be (locally) injective, except when $\cal A$ is precisely a Cartan connection. Indeed, a \textit{torsionless} Cartan connection may provide a submersion between the manifold $\cal M$ and the model space $G/H\,$.

\subsubsection{Vielbeins}

Let us now assume that $\cal M$ is a pseudo-Riemannian manifold. World indices are lowered and raised via the metric $g_{\mu\nu}$ and its inverse. Similarly, tangent indices are lowered and raised via the Minkowski metric $\eta_{ab}$ and its inverse $g^{\mu\nu}$.
A coframe basis is orthonormal if $g^{\mu\nu}e_\mu^ae_\nu^b=\eta^{ab}\,$, in which case it is called a ``vielbein'' basis. Then it is natural to require that the tangent tensors transform only under infinitesimal local Lorentz transformations, via the corresponding tensor representation of $\mathfrak{o}(n-1,1)\,$. Conversely, any vielbein basis endows the manifold $\cal M$ with a metric $g_{\mu\nu}=\eta_{ab} e_\mu^ae_\nu^b\,$, so that the vielbein formulation contains the metric one.

The vanishing of the total covariant derivative of the Minkowski metric,
\begin{equation}
\mbox{\DH}_\mu\eta_{ab}=0\,,
\label{secondvp}
\end{equation}
is called the ``second vielbein postulate'' and is strictly equivalent to the vanishing of the total covariant derivative of the metric, \DH$_\mu g_{\nu\rho}=0\,$, due to the first vielbein postulate. Thus (\ref{secondvp}) also says that the covariant derivative $\nabla$ with respect to the world indices is ``metric compatible''
\begin{equation}
\nabla_\mu \,g_{\nu\rho}\,=\,\partial_\mu\, g_{\nu\rho}\,-\,\Gamma_\mu{}^\sigma{}_\nu\, g^{}_{\sigma\rho}\,-\,\Gamma_\mu{}^\sigma{}_\rho\, g^{}_{\nu\sigma}\,=\,0\,.
\label{metricompat}
\end{equation}
Moreover, a connection one-form $\omega$ satisfying (\ref{secondvp}) must be antisymmetric in the tangent indices: it is the ``spin'' connection $\omega:=\omega^{ab}\textsc{M}_{ab}\,$, where $\textsc{M}_{ab}$ is a basis of the Lorentz algebra $\mathfrak{o}(n-1,1)\,$.
Finally, if the torsion two-form vanishes, $T^a=0\,$, then the spin connection $\omega$ can be expressed in terms of the vielbeins,
\begin{equation}
\omega_\mu\,{}_{ab}=e_a^\nu e_b^\rho\Big(\omega_{[\mu\,\nu]\rho}-\omega_{[\mu\,\rho]\nu}-\omega_{[\nu\,\rho]\mu}
\Big)\,,\quad \omega_{[\mu\,\nu]\rho}\,:=\,(e_\rho)_a\,\partial_{[\mu}e^a_{\nu]}\,,
\end{equation}
and the affine connection $\Gamma$ becomes the Levi-Civita connection on $\cal M\,$, the components of which are the Christoffel symbols
\begin{equation}
\Gamma_\mu{}^\rho{}_\nu\,=\,\frac12\,g^{\rho\sigma}\,\Big(\partial_\mu\, g_{\nu\sigma}+\partial_\nu\, g_{\mu\sigma}-\partial_\sigma\, g_{\mu\nu}\Big)\,.
\end{equation}
Following Klein, the Minkowskian geometry is obtained by considering the flat space-time ${\mathbb R}^{n-1,1}$ as the homogeneous space $IO(n-1,1)/O(n-1,1)\,$. Again, since any pseudo-Riemannian manifold is locally isometric to the Minkowski space-time, it is indeed natural to consider the Cartan connection ${\cal A}$ taking values in the Poincar\'e Lie algebra. Contact is made with pseudo-Riemannian geometry via the vielbein postulates.

The cosmological constant $\Lambda$ can be introduced in this setting. Let us assume that $\Lambda<0$ for definiteness. In this context, the manifold is locally isomorphic to the homogeneous space-time $$AdS_n\cong O(n-1,2)/O(n-1,1)\,,$$ then it is natural to consider the one-form ${\cal A}\,:=\,e^a\,\textsc{P}_a\,+\,\omega^{bc}\,\textsc{M}_{bc}\,$,
where $\{\textsc{P}_a,\textsc{M}_{bc}\}$ now span the $AdS_n$ isometry algebra $\mathfrak{o}(n-1,2)\,$.
The curvature two-form (\ref{curvtwof}) is then related to the torsion and Riemann curvature by 
${\cal F}\,=\,T^a\,\textsc{P}_a\,+\,\overline{R}^{bc}\,\textsc{M}_{bc}\,$, 
where the two-form $\overline{R}^{bc}$ is related to the Riemann curvature and vielbeins through $\overline{R}^{bc}={R}^{bc}-\Lambda \,e^b e^c\,$. Notice that the $AdS_n$ space-time indeed corresponds to the ``flat'' solution ${\cal F}=0\,$.
Again, in Cartan's formulation Einstein's gravity ressembles to Yang-Mills' theory. Nevertheless, there are important differences\footnote{The reader may find more comments on these subtle points in many places, say \textit{e.g.} in the section 2 of the review \cite{BCIV}, in the lecture 2 of the notes \cite{Zanelli:2002qm}, in the chapter 3 of the book \cite{Ortin} or in the section 3 of the report \cite{Hehl:1994ue}.} that should be kept in mind. Geometrically, the decisive distinction is that the Cartan ``connection'' is \textit{not} a Ehresmann connection of a principal $G$-bundle.

The formulation of Cartan allows to deal with various distinct geometries in a unified framework. For instance, Riemannian geometry is formulated as a principal bundle with the Lorentz group as fibre, endowed with a Cartan connection taking values in the isometry algebra of a maximally symmetric space-time. But of course, this game can be played for various Lie (super)groups, say conformal (or super Poincar\'e) group, thereby leading to the various known gravity theories, such as conformal (or super) gravity.

\subsection{Perturbative analysis around a background space-time}\label{perturbative}

A retrospective look at gravity is provided by looking at the previous subsection through the glass of a weak field perturbative expansion around a solution of vacuum Einstein equations where the coframes and the metric are given by the sum of the background and the perturbation. In practice, here only Minkowski and (anti) de Sitter spacetimes will be considered as backgrounds and the expansion will be performed till first order only. 

\subsubsection{Moving frames}

As a start, the subsection \ref{movingfr} on differential geometry (no metric is assumed) is re-examined from a perturbative look.
For later convenience, the full coframe will be written with a capital letter $E_\mu^a$, while $e_\mu^a$ will actually stand for the small perturbation of the flat background, \begin{equation}
E_\mu^a=\delta_\mu^a+e_\mu^a\,.
\label{fullcoframe}
\end{equation}
The Cartan connection reads 
\begin{equation}
{\cal A}_\mu\,=\,E_\mu^a\,\textsc{P}_a+\omega_\mu{}^b{}_c\,\textsc{M}_b{}^c\,=\,\textsc{P}_\mu+\Omega_\mu\,,
\label{Cartanconnpert}
\end{equation}
where $\Omega$ denotes the perturbation
\begin{equation}
\Omega_\mu\,=\,e_\mu^a\,\textsc{P}_a\,+\,\omega_\mu{}^b{}_c\,\textsc{M}_b{}^c\,,
\end{equation}
taking values in $\mathfrak{igl}(n)\,$.
The infinitesimal local affine transformations $\delta_\varepsilon{\cal A}=d\varepsilon+i\,[{\cal A},\varepsilon]$ with gauge parameter $\varepsilon\,=\,\varepsilon^a\,\textsc{P}_a\,+\,\varepsilon^b{}_c\,\textsc{M}_b{}^c$ read, in terms of the perturbation, 
\begin{equation}
\delta_\varepsilon e_\mu^a\,=\,\partial_\mu\varepsilon^a\,+\,\omega_\mu{}^a{}_b\,\varepsilon^b\,-\,\varepsilon^a{}_b\, e_\mu^b
=\,\partial_\mu\varepsilon^a\,-\,\delta_\mu^b\,\varepsilon^a{}_b\,+\,\mbox{linear}\,.
\label{locPoinc}
\end{equation}
where ``linear'' stands for terms at least linear in the coframe and connections.
The form of the gauge transformations (\ref{locPoinc}) in the perturbation implies that it is possible to impose the ``soldering gauge'' $e^a_{\mu}=0\,$. In other words, the perturbation of the coframe can be completely removed by using the gauge freedom associated to the local general linear transformations. In some sense, any such gauge-fixing is only partial because it is preserved by the residual gauge transformations (\ref{locPoinc}) where the local general linear parameter is determined in terms of the local translation parameter by
\begin{equation}
\varepsilon^\mu{}_\nu\,=\,\partial_\nu\varepsilon^\mu\,+\,\mbox{linear}\,,
\label{residual}
\end{equation}
where the tangent indices have been converted by making use of the coframe.

The curvature two-form $\cal F$ of the one-form $\cal A$ is given by (\ref{curvtwof}) and (\ref{torsionRiemann}).
Moreover, it transforms under the adjoint action of the infinitesimal local affine transformations (\ref{locPoinc}). The background is flat, therefore the variation of the curvature two-form is preserved under the gauge transformations (\ref{locPoinc}) at order zero in the perturbation. This is important because it implies that the torsion constraint
\begin{equation}
T^a_{\mu\nu}=0\quad\Longleftrightarrow\quad\partial^{}_{[\mu}e_{\nu]}^a\,=\,e_{[\mu}^b\,\omega^{}_{\nu]}{}^a{}_b\,.
\label{antivielbpostpert}
\end{equation}
and the Riemann tensor are preserved by the local affine transformations (\ref{locPoinc}) at lowest order.

\subsubsection{Vielbeins}\label{metricgauge}

Considering Minkowski space-time as background for simplicity, the metric takes the form
\begin{equation}
g_{\mu\nu}=\eta_{\mu\nu}+h_{\mu\nu}\,,
\label{fullmetric}
\end{equation}
where $h_{\mu\nu}$ is a perturbation.
Although the distinction between world and tangent indices is meaningless around Minkowski space-time since both indices are transformed into each other via the identity matrix (either $\delta_\mu^a$ or $\delta^\mu_a$ in Cartesian coordinates), sometimes one will keep the distinction in order to clarify the perturbative reconstruction of Riemannian geometry
from the Cartan formulation.
The orthogonality condition for (\ref{fullcoframe}) and (\ref{fullmetric}) implies that the rank-two symmetric tensor field $h_{\mu\nu}$ is defined in terms of the perturbation $e_\mu^a$ as
\begin{equation}
h_{\mu\nu}\,=\,2\,\delta_{a(\mu}\,e_{\nu)}^a\,+\,\eta_{ab}\, e_\mu^a\,e_\nu^b\,.
\label{he}
\end{equation}
In other words, at lower order the metric is equal to the symmetric part of the vielbein.
The Cartan connection (\ref{Cartanconnpert}) now takes values in the Poincar\'e algebra $\mathfrak{io}(n-1,1)\,$.
The form of the gauge transformations (\ref{locPoinc}) at order zero in the perturbation imply that it is possible to impose the ``metric gauge'' \begin{equation}
\delta^{}_{a[\mu}e^a_{\nu]}=0\quad\Longleftrightarrow\quad h_{\mu\nu}\,=\,2\,\delta_{a\mu}\,e_{\nu}^a\,+\,\eta_{ab}\, e_\mu^a\,e_\nu^b\,.
\label{mgauge}
\end{equation}
at lowest order. In other words, the ``antisymmetric component'' of the vielbein can be completely removed by using the gauge freedom associated to the local Lorentz transformations. The metric gauge is only a partial gauge fixing because it is preserved by the residual gauge transformations (\ref{locPoinc}) where the local Lorentz parameter is determined in terms of the local translation parameter by
\begin{equation}
\varepsilon_{\mu\nu}\,=\,\partial_{[\nu}\varepsilon_{\mu]}\,+\,{\cal O}(h)\,.
\label{residualP}
\end{equation}
Notice that the metric gauge (\ref{mgauge}) is not covariant with respect to the full diffeomorphisms and it must only be understood as a way to make contact with the metric formulation in the first stages of the perturbative expansion.

As can be checked explicitly, the torsion constraint (\ref{antivielbpostpert}) together with the metric gauge (\ref{mgauge}) allow to express the spin connection in terms of the rank-two symmetric tensor field as follows
\begin{equation}
\omega_{\mu\,[\nu\rho]}\,=\,2\,h_{\mu[\nu,\,\rho]}\,+\,{\cal O}(h^2)\,,
\label{spinconnn}
\end{equation}
where the comma stands for the partial derivative.
Substituting the expression (\ref{spinconnn}) into the residual gauge transformations (\ref{locPoinc}) where the gauge parameter is given by (\ref{residualP}) and the vielbein is defined in terms of $h_{\mu\nu}$ via (\ref{mgauge}) leads to the following gauge transformations for the metric perturbation
\begin{eqnarray}
\delta h_{\mu\nu}&=& 2\,\partial_{(\mu}\varepsilon_{\nu)}\,
+\,\varepsilon^\rho \partial_\rho h_{\mu\nu}\,+\,2\, \partial_{(\mu}\varepsilon_{}^\rho\, h_{\nu)\rho}\,+\,{\cal O}(h^2)\,,
\nonumber\\
&=&2\,\partial_{(\mu}\xi_{\nu)}\,
-\,\Big(2\,\partial_{(\mu}h_{\nu)\rho}-\partial_\rho h_{\mu\nu}\Big)\,\xi^\rho\,+\,{\cal O}(h^2)\,\,,
\label{covder}
\end{eqnarray}
where $\varepsilon_\nu:=\eta_{\nu\sigma}\varepsilon^\sigma$ and the local translation parameter $\varepsilon^\rho\,$, have been distinguished from the gauge parameters $\xi_\nu:=g_{\nu\sigma}\varepsilon^\sigma=\varepsilon_\nu+h_{\nu\sigma}\varepsilon^\sigma$ and $\xi^\rho:=g^{\rho\sigma}\xi_\sigma=\varepsilon^\sigma\,$, due to the conventions followed in this subsection. They have been chosen in such a way that in the first and second line of (\ref{covder}) one recognises, respectively, the Lie derivative of the metric $g_{\mu\nu}=\eta_{\mu\nu}+h_{\mu\nu}$ and the symmetrised covariant derivative of the gauge parameter $\xi_\mu\,$. It is important to stress that they have been reconstructed only by making use of an appropriate gauge choice together with the structure constants of the Poincar\'e algebra, which implicitly appear in the gauge transformations (\ref{locPoinc}) and in the torsion constraint (\ref{antivielbpostpert}).\footnote{Naively, the structure constants of the affine algebra might be expected to be enough in order to recover the diffeomorphisms because they act like general linear transformations on the tangent space. The problem with the local affine symmetries is that the soldering gauge removes all degrees of freedom, which obscures the discussion.} In other words, the infinitesimal diffeomorphisms have been recovered as the local Poincar\'e transformations preserving the metric gauge (at lowest order only). The terms ${\cal O}(h^2)$ in (\ref{covder}) actually vanish and the first order deformation already suggest the exact result. Nevertheless, the terms ${\cal O}(h^2)$ in (\ref{covder}) have been written because this fact is not obvious from the perturbative reconstruction point of view. By analogy, a desirable possibility which cannot be excluded is that such higher-order terms might vanish as well for the higher-spin transformations (\ref{nonabeliangaugetransfos}).

Remark that the linearised Riemann tensor is given by
\begin{equation}
\,R_{\mu\nu\,\,\sigma\rho}\,=\,2\,\partial_{[\mu}\Gamma_{\nu]\,\sigma\rho}\,=\,2\,\partial_{[\mu}\omega_{\nu]\,\sigma\rho}\,
=\,2\,\partial_{[\mu}h_{\nu][\sigma,\rho]}\,,
\end{equation}
due to (\ref{spinconnn}).
It can be checked to be gauge invariant at lowest order, \textit{i.e.} under the linearised diffeomorphisms $\delta_{\xi} h_{\mu\nu}\,=\,\partial_{(\mu}\xi_{\nu)}\,$. The vacuum Einstein equation states that the Riemann tensor is traceless on-shell. The space of solutions of the linearised vacuum Einstein equation can be shown to carry a unitary irreducible representation of the Poincar\'e group corresponding to a massless spin-two particle so that the full Einstein equations may be interpreted as the non-linear equations of a non-Abelian gauge theory for a spin-two field.

All the steps of this perturbative discussion can be adapted to the case of anti de Sitter space-time as background by replacing the partial derivatives by covariant derivatives with respect to the background, \textit{etc}. Thus the infinitesimal diffeomorphisms can also be recovered from the local $O(n-1,2)$ transformations preserving some gauge at lower order, and the space of solutions of the linearised vacuum Einstein equation with a negative cosmological constant can be shown to carry a unitary irreducible representation of the pseudo-orthogonal group $O(n-1,2)\,$.

\section{Higher-spin algebras}\label{Higherspinalgebras}

The $AdS_n/CFT_{n-1}$ higher-spin algebras for any dimension $n$ are reviewed from several perspectives in Subsection \ref{AdSCFT}.
Their realisation in terms of a quotient of the universal enveloping algebra for $\mathfrak{o}(n-1,2)$ was given by Eastwood in \cite{Eastwood}. Then  a construction of the same algebra based on the Weyl agebra was given in \cite{Vasiliev:2003ev}.\footnote{The subsections \ref{abstractdef} and \ref{algdef} provide a short summary of the section 5 of \cite{BCIV}. Notice that, for later convenience, the notation is different.} 
In Subsection \ref{Mink}, the In\"{o}n\"{u}-Wigner contraction to the Minkowski higher-spin algebra discussed in \cite{Vasiliev:2005zu,Bekaert:2005ka} is also reviewed in detail.\footnote{Neither extended nor super algebras are adressed here, only the so-called ``simplest'' algebras (not to be confused with their ``minimal'' subalgebras).} For later purpose, a common subalgebra to all these algebras, christened Lorentz higher-spin algebra is briefly introduced in Subsection \ref{Lorentz}.

\subsection{Anti de Sitter / Conformal algebras}\label{AdSCFT}

(Anti) de Sitter space-times are most simply described via their realisations as one-sheeted hyperboloids in the ambient space ${\mathbb R}^{n+1}$. All the present considerations can be adapted to various choices of signature. For definiteness, the case of $AdS_n$ space-time is covered here, so the ambient space ${\mathbb R}^{n-1,2}$ with coordinates $X^A$ ($A=0,1,2,\ldots,n-1,n$) is endowed with a constant metric $\eta_{AB}$ with signature $-++\ldots+-\,$. The ambient indices $A,B,\ldots$ are lowered (or raised) via this constant metric (or its inverse).

\subsubsection{Abstract definition}\label{abstractdef}

Let $A_{n+1}$ be the Weyl algebra (see Appendix \ref{ApWeyl} for more details) presented by the generators $\textsc{X}^A$ and $\textsc{P}^B$ modulo the commutation relations
\begin{equation}
[\,\textsc{X}^A\,,\,\textsc{P}^B\,]\,=\,i\,\eta^{AB}\,.
\label{Heisenbergambient}
\end{equation}
The Hermitian conjugation $^\dagger$ sending the generators $\textsc{X}^A$ and $\textsc{P}^B$ to themselves endows the Weyl algebra $A_{n+1}$ with a structure of $^*$-algebra.
The commutator algebra of this Weyl algebra is defined as the vector space $A_{n+1}$ endowed with minus $i$ times the commutator as Lie bracket. This complex Lie algebra will be denoted by $[A_{n+1}]\,$.

The complex Lie subalgebra of the ``commutator algebra'' $[A_{n+1}]$ that is spanned by the three elements $\textsc{X}^A\textsc{X}_A$, $\frac12(\textsc{X}^A\textsc{P}_A+\textsc{P}^A\textsc{X}_A)$ and $\textsc{P}^A\textsc{P}_A$ 
is isomorphic to the classical Lie algebra $\mathfrak{sp}(2)\,$:
\begin{equation}
\mathfrak{sp}(2)\,\cong\, \mbox{span}_{\mathbb C}\{\textsc{X}^A\textsc{X}_A\,,\, (\textsc{X}^A\textsc{P}_A+\textsc{P}^A\textsc{X}_A)/2\,,\,\textsc{P}^A\textsc{P}_A\}\,\subset\,\,[A_{n+1}]\,.
\label{sp2}
\end{equation}
The centraliser ${\cal C}_{_{A_{n+1}}}(\,\mathfrak{sp}(2)\,)$ of this $\mathfrak{sp}(2)$ subalgebra in $A_{n+1}$ is by definition
the associative subalgebra of the Weyl algebra $A_{n+1}$ spanned by the elements that commute with the three generators of $\mathfrak{sp}(2)\,$. 
This vector space endowed with $-i\,[\,\,\,,\,]$ as Lie bracket
is the commutator algebra $[\,{\cal C}_{_{A_{n+1}}}(\,\mathfrak{sp}(2)\,)\,]\,$.
The elements of the centraliser of $\mathfrak{sp}(2)$ in $A_{n+1}$ which are self-adjoint with respect to the Hermitian conjugation span a real form of the complex Lie algebra $[\,{\cal C}_{_{A_{n+1}}}(\,\mathfrak{sp}(2)\,)\,]\,$. Accordingly, this real form might be written as 
$[\,{\cal C}_{_{A_{n+1}}}(\,\mathfrak{sp}(2)\,)\,]_{\mathbb R}$ but in the recent litterature on higher-spins it is denoted\footnote{The `h' stands for ``higher'' while the `u' and the `1' stand for the fact that this algebra contains the Abelian Lie subalgebra $\mathfrak{u}(1)\cong\mathbb R$ spanned by the unit element. The `2' stands for the $\mathfrak{sp}(2)$ underlying the construction while the remaining entries are related to the pseudo-orthogonal subalgebra $\mathfrak{o}(n-1,2)$ spanned by the elements $\textsc{X}^{[A}\textsc{P}^{B]}\,$.} by 
$\mathfrak{hu}_\infty (\,1|2\,:\,[n-1,2]\,)$ and called the ``off-shell ($AdS_n/CFT_{n-1}$) higher-spin algebra''.

The centraliser ${\cal C}_{_{A_{n+1}}}(\,\mathfrak{sp}(2)\,)$ of $\mathfrak{sp}(2)$ in $A_{n+1}$ possesses two ideals spanned by its elements that can be written as a sum of products between a generator of $\mathfrak{sp}(2)$ (either on the left or on the right) and some element in the Weyl algebra $A_{n+1}\,$, \textit{i.e.}
elements that belong to $\mathfrak{sp}(2)A_{n+1}$ or $A_{n+1}\mathfrak{sp}(2)\,$. Therefore the quotient of the centraliser by any of these ideals is well-defined and denoted by $\overline{\cal C}_{_{A_{n+1}}}(\,\mathfrak{sp}(2)\,)\,$.
The real form $[\,\overline{\cal C}_{_{A_{n+1}}}(\,\mathfrak{sp}(2)\,)\,]_{\mathbb R}$ of its commutator algebra that is spanned by its self-adjoint elements is denoted by $\mathfrak{hu}(\,1|2\,:\,[n-1,2]\,)$ and is called the ``on-shell ($AdS_n/CFT_{n-1}$) higher-spin algebra''.

The complex Lie subalgebra of the higher-spin algebras that is spanned by the elements $\textsc{M}^{AB}:=\textsc{X}^{[A}\textsc{P}^{B]}$ 
is isomorphic to the classical Lie algebra $\mathfrak{o}(n-1,2)\,$, of which higher-spin algebras are infinite-dimensional extensions.

\begin{remark}\label{minimalhs} The linear anti-automorphism of the Weyl algebra $A_{n+1}$ that is induced by the following transformations of the generators: $\textsc{X}^A\mapsto\textsc{X}^A\,$, $\textsc{P}^B\mapsto -\textsc{P}^B$ splits the Weyl algebra as the direct sum $A_{n+1}=A^+_{n+1}\oplus A^-_{n+1}$ where $A_{n+1}^\pm$ is the corresponding eigenspace of eigenvalue $\pm 1\,$. 
This ${\mathbb Z}_2$-grading is essentially the parity in the generators $\textsc{P}^B\,$. The commutator algebra $[A^-_{n+1}]$ of elements that are odd in the momenta is a Lie subalgebra of $[A_{n+1}]$ for which an identical construction to the above one can be performed. This would lead to the real Lie subalgebra $[\,{\cal C}_{A^-_{n+1}}(\,\mathfrak{sp}(2)\,)\,]_{\mathbb R}$ of the 
off-shell higher-spin algebra, which is called the ``minimal'' off-shell higher-spin algebra and is denoted\footnote{The `o' stands for the degenerate $\mathfrak{o}(1)\cong \{0\}\,$. In other words, the Abelian $\mathfrak{u}(1)$ is \textit{not} a finite-dimensional subalgebra of the minimal higher-spin algebras.} by $\mathfrak{ho}_\infty (\,1|2\,:\,[n-1,2]\,)\,$. Analogously, one would obtain the quotient algebra
$[\,\overline{\cal C}_{A^-_{n+1}}(\,\mathfrak{sp}(2)\,)\,]_{\mathbb R}$ denoted by $\mathfrak{ho}(\,1|2\,:\,[n-1,2]\,)$ and called the ``minimal'' on-shell higher-spin algebra.
\end{remark}

\subsubsection{Algebraic realisation}\label{algdef}

In order to have a more explicit handle on the former higher-spin algebras, one can make use of the Moyal star product calculus (see Appendix \ref{ApMoyal} for more details).
Let $S(X,P)$ be a polynomial of ${\mathbb C}[X^A,P_B]$ and $S_W(\textsc{X},\textsc{P})$ be its Weyl ordered polynomial in $A_{n+1}\,$. 
The set of the images of the three basis vectors of the $\mathfrak{sp}(2)$ subalgebra (\ref{sp2}) under the Wigner map is $\{X^AX_A\,,\, X^AP_A\,,\,P^AP_A\}\,$.

One may easily check that the property that $S_W(\textsc{X},\textsc{P})$ commutes with $\mathfrak{sp}(2)$ reads in terms of $S(X,P)\in{\mathbb C}[X^A,P^B]$ as follows:
\begin{equation}
X^A\frac{\partial S}{\partial P^A}=0\,,\quad X^A\frac{\partial S}{\partial X^A}=P^A\frac{\partial S}{\partial P^A}\,,\quad P^A\frac{\partial S}{\partial X^A}=0\,.
\label{sp2singlet}
\end{equation}
The second equation in (\ref{sp2singlet}) follows as a consistency condition from the first and third equations.
For any analytic function such as the polynomial $S(X,P)\in{\mathbb C}[X^A,P_B]\,$, the first equation of (\ref{sp2singlet}) means that the coefficients in its power expansion are $\mathfrak{gl}(n+1)$-irreducible tensors described by two-row Young\footnote{For an introduction to Young diagrams and their use in representation theory, the reader may look at the section 3 of \cite{BCIV} or the section 4 of \cite{Bekaert:2006py} and references therein.} diagrams, that is to say
\begin{equation}
S(X,P)= \sum\limits_{1\leq t\leq r}\,\frac{1}{t!}\,S_{A_1 B_1\mid A_2 B_2\mid\,\ldots\,\mid A_t B_t\mid A_{t+1}\, \ldots\, A_r}(x)\,
P^{A_1}P^{A_2}\ldots P^{A_r} X^{B_1}\,X^{B_2}\ldots X^{B_t}\,,
\end{equation} 
where the coefficients are antisymmetric in each pair
of indices $(A_m,B_m)\,$. 
The second equation of (\ref{sp2singlet}) says that the respective degrees of homogeneity in $X^A$ and $P^B$ must be equal ($r=t$), thus the coefficients in the power expansion of the polynomial $S\in{\mathbb C}[X^A,P_B]$ are $\mathfrak{gl}(n+1)$-irreducible tensors described by rectangular two-row Young diagrams.
In this sense, for analytic functions $S(X,P)\,$, the third equation in (\ref{sp2singlet}) follows as a consistency condition from the first and second equations.

Finally, one should consider the ideal of ${\cal C}_{_{A_{n+1}}}(\,\mathfrak{sp}(2)\,)$ spanned by the elements $S(X,P)$ that obey to the conditions (\ref{sp2singlet}) and that can be written as sums of Moyal products between one quadratic polynomial in the set $\{X^AX_A\,,\, X^AP_A\,,\,P^AP_A\}$ and a polynomial in ${\mathbb C}[X^A,P_B]\,$. A crucial point is that this Moyal product is equal to the pointwise product of these two polynomials plus a term of lower degree such that both terms satisfy (\ref{sp2singlet}) separately, because the Moyal product is $\mathfrak{sp}(2)$-invariant. Consequently, quotienting by the ideal allows to recursively remove all possible traces in the $\mathfrak{gl}(n+1)$-irreducible coefficients of a polynomial $S(X,P)$ obeying (\ref{sp2singlet}).
\begin{lemma}\label{corAdSCFT}
The centraliser ${\cal C}_{_{A_{n+1}}}(\,\mathfrak{sp}(2)\,)$ of the subspace $\mathfrak{sp}(2)$ in the Weyl algebra $A_{n+1}$ is isomorphic to the subspace of ${\mathbb C}[X^A,P^B]$ of polynomials, the coefficients of which are $\mathfrak{gl}(n+1)$-irreducible tensors described by rectangular two-row Young diagrams, endowed with the Moyal product. The quotient $\overline{\cal C}_{_{A_{n+1}}}(\,\mathfrak{sp}(2)\,)$ of the former algebra by the two-sided ideal ${\cal C}_{_{A_{n+1}}}(\,\mathfrak{sp}(2)\,)\cap A_{n+1}\mathfrak{sp}(2)$ is isomorphic to the subalgebra of the latter where the coefficients are traceless (thus $\mathfrak{o}(n-1,2)$-irreducible) tensors.
\end{lemma}
\begin{corollary}\label{Wproducts}
The complex associative algebra spanned by the Weyl-ordered powers of the generators $\textsc{M}^{AB}=\textsc{X}^{[A}\textsc{P}^{B]}$ of $\mathfrak{o}(n-1,2)$ is isomorphic to
the centraliser ${\cal C}_{A_{n+1}}(\,\mathfrak{sp}(2)\,)$ of the subspace $\mathfrak{sp}(2)$ in $A_{n+1}\,$.
\end{corollary}
\proof{The isomorphism follows directly from the translation of these algebras in terms of Lemma \ref{corAdSCFT}. Indeed, in order to show the bijection it is enough to prove the one-to-one correspondence of the symbols. Any $m$th product of polynomials $X^{[A}P^{B]}$ in ${\mathbb C}[X^A,P^B]$ has coefficients which are $\mathfrak{gl}(n+1)$-irreducible tensors described by a rectangular Young diagram made of two rows of lenght $m\,$. Conversely, any polynomial in ${\mathbb C}[X^A,P^B]$ with coefficients which are $\mathfrak{gl}(n+1)$-irreducible tensors described by a rectangular Young diagram made of two rows of lenght $m$ is equal to a sum of $m$ products of polynomials $X^{[A}P^{B]}\,$.}

The handy reformulation of Eastwood algebras \cite{Eastwood} in terms of star product is due to Vasiliev \cite{Vasiliev:2003ev}. More precisely, it goes as follows:
\begin{corollary}\label{tworowYD}
The off and on shell $AdS_n/$ $CFT_{n-1}$ higher-spin algebras are isomorphic to the subspace of ${\mathbb R}[X^A,P^B]$ of real polynomials, the coefficients of which are $\mathfrak{gl}(n+1)\,$, respectively $\mathfrak{o}(n-1,2)\,$, irreducible tensors described by rectangular two-row Young diagrams, endowed with the Moyal bracket.
\end{corollary}

\subsubsection{Geometric realisation: Conformal}
The (one-sheeted) hyperboloid $X^2=-R^2$ endowed with the induced metric is the $AdS_n$ space-time with curvature radius $R\,$. Its conformal ``boundary at infinity'' is the projective light-cone $X^2=0\,$. More precisely, following \cite{Eastwood}, the manifold ${\mathbb R}^{n-1}$ may be conformally compactified as the sphere $S^{n-1}\subset {\mathbb R}{\mathbb P}_n$ of null directions of the quadratic form $X^2\,$. 
More concretely, any paraboloid, endowed with the induced metric and defined as the intersection between the hypercone $\eta_{AB}X^AX^B=0$ and a hyperplane $X^{n-1}-X^n=$ constant ($\neq 0$), may be identified with a Minkowski space-time ${\mathbb R}^{n-2,1}\,$, with coordinates $X^\alpha$ (with $\alpha=0,1,\ldots, n-2$).

The interest of the ambient formulation is that the conformal transformations of the $(\,n-1)$-dimensional Minkowski space-time are induced by the \textit{linear} action of the pseudo-orthogonal group $O(n-1,2)$ on the $(n+1)$-dimensional ambient space.
\begin{lemma}[Dirac]\label{Dirac}

There is a bijective correspondence between the vector space $C(-w,0)$ of smooth functions $\psi(X^\alpha)$ of conformal weight $w$ on the Minkowski space-time ${\mathbb R}^{n-2,1}$ and the vector space of (equivalence classes of) smooth homogeneous functions $\Psi(X^A)$ on the ambient space ${\mathbb R}^{n-1,2}$ of degree $w$ quotiented by the equivalence relation $\Psi(X)=X^2\, \Theta(X)\sim 0$ with $\Theta(X^C)$ of homogenity degree $w-2\,$. The concrete correspondence is
that $\psi(X^\alpha)$ is the evaluation of any representative $\Psi(X^A)$ at $X^2=0$ and $X^{n-1}-X^n=1\,$.

Moreover, the action of the ambient wave operator $\eta^{AB}\partial_A\partial_B$ on representatives $\Psi(X^C)$ of homogeneity degree $(3-n)/2$ defines the action of the d'Alembertian $\Box=\eta^{\alpha\beta}\partial_\alpha\partial_\beta$ on functions $\psi(X^\gamma)$ of conformal weight $(3-n)/2$.
\end{lemma}
This elegant construction, due to Dirac \cite{Dirac:1936fq}, is nicely reviewed in the section 3 of \cite{Eastwood}.
A function $\psi(X^\alpha)$ of conformal weight $(3-n)/2$ on the Minkowski space-time ${\mathbb R}^{n-2,1}$ is called a ``conformal scalar field.''
A basis of the Lie algebra $\mathfrak{o}(n-1,2)$ of the conformal group of the $(\,n-1)$-dimensional Minkowski space-time is represented in ambient space by the vector fields $X_{[A}\partial_{B]}\,$. The vector space $C(\frac{n-3}2,0)$ of the ``off-shell'' conformal scalar fields on ${\mathbb R}^{n-2,1}$ is an $\mathfrak{o}(n-1,2)$-module. In other words, it is an invariant space carrying a (multiplier) representation of the conformal algebra $\mathfrak{o}(n-1,2)\,$. The subspace $C(\frac{n-3}2,0)\,\cap\,\,$Ker$\Box$ of the ``on-shell'' conformal scalar fields that are solutions to the d'Alembert equation is, roughly\footnote{Strictly speaking, it is the space of \textit{positive energy} solutions that should be considered. This rethoric precaution is also taken because further mathematical conditions are necessary in order to have finite norms, \textit{etc}.} speaking, an irreducible unitary $\mathfrak{o}(n-1,2)$-module called a scalar ``singleton'' (or ``Rac'' for $n=4$) and frequently denoted by $D
 (\frac{n-3}2,0)\,$, where the zero indicates that it originates from a trivial representation of the ``little group.''

A symmetry of the complex on-shell conformal scalar field is a linear differential operator $\textsc{T}$ preserving the space of solutions to the d'Alembert equation $\Box\psi=0\,$. More precisely, $\textsc{T}$ must obey to
\begin{equation}
\Box\,\circ\,\textsc{T}\,=\,\textsc{S}\,\circ\,\Box\,,
\label{symgen}
\end{equation}
for some linear differential operator $\textsc{S}$. These symmetries form a subalgebra of the associative algebra of differential operators.
A symmetry $\textsc{T}$ is ``trivial on-shell'' if $\textsc{T}=\textsc{R}\circ\Box$ for
some linear differential operator $\textsc{R}$. Such an on-shell-trivial symmetry is always a symmetry of the on-shell conformal scalar field,
since it obeys (\ref{symgen}) with $\textsc{S}=\Box\circ\textsc{R}\,$.
The algebra of on-shell-trivial symmetries obviously forms a left ideal in the associative algebra of linear differential operators endowed with the
composition $\circ$ as multiplication. Furthermore, it is also a right ideal in the associative subalgebra of symmetries of the on-shell conformal scalar field.
\begin{lemma}[Eastwood]
For any integer $n>2\,$, the complex associative algebra of symmetries of the on-shell conformal scalar field on the Minkowski space-time ${\mathbb R}^{n-2,1}$ is isomorphic to the centraliser ${\cal C}_{_{A_{n+1}}}(\,\mathfrak{sp}(2)\,)\,$.
The quotient of this algebra by the two-sided ideal of on-shell-trivial symmetries
is isomorphic to the quotient $\overline{\cal C}_{_{A_{n+1}}}(\,\mathfrak{sp}(2)\,)$ of the centraliser.
\end{lemma}
\noindent This is nothing but a reformulation of the theorems 2 and 3 of \cite{Eastwood}. In order to make contact with the previous abstract definition of the Weyl algebra in Subsection \ref{abstractdef} one should perform the identification $\textsc{X}^A\mapsto X^A$ and $\textsc{P}^B\mapsto -i\,{\partial}/{\partial X^B}\,$.

Let $\dagger$ stands for the adjoint with respect to the sesquilinear form $\langle\,\,\,\mid\,\,\,\rangle\,$
on the space $L^2({\mathbb R}^{n-1})$ of square-integrable functions on the Minkowski space-time.
The quadratic action for a (complex) free conformal scalar field $\psi\in L^2({\mathbb R}^{n-1})\cap C(\frac{n-3}2,0)$ can be expressed as a quadratic form
\begin{equation}
S[\psi]\,=\,-\frac12\,\langle\,\psi\mid\Box\mid\psi\,\rangle\,,
\label{quadract}
\end{equation}
where the kinetic operator $\Box$ is self-adjoint, $\Box^\dagger=\Box$.
An (infinitesimal) symmetry of the off-shell conformal scalar field  is defined as a linear differential operator $\textsc{T}$ such that the (finite) transformation $\mid\psi\,\rangle\mapsto\exp(\,i\textsc{T})\mid\psi\,\rangle$ preserves the quadratic action (\ref{quadract}). Equivalently, $\textsc{T}$ must be self-adjoint with respect to the sesquilinear form $\langle\,\,\mid\Box\mid\,\,\rangle\,$. More concretely,
\begin{equation}
\Box\,\circ\,\textsc{T}\,=\,\textsc{T}^\dagger\,\circ\,\Box\,.
\label{offshym}
\end{equation}
The symmetries of the off-shell conformal scalar field form a real Lie algebra endowed with $-i$ times the commutator as Lie bracket.
A linear operator $\textsc{T}=\textsc{R}\circ\Box$ is a symmetry of the quadratic action (\ref{quadract}) if $\textsc{R}$ is self-adjoint. Moreover, the Lie subalgebra of such on-shell-trivial symmetries is an ideal in the real Lie algebra of symmetries of the off-shell conformal scalar field.
\begin{corollary}\label{Ecorol}Let $n\in\mathbb N$ be not smaller than three. 
The quotient of 
the real Lie algebra of symmetries of the complex off-shell conformal scalar field on the Minkowski space-time ${\mathbb R}^{n-2,1}$ by the two-sided ideal of on-shell-trivial symmetries
is isomorphic to the on-shell $AdS_n/CFT_{n-1}$ higher-spin algebra.
\end{corollary}
\proof{Any symmetry $\textsc{T}$ of the off-shell conformal scalar field is always a symmetry of the on-shell conformal scalar field with $\textsc{S}=\textsc{T}^\dagger$ in (\ref{symgen}). Thus one should look for the symmetries which are self-adjoint with respect to the sesquilinear form $\int d^{n-1}X\,\psi^*\,\Box\,\psi\,$. Following Lemma \ref{Dirac}, the ambient image of the ``integrand'' $\psi^*(X^\alpha)\,(\partial^\beta\partial_\beta)\,\psi(X^\alpha)$ is the representative 
$\Psi^*(X^A)\,(\partial^B\partial_B)\,\Psi(X^A)$ of homogeneity degree equal to $1-n\,$. Therefore, its integral over any paraboloid of Lemma \ref{Dirac} does not depend on the choice of such paraboloid. The conformal invariance is manifest in this construction. The corollary is proven by combining Corollary \ref{Wproducts} with the observation in \cite{Mik} that any real Weyl-ordered polynomial
in the symmetries of the off-shell conformal scalar field is itself a symmetry.}

\subsubsection{Geometric realisation: Anti de Sitter}
Another interest of the ambient formulation is that the isometries of the $n$-dimensional anti de Sitter space-time are also induced by the linear action of the pseudo-orthogonal group $O(n-1,2)$ on the ambient space ${\mathbb R}^{n-1,2}\,$. Let $x^\mu$ be coordinates on the hyperboloid $X^2=-R^2$ endowed with the induced metric $g_{\mu\nu}\,$. 
\begin{lemma}[Fr\o nsdal]\label{Fr}
The vector space $\Gamma(\,\otimes^r (TAdS_n)\,)$ of tensor fields of rank $r$ on 
the $n$-dimensional anti de Sitter space-time of curvature radius $R$ is isomorphic to the vector space of tensor fields of rank $r$ on the ambient domain $X^2<0$ that are homogeneous functions of fixed degree quotiented by the subspace of longitudinal tensor fields. The concrete correspondence is that a tensor ${\cal T}_{\mu_1\ldots\mu_r}(x^\nu)$ is the evaluation of the pullback of any representative ${\cal T}_{A_1\ldots A_r}(X^B)$ on the hyperboloid $X^2=-R^2\,$.

Moreover, the action of the (ambient) transverse derivative operator, 
\begin{equation}
\nabla_A\,:=\,\partial_A\,-\,\frac{X_A}{X^2}\,X^B\partial_B\,,
\label{AdSconnection}
\end{equation}
on representatives ${\cal T}_{A_1\ldots A_r}(X^B)$ defines the action of the (anti de Sitter) covariant derivative operator $\nabla_\mu$ on tensors ${\cal T}_{\mu_1\ldots\mu_r}(x^\nu)$.
\end{lemma}
\noindent This very useful construction is explained in details in \cite{Fronsdal:1978vb}. Notice that the ambient transverse metric $G_{AB}:=\eta_{AB}-X_AX_B/X^2$ obviously defines the metric $g_{\mu\nu}$ on $AdS_n\,$. It can be easily checked that the connection defined by (\ref{AdSconnection}) is ``metric,'' $\nabla_A G_{BC}$ is longitudinal, and ``without torsion,'' $\nabla_{[A}\nabla_{B]}\Phi$ is longitudinal for any smooth function on the domain $X^2< 0\,$.

Since $AdS_n$ is a curved manifold, it is necessary to generalise some of the previous definitions to the case of an arbitrary pseudo-Riemannian manifold $\cal M$ of dimension $n$ with coordinates $x^\mu$ and Levi-Civita connection $\nabla\,$. For instance, the d'Alembertian on $\cal M$ may be defined as the Laplace-Beltrami operator $\Box:=\nabla^2=g^{\mu\nu}\nabla_\mu\nabla_\nu\,$.
\begin{corollary}\label{Harmonic}
There is a bijection between the vector space $C^\infty(AdS_n)$ of smooth functions $\phi(x^\mu)$ on the $n$-dimensional anti de Sitter space-time of curvature radius $R$ and the vector space of smooth functions $\Phi(X^A)$ on the ambient domain $X^2<0$ of fixed homogeneity degree. The concrete correspondence is that $\phi(x^\mu)$ is the evaluation of any representative $\Phi(X^A)$ on the hyperboloid $X^2=-R^2\,$.

Moreover, the action of the ambient wave operator $\eta^{AB}\partial_A\partial_B$ on functions $\Phi(X^C)$ on ${\mathbb R}^{n-1,2}$ of homogeneity degree $w$ defines the action of the differential operator $\Box-w(w+n-1)/R^2$ on functions $\phi(x^\mu)$ on $AdS_n\,$.
\end{corollary}
\proof{This is precisely analogous to the construction of spherical harmonics on $S^n$ of fixed degree $\ell$ as (the evaluation of) harmonic polynomials in ambient space ${\mathbb R}^{n+1}$ of homogeneity degree equal to $\ell\,$. The proof is a straightforward computation following the recipe of Lemma \ref{Fr}. More concretely, one should merely check the equality $$G^{AB}\nabla_B\nabla_A\,=\,\eta^{AB}\partial_A\partial_B-\frac{1}{X^2}(X^A\partial_A)(X^B\partial_B+n-1)\,.
$$}

The symbol of the commutator between the d'Alembertian with a differential operator $\textsc{T}$ of order $m$ and of symbol $\textsc{T}^{\mu_1\ldots\mu_m}(x)$ is given by the symmetrised covariant derivative of the symbol of $\textsc{T}\,$:
\begin{equation}
[\,\Box\,\stackrel{\circ}{,}\,\textsc{T}\,]\,=\,2\,\nabla^{(\mu_1}\textsc{T}^{\mu_2\ldots\mu_{m+1})}\,\partial_{\mu_1}\ldots\partial_{\mu_{m+1}}\,+\,\mbox{lower order}\,.
\end{equation}
By definition, a ``Killing tensor field'' of a pseudo-Riemannian manifold $\cal M$ is a symmetric tensor field ${\cal T}_{\mu_1\ldots\mu_m}(x^\nu)$ on $\cal M$ such that its symmetrised covariant derivative is equal to zero, $\nabla_{(\mu_1}{\cal T}_{\mu_2\ldots\mu_{m+1})}=0\,$.
The space ${\cal K}({\cal M})$ of Killing tensors on $\cal M$ is endowed with a commutative graded algebra structure via the symmetric product 
(\ref{symmetric product}) of symmetric tensor fields.
The symbol of any differential operators on $\cal M$ commuting with the d'Alembertian is a Killing tensor field of $\cal M\,$. 
Killing tensor fields on spaces of constant curvature have been extensively studied by mathematicians \cite{Killingconstcurvv,Killingambient}.
\begin{lemma}\label{Killg}Let $\cal M$ be any $n$-dimensional constant curvature space-time.

The associative algebra of differential operators on $\cal M$ that commute with the d'Alembertian is filtered by the order.
The graded algebra associated to this filtered algebra is isomorphic to the commutative algebra ${\cal K}({\cal M})$ of Killing tensor fields on $\cal M$ graded by the rank.
\end{lemma}
\proof{The corresponding restriction of the maps (\ref{isomo}) and (\ref{injection}) shows that there is a bijective correspondence between symbols and symmetric tensor fields. The isomorphism of commutative graded algebras is shown if one may associate to any Killing tensor field a differential operators that commute with the d'Alembertian. Any Killing tensor field on a constant curvature space-time is a sum of symmetric product of Killing vector fields \cite{Killingconstcurvv}. All the corresponding composition products commute with the d'Alembertian, so the lemma is proven.}
Let $\dagger$ stands for the adjoint with respect to the sesquilinear form 
\begin{equation}
\langle\,\phi\,\mid\,\psi\,\rangle\,:=\,\int\limits_{\cal M}d^nx\,\sqrt{-g}\,\phi^*(x)\,\psi(x)
\end{equation}
on the space of square-integrable functions on $\cal M\,$. The d'Alembertian is still Hermitian, $\Box^\dagger=\Box\,$. The quadratic action for a complex free scalar field $\phi$ can be expressed as a quadratic form
\begin{equation}
S[\phi\,]\,=\,-\frac12\,\langle\,\phi\mid(\,\Box-m^2)\mid\phi\,\rangle\,,
\label{quadract2}
\end{equation}
where $m^2$ is a real parameter.
An (infinitesimal) symmetry of the complex off-shell scalar field on $\cal M$ is defined as a linear differential operator $\textsc{T}$ that satisfies the equation
\begin{equation}
(\,\Box-m^2)\,\circ\,\textsc{T}\,=\,\textsc{T}^\dagger\,\circ\,(\,\Box-m^2)\,.
\label{offshym2}
\end{equation}
The symmetries of the off-shell scalar field form a real Lie algebra endowed with $-i$ times the commutator as Lie bracket.
\begin{proposition}\label{offhsAdS}
The real Lie algebra of Hermitian symmetries of the complex off-shell scalar field on the $n$-dimensional anti de Sitter space-time is isomorphic to the off-shell $AdS_n/CFT_{n-1}$ higher-spin algebra.
\end{proposition}
\proof{If a linear differential operator $\textsc{T}$ is Hermitian, $\textsc{T}=\textsc{T}^\dagger\,$, then $\textsc{T}$ is a symmetry of the off-shell scalar field if and only it commutes with the d'Alembertian. Hence, the symbol of any Hermitian symmetry of the complex off-shell scalar field on a manifold $\cal M$ must be a Killing tensor field.
The proposition \ref{offhsAdS} is proven by making use of Lemma \ref{Killg} and by noticing that for any Killing tensor $\textsc{T}^{\mu_1\ldots\mu_m}(x)$ of rank $m$ of $AdS_n$ there exists a canonically defined Hermitian operator $\textsc{T}$ of order $m$ commuting with the d'Alembertian and of symbol equal to $\textsc{T}^{\mu_1\ldots\mu_m}(x)\,$. 

One proof of this property follows exactly the same philosophy than the proof of the theorem 2 of \cite{Eastwood}. Accordingly, it makes use of an ambient construction (here, the one of Lemma \ref{Fr} and Corollary \ref{Harmonic}) for the Killing tensor fields, as given in \cite{Killingambient} and as follows directly from Corollary \ref{Wproducts}, because this leads to a manifest correspondence with an element of the off-shell $AdS_n/CFT_{n-1}$ higher-spin algebra. Another proof is based on the fact that any Killing tensor field on a constant curvature space-time is a sum of symmetric product of Killing vector fields \cite{Killingconstcurvv}. All the corresponding Weyl-ordered composition products are Hermitian and commute with the d'Alembertian, so the proposition is proven.}

\begin{remark} The vector space of \textit{real} scalar fields is only preserved by transformations generated by symmetries with \textit{pure imaginary} coefficients. The above-mentioned correspondence $\textsc{X}^A\mapsto X^A$ and $\textsc{P}^B\mapsto -i\,{\partial}/{\partial X^B}$ implies that one is restricted to the real subspace $A^-_{n+1}$ of the Weyl algebra corresponding to ``symmetric'' differential operators. Accordingly, it is the minimal higher-spin algebras reviewed in Remark \ref{minimalhs} that would appear in the analogue of Corollary \ref{Ecorol} and Proposition \ref{offhsAdS} for a real scalar field.
\end{remark}

\subsection{Minkowski algebra}\label{Mink}

The off-shell Minkowski higher-spin algebra is intuitively understood as an In\"{o}n\"{u}-Wigner contraction of the off-shell $AdS$ higher-spin algebra. Nevertheless, for technical reasons it turns out to be convenient to define the former through a filtration of the latter.

\subsubsection{Abstract definition}

Let $A_n$ be the Weyl algebra presented by the generators $\textsc{X}^a$ and $\textsc{P}^b$ ($a,b=0,1,2,\ldots,n-1$)
modulo the corresponding subset of the commutation relations.
The centraliser ${\cal C}_{_{A_n}}(\,\textsc{P}^2\,)$ of the quadratic element $\textsc{P}^2:=\textsc{P}^a\textsc{P}_a$ is a subalgebra of $A_n\,$.
This space may be endowed with a Lie algebra structure $[\,{\cal C}_{_{A_n}}(\,\textsc{P}^2\,)\,]$ via the bracket $-i\,[\,\,\,,\,]\,$.
The real vector space spanned by the self-adjoint elements of the centraliser of $\textsc{P}^2$ in the Weyl algebra $A_n\,$, endowed with the bracket $-i\,[\,\,\,,\,]\,$, is a real form of the complex Lie algebra $[\,{\cal C}_{_{A_n}}(\,\textsc{P}^2\,)\,]\,$. This real Lie algebra is called the ``off-shell Minkowski higher-spin algebra'' \cite{Vasiliev:2005zu,Bekaert:2005ka}. 
The complex Lie subalgebra of the higher-spin algebras that is spanned by the elements $\textsc{P}^a$ and $\textsc{M}^{ab}:=\textsc{X}^{[a}\textsc{P}^{b]}$ 
is isomorphic to the Poincar\'e algebra $\mathfrak{io}(n-1,1)\,$, of which the off-shell Minkowski higher-spin algebra is an infinite-dimensional extension. In order to perform the link with the off-shell $AdS/CFT$ higher-spin algebra and thereby providing an ambient construction, some lemmas are needed.

The Weyl algebra $A_{n+1}$ in the generators $\textsc{X}^A$ and $\textsc{P}^B$ ($A,B=0,1,2,\ldots,n-1,n$) is filtered by the polynomial degree in the generator $\textsc{X}^n\,$. The graded algebra associated to this filtration is related to the Weyl algebra $A_n\,$.
\begin{lemma}\label{filterWeyl}
The Weyl algebra $A_{n+1}$ in the generators $\textsc{X}^A$ and $\textsc{P}^B$ is filtered by the polynomial degree in the generator $\textsc{X}^n\,$. The graded algebra $gr(A_{n+1})$ associated to this filtration is isomorphic to the direct product $A_n\otimes {\mathbb C}[X^n,P^n]$ of the Weyl algebra $A_n$ in the generators $\textsc{X}^a$ and $\textsc{P}^b$ with the polynomial algebra ${\mathbb C}[\textsc{X}^n,P^n]\,$:
\begin{equation}
gr(A_{n+1})\,\cong\, A_n\otimes {\mathbb C}[X^n,P^n]\,.
\label{isofilt}
\end{equation}

Moreover, the centraliser ${\cal C}_{gr(A_{n+1})}(\,\mathfrak{sp}(2)\,)$ of the subspace $\mathfrak{sp}(2)$ in the graded algebra associated to the filtration of $A_{n+1}$  is isomorphic to the direct product ${\cal C}_{_{A_n}}(\,\textsc{P}^2\,)\otimes {\mathbb C}[X^n,P^n]$ of the centraliser ${\cal C}_{_{A_n}}(\,\textsc{P}^2\,)$ of the element $\textsc{P}^2$ in $A_n$ with the algebra ${\mathbb C}[\textsc{X}^n,P^n]\,$.
\end{lemma}
\noindent It should be stressed that the subspace (\ref{sp2}) inside the graded associative algebra $gr(A_{n+1})$ is \textit{not} a Lie subalgebra of the corresponding commutator algebra  $[\,gr(A_{n+1})\,]$, though the notation $\mathfrak{sp}(2)$ is kept in order to remind the reader of its origin.  
\proof{Let $P \in A_{n+1}$ be a polynomial of degree $p$ in $\textsc{X}^n$, $$P(\textsc{X}^A,\textsc{P}^B)\,=\,(\textsc{X}^n)^p\,\,R(\textsc{X}^a,\textsc{P}^A)\,+\,\mbox{lower}\,.$$
Its representative in $gr_p(A_{n+1})$ may be taken to be the polynomial $R(\textsc{X}^a,\textsc{P}^A)\,$.
This proves the isomorphism (\ref{isofilt}) of vector spaces.

Let $Q(\textsc{X}^A,\textsc{P}^B)$ be a second polynomial of the Weyl algebra $A_{n+1}$ of degree $q$ in $\textsc{X}^n\,$,
$$Q(\textsc{X}^A,\textsc{P}^B)\,=\,(\textsc{X}^n)^q\,\,S(\textsc{X}^a,\textsc{P}^A)\,+\,\mbox{lower}\,.$$
It is clear that the product of the two polynomials $P$ and $Q$ is of degree $p+q$ in $\textsc{X}^n\,$,
$$P(\textsc{X}^A,\textsc{P}^B)\,Q(\textsc{X}^A,\textsc{P}^B)\,=\,(\textsc{X}^n)^{p+q}\,\,T(\textsc{X}^a,\textsc{P}^A)\,+\,\mbox{lower}\,,$$
where the representative $T(\textsc{X}^a,\textsc{P}^A)$ is obtained through the product of the two representatives $R$ and $S$ without taking into account the commutation relation $[\,\textsc{X}^n\,,\,\textsc{P}^n]=i\,$.
This proves the isomorphism (\ref{isofilt}) of associative algebras.

The second part of the lemma is proven by considering the representatives of the basis elements of the $\mathfrak{sp}(2)$ subalgebra (\ref{sp2}) of $[A_{n+1}]$: they are, modulo signs, the elements $(\textsc{X}^n)^2\,$, $\textsc{X}^n\textsc{P}_n$ and $\textsc{P}^A\textsc{P}_A\,$. The corresponding elements in $A_n\otimes {\mathbb C}[X^n,P^n]$ are (up to signs) the polynomials $(X^n)^2\,$, $X^nP_n$ and $\textsc{P}^a\textsc{P}_a-(P^n)^2\,$. The first two elements belong to the commutative subalgebra ${\mathbb C}[X^n,P^n]\,$, thus they are central. In conclusion, the only non-trivial commutation condition for ${\cal C}_{gr(A_{n+1})}(\,\mathfrak{sp}(2)$ is with the third element. So the lemma \ref{filterWeyl} is shown because only the part $\textsc{P}^a\textsc{P}_a$ implies a non-trivial commutation condition.}

\begin{remark}\label{remB2}
The Weyl algebra $A_n$ is itself filtered by another degree: the polynomial degree in the ``positions'' $\textsc{X}^a\,$. The degree defined by substracting one to this degree in the positions filters the commutator algebra $[A_n]\,$. Hence, the elements of $A_n$ that are at most of degree one in the generators $\textsc{X}^a$ span a Lie subalgebra of $[A_n]\,$. Accordingly, the elements of the off-shell Minkowski higher-spin algebra that are at most linear in $\textsc{X}^a$ span a Lie subalgebra, which was introduced in the first attempt \cite{Bengtsson:1986bz} of constructing a non-Abelian theory corresponding to the frame-like formalism of \cite{Vframe80}.
\end{remark}

\subsubsection{Algebraic realisation}\label{algreal}

Again one may easily check that the property that the Weyl-ordered polynomial $S_W(\textsc{X}^a,\textsc{P}^b)\in A_n$ commutes with $\textsc{P}^a\textsc{P}_a$ reads in terms of $S(X,P)\in{\mathbb C}[X^a,P^b]$ as follows:
\begin{equation}
P^a\frac{\partial S}{\partial X^a}=0\,.
\label{Minksp}
\end{equation}
For any analytic function such as the polynomial $S(X,P)\in{\mathbb C}[X^a,P^b]\,$, it means that the coefficients in its power expansion are $\mathfrak{gl}(n)$-irreducible tensors described by two-row Young diagrams.
\begin{lemma}\label{centralfilt}
The centraliser ${\cal C}_{A_{n+1}}(\,\mathfrak{sp}(2)\,)$ of $\mathfrak{sp}(2)$ in the Weyl algebra $A_{n+1}$ in the generators $\textsc{X}^A$ and $\textsc{P}^B$ is filtered by the polynomial degree in the generator $\textsc{X}^n\,$. The following three algebras are isomorphic:
\begin{itemize}
	\item[$\bullet$] The graded algebra $gr\Big({\cal C}_{A_{n+1}}(\,\mathfrak{sp}(2)\,)\Big)$ associated to this filtration.
	\item[$\bullet$] The subspace of the algebra ${\mathbb C}[X^a,P^b]$ spanned by all polynomials, the coefficients of which are $\mathfrak{gl}(n)$-irreducible tensors described by two-row Young diagrams, endowed with the Moyal product.
  \item[$\bullet$] The centraliser ${\cal C}_{_{A_n}}(\,\textsc{P}^2\,)$ of the element $\textsc{P}^a\textsc{P}_a$ in the Weyl algebra $A_n$ in the generators $\textsc{X}^a$ and $\textsc{P}^b\,$.
\end{itemize}
\end{lemma}
\proof{The lemma \ref{corAdSCFT} states that the polynomial of ${\mathbb C}[X^A,P^B]$ associated to any Weyl-ordered polynomial of ${\cal C}_{_{A_{n+1}}}(\,\mathfrak{sp}(2)\,)$ has coefficients which are $\mathfrak{gl}(n+1)$-irreducible tensors described by rectangular two-row Young diagrams. The branching rules of the restriction of $GL(n+1)$ to $GL(n)$ implies that the leading part in the variable $X^n$ of any such polynomial (i) is independent of the variable $P^n\,$, (ii) has coefficients which are $\mathfrak{gl}(n)$-irreducible tensors described by two-row Young diagrams, and (iii) has homogeneity degree in $X^n$ equal to the difference between the respective homogeneity degrees in $P^b$ and $X^a$. 
Following the procedure of the proof of Lemma \ref{filterWeyl}, the significant part in the representative of any basis element of $gr\big(\,{\cal C}_{_{A_{n+1}}}(\,\mathfrak{sp}(2)\,)\,\big)$ is a polynomial in ${\mathbb C}[X^a,P^b]$ with coefficients which are $\mathfrak{gl}(n)$-irreducible tensors described by two-row Young diagrams.}
\begin{corollary}\label{coeffgln}
The off-shell Minkowski higher-spin algebra in $n$ dimensions is isomorphic to the subspace of ${\mathbb R}[X^a,P^b]$ of real polynomials, the coefficients of which are $\mathfrak{gl}(n)$-irreducible tensors described by two-row Young diagrams, endowed with the Moyal bracket.
\end{corollary}
\noindent Although the previous corollary follows directly from the condition (\ref{Minksp}), it may also be understood from the filtration of the off-shell $AdS/CFT$ higher-spin algebra via Lemma \ref{centralfilt}:
\begin{proposition}The off-shell $AdS/CFT$ higher-spin algebra is filtered by the polynomial degree in one of the time-like generator. The graded algebra associated to this filtration is isomorphic to the off-shell Minkowski higher-spin algebra.
\end{proposition}
Another useful corollary of Lemma \ref{centralfilt} is:
\begin{corollary}\label{WproductsM}
The complex associative algebra spanned by the Weyl-ordered powers of the generators $\textsc{P}^a$ and $\textsc{M}^{ab}=\textsc{X}^{[a}\textsc{P}^{b]}$ of $\mathfrak{io}(n-1,1)$ is isomorphic to
the centraliser ${\cal C}_{A_n}(\,\textsc{P}^2\,)$ of the element $\textsc{P}^2$ in $A_n\,$.
\end{corollary}
\proof{The isomorphism follows directly from the translation of both algebras in terms of Lemma \ref{centralfilt}. Indeed, any $m$th power in the variable $P^a$ times any $p$th power in the polynomial $X^{[a}P^{b]}$ in ${\mathbb C}[X^a,P^b]$ has coefficients which are $\mathfrak{gl}(n)$-irreducible tensors described by a Young diagram made of two rows, the first one of lenght $m+p$ and the second one of length $p\,$. A similar proof was already given in \cite{Bekaert:2005ka}.}

\begin{remark}\label{traceconstr} The In\"{o}n\"{u}-Wigner contraction of the \textit{on}-shell $AdS$ higher-spin algebra is not so easy to perform because it involves the subtle\footnote{Notice that the representatives of the basis vectors of $\mathfrak{sp}(2)$ are (up to signs) the polynomials $(X^n)^2\,$, $X^nP_n$ and $P^aP_a-(P^n)^2\,$. Firstly, some of them explicitly depend on $P^n$ and this is tricky to reconcile with the approach of Lemma \ref{centralfilt}. Secondly, a naive factorisation of all terms proportional to $(X^n)^2$ seem to lead to the algebra mentioned in Remark \ref{remB2}. Thirdly and most importantly, the on-shell $AdS$ higher-spin algebra is not filtered because the degree in $X^n$ is not well defined for equivalence classes because it depends on the choice of representative.} factorisation of an ideal and the related choice of trace conditions. Till now, similar technical problems seem to prevent a satisfactory definition of an on-shell Minkowski higher-spin algebra which, in the light of Corollary \ref{Ecorol}, might be related to the property that the singleton module of the $AdS_n$ isometry algebra does not admit a flat space-time limit. All these obstacles are 
 the main reasons why constrained higher-spin theories are not covered here.
\end{remark}

\subsubsection{Geometric realisation}

The geometric realisation of the off-shell $AdS_n/CFT_{n-1}$ higher-spin algebra from Proposition \ref{offhsAdS} indicates that there must exist an In\"{o}n\"{u}-Wigner contraction via the flat space-time limit $R\rightarrow\infty\,$. The quotient in the previous subsection is nothing but the algebraic translation of the geometrical property that, in a compact neighborhood of the point $(X^a,X^n)=(0,R)\,$, the hyperboloid $X^AX_A=-R^2$ ``ressembles'' to the hyperplane $X^n=R$ when $R\rightarrow\infty\,$. Moreover, in this limit $\partial/\partial X^n\sim 1/R\,$. The filtration corresponds to the fact that it is the term with the highest power of $X^n\sim R$ which ``dominates''. 

Actually, Lemmas \ref{filterWeyl} and \ref{centralfilt} imply that the elements of the centraliser ${\cal C}_{_{A_n}}(\,\textsc{P}^2\,)$ may be obtained from the elements of the centraliser ${\cal C}_{_{A_{n+1}}}(\,\mathfrak{sp}(2)\,)$ by evaluating them at $X^n=1$ and $P^n=0\,$.
\begin{proposition}\label{offhsMink}
The real Lie algebra of Hermitian symmetries of the complex off-shell scalar field on the Minkowski space-time is isomorphic to the off-shell Minkowski higher-spin algebra.
\end{proposition}
\noindent The truth of this property follows geometrically as the flat space-time limit of Proposition \ref{offhsAdS}. A more direct and rigorous proof of this proposition goes exactly along the same lines than the one of Proposition \ref{offhsAdS} by making use of Corollary \ref{WproductsM}. To some extent, Property \ref{offhsMink} is a mere reformulation of the definition itself.

\subsection{Lorentz algebra}\label{Lorentz}

The various definitions of Subsection \ref{AdSCFT} for the Weyl algebra $A_{n+1}$ presented by the generators $\textsc{X}^A$ and $\textsc{P}^B$ modulo the commutation relations (\ref{Heisenbergambient}) can be applied to the case of the Weyl subalgebra $A_n$ generated by $\textsc{X}^a$ and $\textsc{P}^b$ by considering the complex Lie subalgebra of the commutator algebra $[A_n]$ that is spanned by the three elements $\textsc{X}^a\textsc{X}_a$, $\frac12(\textsc{X}^a\textsc{P}_a+\textsc{P}^a\textsc{X}_a)$ and $\textsc{P}^a\textsc{P}_a$ and which
is isomorphic to the classical Lie algebra $\mathfrak{sp}(2)\,$.
The real vector space spanned by the self-adjoint elements of the centraliser of this $\mathfrak{sp}(2)$ subalgebra in $A_n$ endowed with a Lie algebra structure via the bracket $-i\,[\,\,\,,\,]$ is a real form of the complex Lie algebra 
$[\,{\cal C}_{_{A_n}}(\,\mathfrak{sp}(2)\,)\,]$ which should be denoted by $\mathfrak{hu}_\infty (\,1|2\,:\,[n-1,1]\,)$ and which will be called ``off-shell Lorentz higher-spin algebra''. Its name originates from the corollary \ref{Wproducts} in the sense that the complex associative algebra spanned by the Weyl-ordered powers of the basis elements $\textsc{M}^{ab}=\textsc{X}^{[a}\textsc{P}^{b]}$ of the Lorentz algebra $\mathfrak{o}(n-1,1)$ is isomorphic to
the centraliser ${\cal C}_{A_{n}}(\,\mathfrak{sp}(2)\,)\,$. 
Analogously, the real Lie algebra $\mathfrak{hu}(\,1|2\,:\,[n-1,1]\,)$ will be called ``on-shell Lorentz higher-spin algebra''.
The Corollary \ref{tworowYD} implies that the on(off)-shell Lorentz higher-spin algebra is isomorphic to the subspace of ${\mathbb R}[X^a,P^b]$ of real polynomials, the coefficients of which are $\mathfrak{gl}(n)\,$, respectively $\mathfrak{o}(n-1,1)\,$,  irreducible tensors described by rectangular two-row Young diagrams, endowed with the Moyal bracket.
By construction, the following proposition should be clear:
\begin{proposition}
The Lorentz higher-spin algebras are subalgebras of the (corresponding) $AdS/CFT$ and Minkowski higher-spin algebras. 
\end{proposition}
The interest of this proposition is that it enables to formulate in a unified fashion the torsion constraints in higher-spin gauge field theories.

\section{Frame-like formulation}\label{frameformul}

The modern view\footnote{For instance, the leitmotiv behind the ``non-commutative geometry'' programme is the Gelfand representation theorem which
emphasises the dual role of the commutative $C^\ast$ algebra of functions on the manifold.}
on the intimate relation between algebra and geometry shifts the focus from the manifold itself to its ``dual,'' the vector space of functions on the manifold. Symmetry transformations may be characterised by their action on the coordinates. Looking at the action of the symmetry group on the dual space, a smooth change of coordinates is generated by a first-order linear differential operator.
Therefore, a higher-order linear differential operator does \textit{not} generate coordinate transformations.
For instance, an isometry generator is a first-order linear differential operator corresponding to a Killing vector field.
But the higher-derivative symmetries discussed in the previous section are powers of such isometry generators, that is, higher-order linear differential operators. Thus they do not generate coordinate transformations and so they should fit in some sort of ``generalisation'' of the Erlangen programme, as discussed in Subsection \ref{geometricp}. The corresponding Cartan-like formulation reproduces the frame-like formulation of higher-spin gauge fields, reviewed accordingly in Subsection \ref{conncurvconstr}. The unconstrained theory linearised around a flat background is discussed in more details in Subsection \ref{framelggtransf}.

\subsection{Geometric perspective}\label{geometricp}

On purely esthetic ground, one may desire to try to reformulate the known frame-like higher-spin constructions along the lines of Cartan's generalisation of Klein's programme.
To start with, one should look for the analogue of Klein's view of homogeneous geometries. The off-shell\footnote{This discussion should allow a proper generalisation for the on-shell higher-spin algebras as well, by considering unitary modules of on-shell scalar fields. The reasons behind the restriction to the off-shell case were explained in Remark \ref{traceconstr}.} higher-spin algebras discussed in the previous sections suggest a possible generalisation: it is natural to keep considering a quotient space $G/H$ but one should now focus on the Hilbert space of square-integrable functions on $G/H\,$. The geometrical concept of (infinitesimal) displacements should be extended to include (higher-derivative) Hermitian operators. 

The vector space $C^\infty(G/H)$ of smooth functions on the homogeneous space $G/H$ is a module of the Lie algebra $\mathfrak{g}$ of infinitesimal symmetries of the homogeneous space. Therefore, $C^\infty(G/H)$ is a module of the universal enveloping algebra ${\cal U}(\mathfrak{g})\,$.\footnote{An introduction to universal enveloping algebras for physicists is provided in \cite{X}.} The elements of the Lie algebra $\mathfrak{g}$ are realised on $C^\infty(G/H)$ as vector fields of $\Gamma(\,T(G/H)\,)\,$, hence the associative algebra ${\cal U}(\mathfrak{g})$ algebra is realised on $C^\infty(G/H)$ as linear differential operators. The quotient of the universal enveloping algebra ${\cal U}(\mathfrak{g})$ by the annihilator $Ann(\,C^\infty(G/H)\,)$ of the ${\cal U}(\mathfrak{g})$-module $C^\infty(G/H)$ is isomorphic to this realisation in terms of differential operators.
The $\mathfrak{g}$-submodule $L^2(G/H)$ of square-integrable functions on $G/H$ is unitary. Correspondingly, this selects a real form of the commutator algebra $[\,{\cal U}(\mathfrak{g})\,]$ such that $L^2(G/H)$ is a unitary module of this real Lie algebra. Concretely, one may define an (off-shell) ``higher-spin algebra for a homogeneous geometry'' as the real Lie algebra of Hermitian differential operators spanned by the symmetrised products of the vector fields realising the Lie algebra $\mathfrak{g}$ on $C^\infty(G/H)\,$. This realisation is isomorphic to a real form of the quotient ${\cal U}(\mathfrak{g})/Ann(\,C^\infty(G/H)\,)$ endowed with the commutator bracket.
This quotient can be computed explicitly when the symmetry group $G$ is actually a matrix group, because its elements may be realised as vector fields at most linear in some Cartesian coordinates.
For instance, the corollaries \ref{Wproducts} and \ref{WproductsM} imply the following result:
\begin{proposition}
The quotient of the universal enveloping algebra ${\cal U}(\mathfrak{g})$ of the Lie group $G$ by the annihilator $Ann(\,C^\infty(G/H)\,)$ of its realisation on the space of smooth functions on the coset space $G/H$ is isomorphic to the associative algebra
\begin{itemize}
	\item ${\cal C}_{A_{n+1}}(\,\mathfrak{sp}(2)\,)$ for $G=O(n-1,2)$ and $H=O(n-1,1)\,$,
	\item ${\cal C}_{A_n}(\,\textsc{P}^2\,)$ for $G=IO(n-1,1)$ and $H=O(n-1,1)\,$,
\end{itemize}
\end{proposition}
\noindent The corresponding ($AdS/CFT$ and Minkowski) higher-spin algebras defined in the previous section therefore agree with the alternative definition of higher-spin algebra (for a homogeneous geometry) presented in the present section. 

One may also define the ``isotropy higher-spin subalgebra for a homogeneous geometry'' as the real Lie algebra of Hermitian differential operators spanned by the symmetrised products of the vector fields realising the Lie subalgebra $\mathfrak{h}$ on $C^\infty(G/H)\,$. Analogously, this realisation is isomorphic to a real form of the quotient ${\cal U}(\mathfrak{h})/Ann(\,C^\infty(G/H)\,)$ endowed with the commutator bracket. In the previous cases, it is isomorphic to the Lorentz higher-spin algebra.

\subsection{Connections, curvatures and constraints}\label{conncurvconstr}

The main point of Cartan's generalisation of Klein's view on geometry was the combination of two ingredients: (i) a one-form taking values in the symmetry algebra and containing the solder form, and (ii) the vielbein postulates and torsion constraints on the Cartan connection. 
It is fair to say that the analogue of the first ingredient is nicely suggested by the higher-spin algebras, but the geometrical meaning of the second ingredient remains mysterious for higher-spins. Physically, it comes from the requirement that unphysical degrees of freedom should be removed in order to make contact with the metric-like formulation.

Following the decisive observation of Fradkin and Vasiliev \cite{Fradkin:1987ks,Vframe87}, it is suggestive to generalise the Cartan connection to higher-spin gauge fields by considering a one-form $\cal A$ taking values in a higher-spin (super)algebra and containing the genuine Cartan connection.
In the particular case of the (un)constrained frame-like formulation one considers that $\cal A$ takes values in the (off)on-shell higher-spin algebra. Concretely, this is conveniently realised by using the algebraic realisation of the corresponding Weyl algebra, \textit{i.e.} the one-form reads ${\cal A}=dx^\mu{\cal A}_\mu(x,X,P)$ and satisfies some algebraic conditions in the dependence on the auxiliary variables $X$ and $P$ (see the conditions imposed in the correspond subsection ``algebraic realisation''). These conditions are somehow the generalisation of the second vielbein postulate which implies that the Ehresmann connection one-form $\omega^{ab}$ is antisymmetric, \textit{i.e.} it is the spin connection.
The Moyal product on the polynomial space in the capital letters $X$ and $P$ is denoted by a big star {\footnotesize$\bigstar$}\,.
The curvature of the Cartan-like connection $\cal A$ is defined by ${\cal F}:=d{\cal A}+{\cal A}\,\mbox{\footnotesize$\bigstar$}\,{\cal A}\,$. 

The generalisation of the torsion constraint could be the requirement that the two-form ${\cal F}$ is of homogeneity degree in $X^a$ equal to the homogeneity degree in $P^b\,$, that is to say, in the case of constant-curvature space-time algebras that it takes values in the Lorentz higher-spin subalgebra (defined in Subsection \ref{Lorentz}). More generally, a Cartan-like connection $\cal A$ could be called ``torsionless'' if its curvature $\cal F$ takes values in the isotropy higher-spin subalgebra. The Cartan-like connection should presumably extend the concept of parallel transport of points in the model space to the case of functions on it. In this perpective, a torsionless Cartan-like connection might be necessary in order to identify locally the manifold with the model space and/or their dual spaces. The genuine geometrical interpretation of the higher-spin algebra valued one-form $\cal A$ remains elusive and deserves further study.

Let the exterior covariant-like derivative be denoted by
$${\cal D}:=d+i\,[{\cal A}\stackrel{\bigstar}{,}\,\,]_{\pm}$$
where $[\,\,\stackrel{\bigstar}{,}\,\,]_{\pm}$ is the graded star-commutator.
Under the infinitesimal transformations 
\begin{equation}
\delta_\epsilon{\cal A}={\cal D}\,\epsilon\,,
\label{adjointgtransf}
\end{equation}
where $\epsilon(x,X,P)$ takes values in the corresponding higher-spin algebra, the curvature two-form $\cal F$ transforms under the adjoint action of the higher-spin algebra,
$\delta_\varepsilon{\cal F}=i\,[{\cal F}\stackrel{\bigstar}{,}\varepsilon]\,$.
Let $\stackrel{(0)}{\cal A}$ be a flat background in the sense that $\stackrel{(0)}{\cal F}=0\,$, and $\Omega$ the perturbation in the sense that \begin{equation}
{\cal A}=\,\stackrel{(0)}{\cal A}+\,\Omega\,.
\label{aomega}
\end{equation}
The exterior covariant differential with respect to the flat background is denoted by
$\stackrel{(0)}{\cal D}:=d+i\,[\stackrel{(0)}{\cal A}\,\stackrel{\bigstar}{,}\,\,]_{\pm}\,$. Then ${\cal F}=\,\stackrel{(0)}{\cal D}\Omega+\Omega\,\mbox{\footnotesize$\bigstar$}\,\Omega\,$, so that the linearised curvature two-form $\stackrel{(0)}{\cal F}:=\,\stackrel{(0)}{\cal D}\Omega$ is invariant under the linearised transformations
\begin{equation}
\stackrel{(0)}{\delta_\epsilon}\Omega(x,X,P)\,=\,\,\stackrel{(0)}{\cal D}\epsilon(x,X,P)\,.
\label{frameabeliangaugetransfo}
\end{equation}
This property implies that the torsion constraint is preserved at lowest order under gauge transformations where the parameter takes any value in the higher-spin algebra. 

Since free higher-spin gauge fields are known to propagate consistently on constant-curvature space-times, it is natural to assume that the background for $\cal A$ corresponds to one of this space-time (like in the spin-two case). 
The gauge theory of an infinite tower of free gauge fields including all integer spins with multiplicity one is recovered by imposing the torsion constraints at linearised order in the perturbation, as is shown in the next section for the simpler case of Minkowski space-time.
An open issue is the status of the torsion constraints at higher order in the perturbation since they are preserved only by a subgroup of the non-Abelian transformations (\ref{adjointgtransf}). As mentioned above, the torsion constraints require the curvature to take values in the Lorentz higher-spin algebra. This requirement is not preserved by the adjoint action of the whole higher-spin algebra because the Lorentz higher-spin algebra is not an ideal. Two natural ways of circumventing this problem arise: either (1) one deforms the gauge symmetries \textit{and} the torsion constraints in such a way that they remain compatible with a natural adjoint action of the higher-spin algebra, or (2) one keeps the same definition of the torsion constraints but one faces the fact that the gauge symmetries are deformations of the natural adjoint action of the entire higher-spin algebra.

``Breaking'' part of the undeformed gauge symmetries at higher order is dangerous because it would lead to the propagation of unphysical degrees of freedom. Actually, this issue is very subtle since, in the example of gravity, local translations (\textit{i.e.} translations on the tangent space) are somehow exchanged with diffeomorphisms (\textit{i.e.} translations on the base space) precisely with the help of the torsion constraint.
Although the ``doubling of oscillators'' by Vasiliev, which has been so successful, seems to correspond to the first way of circumventing the problem, the analogy with gravity would suggest to investigate the second possibility by looking for some (maybe higher-derivative) analogue of the diffeomorphisms.
The present paper is not conclusive on this issue because only the linearised torsion constraints are used. Hence the analysis does not discriminate between both possibilities. Nevertheless, a suggestive generalisation of the diffeomorphisms can already be obtained at this order, as mentioned in the introduction.

\subsection{Minkowski background and Abelian gauge transformations}\label{framelggtransf}

Although the Abelian frame-like theory was initially developed with trace constraints and around an (anti) de Sitter background \cite{Vframe87}, only the unconstrained frame-like formulation around Minkowski space-time will be discussed here for the sake of simplicity.

Since the spin connection vanishes and the coframe is the identity for the Minkowski background, it corresponds to the one-form
\begin{equation}
\stackrel{(0)}{\cal A}\,=\,dx^\mu P_\mu
\label{A0}
\end{equation}
where, from now on, one does not distinguish between world and tangent indices, since one works around Minkowski space-time. Consequently, the exterior covariant differential with respect to the flat background acts as 
\begin{equation}
\stackrel{(0)}{\cal D}\,=d+dx^\mu\partial/\partial X^\mu=dx^\mu\Big(\frac{\partial}{\partial x^{\mu}}+\frac{\partial}{\partial X^{\mu}}\Big)\,.
\label{D0} 
\end{equation} 

The unconstrained frame-like formulation corresponds to a one-form $\Omega(x,X,P)$ taking values in the off-shell Minkowski higher-spin algebra defined in Subsection \ref{Mink}. Let $\cal Y$ denote the projector on the off-shell Minkowski higher-spin subalgebra. In the spirit of Corollary \ref{coeffgln}, the operator $\cal Y$ could be called the ``Young projector.'' The algebraic condition (\ref{Minksp}) that the one-form $\Omega(x,X,P)$ must obey can be written as
\begin{equation}
{\cal Y}\Omega=\Omega\quad\Longleftrightarrow\quad P^a\frac{\partial \Omega}{\partial X^a}=0\,.
\label{Minkspcd}
\end{equation}
The general solution of (\ref{Minkspcd}) takes the form
\begin{eqnarray}
&&\Omega(x,X,P)=e+\omega\,,\nonumber\\
e&:=&\Omega(x,X=0,P)=\sum\limits_r \,\frac{1}{r!}\,e^{a_1 a_2\,\ldots a_r}(x)\,P_{a_1}P_{a_2}\ldots P_{a_r}\,,\nonumber\\
\omega&:=& \sum\limits_{1\leq t\leq r}\,\frac{1}{r!\,t!}\,\omega^{a_1 b_1\mid a_2 b_2\mid\,\ldots\,\mid a_t b_t\mid a_{t+1}\, \ldots\, a_r}(x)\,
P_{a_1}P_{a_2}\ldots P_{a_r} X_{b_1}\,X_{b_2}\ldots X_{b_t}\,,
\label{powerexpomega}
\end{eqnarray} 
where the one-forms $\omega_{a_1 b_1\mid\,\ldots\,\mid a_t b_t\mid a_{t+1}\mid\, \ldots\,\mid a_r}$ are antisymmetric in each pair
of indices $(a_m,b_m)$ so that, without loss of generality, the coefficients may be taken to be $\mathfrak{gl}(n)$-irreducible tensors described by two-row Young diagrams, as stated in Corollary \ref{coeffgln}. 
The one-forms $e^{a_1 \ldots a_r}$ are the generalisations of the $U(1)$ connection ($r=0$) and of the coframe ($r=1$) for higher-spins ($r\geq 2$) in which context they are called the ``frame-like'' fields, while $\omega^{a_1 b_1\mid\,\ldots\,\mid a_t b_t\mid a_{t+1}\mid\, \ldots\,\mid a_r}$ ($1\leq t\leq r$) generalise the spin connection ($t=r=1$) and will be called ``extra''\footnote{In the case of the constrained frame-like formulation, a further distinction for the auxilliary fields is sometimes drawn between the ``Lorentz-like connections'' ($t=1$) and the other ``extra'' fields.} fields of order $t\,$.

The torsion constraint is equivalent to the following algebraic condition
\begin{equation}
X^a\frac{\partial \cal F}{\partial P^a}=0\,,
\label{torsionHS}
\end{equation}
since, together, (\ref{Minkspcd}) and (\ref{torsionHS}) imply the consistency conditions
\begin{equation}
P^a\frac{\partial {\cal F}}{\partial X^a}=0\,,\quad X^a\frac{\partial \cal F}{\partial X^a}=P^a\frac{\partial \cal F}{\partial P^a}\,. 
\end{equation}
Therefore, following the discussion of Subsection \ref{Lorentz}, the two-form $\cal F$ indeed takes values in the Lorentz higher-spin algebra only:
\begin{equation}
{\cal F}= \sum\limits_{r}\,\frac{1}{r!}\,{\cal F}^{a_1 b_1\mid a_2 b_2\mid\,\ldots\,\mid a_r b_r}(x)\,
P_{a_1}P_{a_2}\ldots P_{a_r} X_{b_1}\,X_{b_2}\ldots X_{b_r}\,.
\label{tctrt}
\end{equation}
The components of the linearised curvature two-form $\stackrel{(0)}{\cal F}=\frac12 \stackrel{(0)}{\cal F}_{\mu\nu}dx^\mu dx^\nu$ read
\begin{equation}
\stackrel{(0)}{\cal F}_{\mu\nu}=\Big(\frac{\partial}{\partial x^{[\mu}}+\frac{\partial}{\partial X^{[\mu}}\Big)\Omega^{}_{\nu]}(x,X,P)\,.
\label{tcstrt}
\end{equation}
In the light of the development (\ref{powerexpomega}), the linearisation of the torsion constraint (\ref{tctrt}) allows to express all extra fields of order $t$ as linear combinations of $t$ partial derivatives of the coframe, as is reviewed in the next section. Of course, the name of the extra fields arise from the property that they are only auxiliary fields. In conclusion only the frame-like one-form $e$ contains some physical degrees of freedom.
Nevertheless, the following gauge arbitrariness remains from (\ref{frameabeliangaugetransfo}) evaluated at $X=0$ by making use of (\ref{D0} ),
\begin{equation}
\stackrel{(0)}{\delta_\epsilon}e(x,P)=d\epsilon(x,P)+\epsilon^{\,\prime}(x,P)\,,
\label{cofrggtransf}
\end{equation} 
where the one-form 
\begin{equation}
\epsilon^{\,\prime}(x,P):=dx^\mu\epsilon_\mu^{\,\prime}(x,P)\,,
\end{equation}
is defined in terms of the zero-form
\begin{equation}
\epsilon(x,X,P):=\epsilon(x,P)+X^\mu\,\epsilon^{\,\prime}_\mu(x,P)+{\cal O}(X^2)\,.
\end{equation}
Notice that the one-form $\epsilon^{\,\prime}(x,P)$ obeys to the irreducibility condition $P^\mu\,\epsilon_\mu^{\,\prime}(x,P)=0\,$, as follows from $P^\mu\partial \epsilon/\partial X^\mu=0\,$. In the spin-two case, $\epsilon_\mu^{\,\prime}(x,P)=P^\nu\epsilon_{\mu\nu}^{\,\prime}(x)$ and  the antisymmetric tensor field $\epsilon_{\mu\nu}^{\,\prime}(x)$ corresponds to the local Lorentz parameter.
The gauge transformations (\ref{cofrggtransf}) allow several partial fixations, one of which leads to the unconstrained metric-like formulation.
Besides technical complications, all the previous steps work if one starts with an $AdS$ background and the corresponding $AdS/CFT$ higher-spin algebras (see \textit{e.g.} \cite{Vasiliev:2004qz,BCIV,Vframe87,Sagnotti:2005,Engquist:2007yk,Vasiliev:2003ev} for more details).

\section{Metric-like formulation}\label{metricformul}

Looking for a geometrical interpretation for the infinite collection of symmetric tensor gauge fields $\varphi_{\mu_1\ldots\mu_s}(x)$ of the metric-like gauge theory, it might be convenient to summarise this spectrum into a single function on the tangent bundle $T{\mathbb R}^{n-1,1}\,$:
\begin{equation}
\varphi(x,p)\,\,=\,\sum\limits_{s}\,\frac{1}{s\,!}\,\,\varphi_{\mu_1\ldots\mu_s}(x)\,p^{\mu_1}\ldots p^{\mu_s}\,, 
\label{varphiup}
\end{equation}
This trick allows to write very compact expressions for the gauge transformations, \textit{etc}. Moreover, if the coordinates $p^\mu$ on the fibre of $T{\mathbb R}^{n-1,1}$ are replaced by commuting creation oscillators $(a^\mu)^\dagger\,$, then the function (\ref{varphiup}) is interpreted as a string field (see \textit{e.g.} \cite{Bengtsson:1987jt,Buchbinder:2006eq}). Using the Minkowski metric, one may of course equivalently summarise the spectrum into a function on the cotangent bundle $T^*{\mathbb R}^{n-1,1}\,$, as in (\ref{varphi}). This is even more convenient and suggestive for the present purpose since the cotangent space is a symplectic manifold, hence it allows quantisation.\\

The linearised metric-like theory arising from the unconstrained frame-like around Minkowski space-time is obtained in Subsection \ref{mlikeAbelian} following the lines of \cite{Vframe87,Vasiliev:2003ev} reviewed \textit{e.g.} in \cite{Vasiliev:2004qz,BCIV,Engquist:2007yk}. The aim of the subsection \ref{proof} is to provide the proof of the result stated in the introduction of the paper. Various suggestive features of this result are examined in Subsection \ref{remarks}.

\subsection{Minkowski background and Abelian gauge transformations}\label{mlikeAbelian}

The aim is to recover the metric-like unconstrained  formulation from the frame-like one, so the perturbation $\Omega(x,X,P)$ in (\ref{aomega}) is taken to be a one-form taking values in the off-shell Minkowski higher-spin algebra, as in Subsection \ref{framelggtransf}. 

At lowest order, the coordinates $p^\mu$ on the fibre of the cotangent bundle can be identified with the coordinates $P^a$ from the Weyl bundle via the Minkowski coframe $\delta_\mu^a\,$.
The metric-like field $\varphi$ is defined in terms of the frame-like field as
\begin{equation}
\varphi(x,p)\,:=\,\lambda\, p_\mu \,\eta^{\mu\nu} e_\nu(x,P^a=\lambda^{-1}\delta^a_\mu p^\mu)\,,
\label{sympart}
\end{equation}
thereby generalising the orthogonality condition (\ref{he}) at linearised order. Indeed, the definitions (\ref{varphi}) and (\ref{powerexpomega}) implies that the relation (\ref{sympart}) reads in components, $$\varphi^{\mu_1\ldots\mu_s}\,=\,s\,\lambda^{2-s}\,\eta^{\nu(\mu_1}e_\nu{}^{\mu_2\ldots\mu_s)}\,$$
The constant $\lambda$ (introduced in Subsection \ref{nonabsym}) has been inserted in the definition (\ref{sympart}) for dimensional reason.
Let us define the metric-like gauge parameter by 
\begin{equation}
\varepsilon(x,p)\,:=\,\lambda\,\,\epsilon(x,X=0,P=\lambda^{-1}p)\,,
\end{equation}
hence $\varepsilon^{\mu_1\ldots\mu_r}\,=\,\lambda^{1-r}\,\epsilon^{\mu_1\ldots\mu_r}\,$.
The Abelian gauge transformation (\ref{cofrggtransf}) reproduces, in the metric-like formulation, the form of Fr\o nsdal's gauge transformation,
\begin{equation}
\stackrel{(0)}{\delta_\varepsilon}\varphi(x,p)\,=\,\Big(p^\mu\frac{\partial}{\partial x^\mu}\Big)\,\varepsilon(x,p)\,,
\label{abeliangaugetransfo}
\end{equation}
since the one-form $\epsilon^{\,\prime}$ obeys to the condition $P^\mu\epsilon_\mu^{\,\prime}(x,P)=0\,$.
The freedom in this one-form $\epsilon^{\,\prime}$ is precisely enough for imposing the gauge fixing condition 
\begin{equation}
\Big(dx^\mu\frac{\partial}{\partial P^\mu}\Big)\,e(x,P)=0\,,
\label{mlikegfix}
\end{equation}
generalising the metric gauge (\ref{mgauge}) in gravity.
The metric-like gauge (\ref{mlikegfix}) means that the only non-vanishing components of the frame-like field are the components of the metric-like field.
Actually, the metric-like gauge can be extended to the following condition
\begin{equation}
{\cal Y}\Big(X^\mu\Omega_\mu(x,X,P)\Big)=0\,.
\label{mlikextended}
\end{equation}
It is straightforward to check that the part of homogeneity degre one in $X$ of the equation (\ref{mlikextended}) is indeed equivalent to (\ref{mlikegfix}).
Analogously to the particular spin-two case, the torsion constraint (\ref{tcstrt}) together with the gauge condition (\ref{mlikextended}) imply that 
\begin{equation}
\omega_{\mu\,\, \nu_1 \rho_1\mid \nu_2 \rho_2\mid\,\ldots\,\mid \nu_t \rho_t\mid \nu_{t+1}\, \ldots\, \nu_r}
\,=\,\lambda^{r-2}\,\,\partial_{[\rho_t}\ldots\partial_{[\rho_2}\partial_{[\rho_1}\varphi_{\nu_1] \nu_2]\,\ldots \nu_t]\nu_{t+1}\, \ldots\, \nu_r\mu}
+{\cal O}(\varphi^2)\quad (t\leq r)\,.
\label{soltorsion}
\end{equation}

In turn, this implies that the non-vanishing components of the linearised curvature two-form
are the curvature tensors introduced in \cite{Weinberg:1965rz} and investigated in \cite{deWit:1979pe,DuboisVH},
\begin{equation}
\stackrel{(0)}{\cal F}_{\mu_1 \nu_1\mid \mu_2 \nu_2\mid\,\ldots\,\mid \mu_s \nu_s}
\,=\,\lambda^{s-2}\,\,\partial_{[\mu_s}\ldots\partial_{[\mu_2}\partial_{[\mu_1}\varphi_{\nu_1] \nu_2]\,\ldots \nu_s]}\,,
\end{equation}
generalising the electromagnetic fieldstrength ($s=1$) and linearised Riemann tensors ($s=2$). The vacuum field equations generalising the ones of Bargmann and Wigner \cite{Bargmann:1948} (for $n=4$) state that, on-shell, the linearised curvature tensor is traceless and divergenceless. The space of solutions of these equation can be shown to carry a unitary irreducible representation of the Poincar\'e group corresponding to a massless symmetric tensor field of rank $s$ (see \textit{e.g.} \cite{Bekaert:2006py} and references therein).

Similarly to the spin-two case, the metric-like gauge is only a partial gauge fixing because it is preserved by gauge transformations (\ref{frameabeliangaugetransfo})
which do not affect (\ref{soltorsion}). This is satisfied if and only if the right-hand-side of (\ref{frameabeliangaugetransfo}) belongs to the off-shell Lorentz higher-spin subalgebra.
A simpler way to address this is to impose explicitly the invariance of the gauge condition (\ref{mlikextended}) under (\ref{frameabeliangaugetransfo}) taking into account  (\ref{D0}):
\begin{equation}
0\,=\,\stackrel{(0)}{\delta_\epsilon}\,{\cal Y}\Big(X^\mu\Omega_\mu(x,X,P)\Big)={\cal Y}\Big[X^\mu\Big(\frac{\partial}{\partial x^\mu}+
\frac{\partial}{\partial X^\mu}\Big)\epsilon(x,X,P)\Big]=0\,.
\end{equation}
Using the fact that $X^\mu\partial\epsilon/\partial X^\mu$ automatically takes values in the higher-spin algebra, one gets that the general solution of 
\begin{equation}
\Big(X^\mu\frac{\partial}{\partial X^\mu}\Big)\epsilon(x,X,P)\,=\,-\,{\cal Y}\Big(X^\mu \frac{\partial}{\partial x^\mu}\Big)\epsilon(x,X,P)\,.
\end{equation}
is
\begin{equation}
\epsilon(x,X,P)=\,\epsilon(x,X=0,P)-\int\limits_0^1 \frac{du}{u}\,\Big[{\cal Y}\Big(X^\mu\frac{\partial}{\partial x^\mu}\Big)\epsilon\,\Big](x,uX,P)
\end{equation} 
More explicitly, this can be solved for the components
\begin{equation}
\epsilon(x,X,P)=\sum\limits_{t\leq r} \,\frac{(-1)^t}{t\,!\,r\,!}\,\,\partial_{[\nu_t}\ldots\partial_{[\nu_2}\partial_{[\nu_1}\epsilon_{\mu_1]\mu_2]\,\ldots\,\mu_t]\mu_{t+1}\,\ldots\,\mu_r}(x)\,
P^{\mu_1}\ldots P^{\mu_r} X^{\nu_1}\,\ldots X^{\nu_t}\,,
\label{powerexpomegaeps}
\end{equation}
as can be checked by direct but tedious computation.

\subsection{Deforming the gauge transformations}\label{proof}

Although the proof of the results presented in Subsection \ref{nonabsym} has been obtained thanks to the experience gained in the BRST techniques of \cite{Barnich:1993vg,Bekaert:2006us}, this machinery is not introduced in this paper and the results are formulated along the lines of \cite{Berends:1984rq}. Nevertheless, a prerequisite is the concept of jet (space, bundle, ...) which is reviewed in the Appendix \ref{Jspace} where the corresponding notation is introduced.

The subsection \ref{generalset} is devoted to the general setting while subsection \ref{defo} addresses the case under consideration here: the non-Abelian higher-spin gauge symmetries.

\subsubsection{Gauge structure in jet language}\label{generalset}

Let $\chi$ stands collectively for some gauge fields taking values in the vector space $V$, and let $\eta$ denote the gauge parameters which parametrise the vector space $W\,$.

An infinitesimal gauge transformation of parameter $\eta\,$,
\begin{equation}
\delta_\eta\,:=\,\Delta(x,[\chi],[\eta])\,\,\frac{\partial}{\partial \chi}\,,
\label{deltaeta}
\end{equation}
is, for the jet bundle ${\cal J}^\infty({\cal M}\times V)\,$, an evolutionary vector field taking values in the dual space $(\,J^\infty W\,)^*\,$.
In other words, its characteristic $\delta_\eta\,\chi=\Delta(x,[\chi],[\eta])$ is a (pseudo)local function of the gauge field variable $\chi$ which is linear in the coordinates $[\eta]$ of the jet space $J^\infty W\,$.
Equivalently, the following map from the jet space $J^\infty W$ of the gauge parameters into the space of derivations on the algebra of (pseudo)local functions of the gauge field,
\begin{equation}
\delta_\bullet\,:\,J^\infty W\rightarrow Der\Big(\,C^\infty\left(\,{\cal J}^\infty({\cal M}\times V)\,\right)\,\Big)\,:\,\,[\,\eta\,]\,\mapsto\,\delta_\eta\,,
\label{deltabullet}
\end{equation}
is linear.
By a slight abuse of notation, the infinite prolongation of the evolutionary vector field (\ref{deltaeta})
is also denoted by $\delta_\eta\,$.
Translation invariance implies that the function $\Delta$ can be assumed to be independent of the position $x\,$, as will be done from now on.

Let $\eta(x,[\chi],[\Lambda])$ denote a $W$-valued pseudolocal function of the field variables $\chi$ and $\Lambda$ taking values, respectively, in the vector spaces $V$ and $U$.
The infinite prolongation of this pseudolocal function defines a map from the jet bundle ${\cal J}^\infty({\cal M}\times U)$ to the jet space $J^\infty W$ via the identification
$$\partial_{\mu_1}\ldots\partial_{\mu_k}\eta:=\partial_{\mu_1}^T\ldots\partial_{\mu_k}^T\eta(x,[\chi],[\Lambda])\,.$$
By definition, the gauge transformations are ``irreducible'' if there is no translation invariant non-vanishing pseudolocal function,
say $\eta([\chi],[\Lambda])$, which is solution of the equation $\delta_\eta\,=0\,$.
The gauge transformations are said to ``close off-shell'' if
\begin{equation}
\Big({\delta}_{\eta_1}{\delta}_{\eta_2}-{\delta}_{\eta_2}{\delta}_{\eta_1}\Big)\,\chi\,=\,{\delta}_{\eta_3}\,\chi\,,
\label{offshellclosure}
\end{equation}
where $\eta_3=\{\eta_1,\eta_2\}$ stands for the new parameter coordinates corresponding to the commutator of two gauge transformations, in the sense that
$$\partial_{\mu_1}\ldots\partial_{\mu_k}\eta_3:=\partial_{\mu_1}^T\ldots\partial_{\mu_k}^T\{\eta_1,\eta_2\}\,.$$ In other words,
the bracket
\begin{equation}
\{\,\,,\,\}:(J^\infty W)^{\wedge 2}\rightarrow \Gamma\big(\,{\cal J}^\infty({\cal M}\times V)\,\big)\otimes J^\infty W\,,
\label{bracket}
\end{equation}
is a linear map which encodes the structure of the gauge algebra and which is defined as the pullback (when it exists) by the linear map (\ref{deltabullet}) of the commutator bracket (for the Lie algebra of derivations on the algebra of pseudolocal functions of the gauge fields).
\begin{lemma}\label{Liealgg}
Let $\delta_\eta$ be some irreducible gauge transformations which close off-shell.

If the gauge parameter $\{\eta_1,\eta_2\}$ corresponding to the commutator of two gauge transformations with parameters $\eta_1$ and $\eta_2$ is independent of the position $x$ and the gauge field variables $\chi$, then the linear map
$\{\,\,,\,\}:(J^\infty W)^{\wedge 2}\rightarrow J^\infty W$
defines a Lie bracket in the jet space $J^\infty W$ of gauge parameters.
\end{lemma}
\noindent The proof is obvious because the Jacobi identity is induced from the one for the commutator. This property was underlined in \cite{Berends:1984rq}.

The perturbative expansion of pseudolocal functions (which will always be assumed to be formal power series in $[\chi]$) in powers of the gauge field $\chi$ will be considered now. Accordingly, the evolutionary field (\ref{deltaeta}) may be expanded as
\begin{equation}
\delta_\eta\,=\,\stackrel{(0)}{\delta_\eta}\,+\,\stackrel{(1)}{\delta_\eta}\,+\,\stackrel{(2)}{\delta_\eta}\,+\,\ldots\,,
\label{defogaugetransfos}
\end{equation}
where $\stackrel{(n)}{\delta_\eta}\chi=\,\stackrel{(n)}{\Delta}(\,[\chi],[\eta]\,)$ is  of homogeneity degree $n\in\mathbb N$ in the coordinates $[\chi]$ of the jet space $J^\infty V\,$.
Therefore,
\begin{equation}
\stackrel{(n)}{\delta}_{\eta_1}\stackrel{(0)}{\delta}_{\eta_2}\chi=0\,,
\label{deltadeltazero}
\end{equation}
for any $n\in\mathbb N\,$.
The commutator of two gauge transformations (\ref{defogaugetransfos}) is assumed to close as in (\ref{offshellclosure}).
The bracket (\ref{bracket}) should also be expanded in powers of the gauge field,
$$
\{\,\,,\,\}\,=\,\{\,\,,\,\}_{_{(0)}}+\{\,\,,\,\}_{_{(1)}}+\{\,\,,\,\}_{_{(2)}}+\ldots
$$
Then the lowest non-trivial relation in the weak field expansion of (\ref{offshellclosure}) is
\begin{equation}
\Big(\stackrel{(0)}{\delta}_{\eta_1}\stackrel{(1)}{\delta}_{\eta_2}
-\stackrel{(0)}{\delta}_{\eta_2}\stackrel{(1)}{\delta}_{\eta_1}\Big)\,\chi\,=
\,\,\stackrel{(0)}{\delta}_{\{\eta_1,\eta_2\}_{(0)}}\chi\,,
\label{defogaugecomm}
\end{equation}
due to the identity (\ref{deltadeltazero}).
The gauge transformation (\ref{defogaugetransfos}) at all orders is interpreted as a non-Abelian deformation of the linearised transformation.
The bracket $\{\,\,,\,\}_{_{(0)}}$ shows up at linear order and is assumed to be non-trivial. It will be referred to as ``lowest order bracket'' in the sequel.
If the lowest order bracket satisfies the Jacobi identity, then Lemma \ref{Liealgg} suggests that it might already be the correct Lie bracket at all orders (as is the case for spin one and two gauge theories \cite{Bekaert:2006us}).

A deformation
\begin{equation}
\delta_\eta\,\chi\,=\,\stackrel{(0)}{\delta_{\eta^\prime}}\chi^\prime 
\end{equation}
of the linearised gauge transformation $\stackrel{(0)}{\delta_{\eta}}\chi$ that corresponds to a mere redefinition
of the gauge fields $\chi\mapsto \chi^\prime\big([\chi],[\eta]\big)$ and the gauge parameters $\eta\mapsto \eta^\prime\big([\chi],[\eta]\big)$ 
is said to be trivial. 
Let the power expansion of the redefined gauge field read
\begin{eqnarray}
\chi^\prime\,=\,\chi\,+\,Q\big(\,[\chi]\,\big)\,+\,\ldots\,,
\end{eqnarray}
where $Q$ is a quadratic form on the jet space $J^\infty V\,$.
Then the first order deformation of a mere redefinition of the gauge field is equal to
\begin{equation}
\stackrel{(1)}{\delta_{\eta}}\chi\,=\,\,\stackrel{(0)}{\delta_{\eta}}Q([\chi])\,, 
\end{equation}
and does not deform the gauge algebra.
Let the power expansion of the redefined gauge parameters read
\begin{eqnarray}
\eta^\prime\,=\,\eta\,+\,\stackrel{(1)}{N}\big(\,[\chi],[\eta]\,\big)\,+\,\stackrel{(2)}{N}\big(\,[\chi],[\eta]\,\big)\,+\,\ldots\,,
\end{eqnarray}
then (\ref{defogaugecomm}) implies that the lowest order bracket reads
$$
\{\eta_1,\eta_2\}_{(0)}\,=\,\,\stackrel{(1)}{N}\Big(\,\Big[\stackrel{(0)}{\Delta}([\eta_1])\Big],\,\Big[\eta_2\Big]\,\Big)-\stackrel{(1)}{N}\Big(\,\Big[\stackrel{(0)}{\Delta}([\eta_2])\Big],\,\Big[\eta_1\Big]\,\Big)
$$
and is manifestly linear in the images $\stackrel{(0)}{\Delta}([\eta])$ of the linearised gauge transformations.
The following known property follows:
\begin{lemma}\label{trivdefgalg}
A non-Abelian deformation of some linearised gauge transformation is trivial if and only if
the lowest order bracket is at least linear in the images of the linearised gauge transformation. 
\end{lemma}

\subsubsection{Deformation of the Abelian gauge algebra}\label{defo}

In the case under consideration (a higher-spin gauge theory), the gauge transformations at lowest order are given by (\ref{frameabeliangaugetransfo}) in the frame-like formulation
and by (\ref{abeliangaugetransfo}) in the metric-like formulation.
The main question addressed in the present paper is: What is the counterpart of\footnote{Actually, consistency with the torsion constraint at this order would require an extra term in the right-hand-side, proportional to the linearised curvature. However such terms do not modify the gauge algebra at this order, therefore their presence would not modify the conclusions of the perturbative analysis.}
\begin{equation}
\stackrel{(1)}{\delta_\epsilon}\Omega(x,X,P)\,=\, i\,[\,\Omega(x,X,P)\,\stackrel{\bigstar}{,}\,\epsilon(x,X,P)\,]
\label{delta1frlike}
\end{equation}
in the metric-like formalism? In principle, the only thing to do is to insert the relation (\ref{soltorsion}) inside (\ref{powerexpomega}) and also to use (\ref{powerexpomegaeps}) in order to compute explicitly the deformation
(\ref{delta1frlike}) which, finally, should be evaluated at $X=0$ and $P=p/\lambda \,$. However, this direct approach is technically cumbersome.
The tactic of the following proof is to circumvent this obstacle by, firstly, computing instead the metric-like deformation of the gauge algebra arising from the frame-like one and by, secondly, looking for the corresponding deformation of the metric-like gauge transformations. 
The answer is given in Subsection \ref{nonabsym} and takes the form
\begin{equation}
\stackrel{(1)}{\delta_\varepsilon}\varphi(x,p)\,=\,\frac{\,i}{\lambda}\,[ \,\varphi(x,p)\,\stackrel{\star}{,}\,\varepsilon(x,p)\,]\,+\,{\cal K}\,.
\label{defogaugetransfo}
\end{equation}
Now comes the detailed proof of this result.

In the unconstrained metric-like formulation, the gauge field variables are denoted by $\varphi$ while the gauge parameter variables by $\varepsilon\,$. They are coordinates for the vector spaces $V\cong W\cong\vee({\mathbb R}^{n*})\,$. The linearised gauge transformations are defined by (\ref{abeliangaugetransfo}).
\begin{lemma}\label{Hgamma}Consider the jet spaces $J^\infty\vee({\mathbb R}^{n*})\,\cong\, (\,\vee(\,{\mathbb R}^{n*})\,)^{\otimes 2}$ of coordinates $[\varphi]$ and $[\varepsilon]$ for two collections of symmetric tensors $\varphi$ and $\varepsilon\,$.

There exists a linear and invertible change of coordinates $[\varphi]\rightarrow (K,F)$ and $[\varepsilon]\rightarrow (D,E)$ 
where the variables $K$ stand for the curvature tensors and their partial derivatives, the variable $D$ for the gauge parameters of any rank $r$ up to their rth exterior derivatives, while the other variables span their linear complement in their respective jet spaces.

The new independent variables $K$, $F$, $D$ and $E$ of the jet spaces are such that 
\begin{itemize}
 \item the linearised gauge transformation provide a one-to-one correspondence between the coordinates $F$ and $E$,
in the sense that $\stackrel{(0)}{\delta_\varepsilon}F=E\,$.
 \item the variables $K$ are gauge invariant at lowest order: $\stackrel{(0)}{\delta_\varepsilon}K=0\,$.
 \item there is no non-vanishing linear combination of the variables $D$ which is the variation of something under the linearised gauge transformation.
\end{itemize}
\end{lemma}
\noindent This is essentially the content of the theorem 1 in \cite{Bekaert:2005ka} and the idea of its proof. More explicitly, the change of coordinates in the jet space for the gauge parameters is
\begin{equation}
\partial_{\nu_t}\ldots\partial_{\nu_2}\partial_{\nu_1}\varepsilon_{\mu_1\mu_2\,\ldots\,\mu_t\mu_{t+1}\,\ldots\,\mu_r}
\,=\,\partial_{[\nu_t}\ldots\partial_{[\nu_2}\partial_{[\nu_1}\varepsilon_{\mu_1]\mu_2]\,\ldots\,\mu_t]\mu_{t+1}\,\ldots\,\mu_r}\,+\,\stackrel{(0)}{\delta_\varepsilon}(\mbox{something})\,.
\label{trivialdefredef}
\end{equation}
A decomposition of the jet space $J^\infty(V\oplus W)$ as in Lemma \ref{Hgamma} is very convenient because it also implies the
\begin{corollary}\label{notdeformginv}
Any first order deformation $\stackrel{(1)}{\delta_\varepsilon}\varphi$ of the linearised gauge transformation $\stackrel{(0)}{\delta_\varepsilon}\varphi$ that does not deform the gauge algebra at lowest order,
\begin{equation}
\Big(\stackrel{(0)}{\delta}_{\varepsilon_1}\stackrel{(1)}{\delta}_{\varepsilon_2}
-\stackrel{(0)}{\delta}_{\varepsilon_2}\stackrel{(1)}{\delta}_{\varepsilon_1}\Big)\,\varphi\,=0\,,
\label{nodeform} 
\end{equation}
is equal to, modulo trivial redefinitions of the gauge field, a function which depends on the gauge field $\varphi$ (and their derivatives) only through quantities which are strictly invariant under the linearised gauge transformations.
\end{corollary}
\proof{The proposition will be shown to hold for any gauge theory such that the set of partial derivatives of the gauge parameter can be decomposed into two sets, as in Lemma \ref{Hgamma} for symmetric tensor gauge fields.

By definition, $$\stackrel{(1)}{\delta_\varepsilon}\varphi\,=\,\stackrel{(1)}{\Delta}\Big(\,[\varphi]\,,\,[\varepsilon]\,\Big)\,,$$
where $\stackrel{(1)}{\Delta}:J^\infty V\otimes J^\infty W\rightarrow\mathbb K$ is a linear map, \textit{i.e.} a bilinear pseudolocal function of the gauge fields and parameters.
By the hypothesis on the gauge theory structure, this bilinear form can be decomposed as the sum of three terms
\begin{equation}
\stackrel{(1)}{\delta_\varepsilon}\varphi\,=\,\stackrel{(1)}{\Delta}_1\Big(\,K\,,\,D\,\Big)+ \stackrel{(1)}{\Delta}_2\Big(\,K\,,\,E\,\Big)
+\stackrel{(1)}{\Delta}_3\Big(\,F\,,\,D\,\Big)+\stackrel{(1)}{\Delta}_4\Big(\,F\,,\,E\,\Big)\,.
\label{firstorderdefo} 
\end{equation}
The condition (\ref{nodeform}) becomes $$\stackrel{(1)}{\Delta}_3\Big(\,E_1\,,\,D_2\,\Big)-\stackrel{(1)}{\Delta}_3\Big(\,E_2\,,\,D_1\,\Big)+\stackrel{(1)}{\Delta}_4\Big(\,E_1\,,\,E_2\,\Big)-\stackrel{(1)}{\Delta}_4\Big(\,E_2\,,\,E_1\,\Big)=0\,,$$
due to the relations $\stackrel{(0)}{\delta_\varepsilon}K=0$ and $\stackrel{(0)}{\delta_\varepsilon}F=E\,$.
Since the variables $D$ and $E$ are independent this can be satisfied if and only if $\stackrel{(1)}{\Delta}_3=0$ and $\stackrel{(1)}{\Delta}_4$ are symmetric.
Therefore, the first order deformation (\ref{firstorderdefo}) can be put in the form 
$$
\stackrel{(1)}{\delta_\varepsilon}\varphi\,=\,\stackrel{(1)}{\Delta}_1\Big(\,K\,,\,D\,\Big)+ \stackrel{(0)}{\delta_\varepsilon}\Big[\stackrel{(1)}{\Delta}_2\Big(\,K\,,\,F\,\Big)
+\frac12\,\stackrel{(1)}{\Delta}_4\Big(\,F\,,\,F\,\Big)\Big]\,.
$$
This deformation satisfies the conclusion of Corollary \ref{notdeformginv}.
}
Together the lemma \ref{Hgamma} and its corollary \ref{notdeformginv} imply that the component of the gauge transformation at linear order that does not modify the gauge algebra at lowest order must precisely be linear in the curvature tensors, like $\cal K$ in (\ref{nonabeliangaugetransfos}), modulo redefinitions of the gauge fields. Consequently, it is enough to find \textit{any} first order deformation of the gauge transformation which give rise to the right lowest order bracket to finish the proof.

In the unconstrained frame-like formulation, the gauge field variable is the one-form $\Omega$ in (\ref{powerexpomega}) and the gauge parameter coordinate is the zero-form $\epsilon(X,P)$ taking values in the Minkowski higher-spin algebra seen as a subalgebra of ${\mathbb R}[[X,P]]\,$. The linearised gauge transformations are defined by (\ref{abeliangaugetransfo}) and their first order deformation by (\ref{delta1frlike}). 
The gauge parameter coordinates $\varepsilon(p)\in {\mathbb R}[[p]]$ of the metric-like formulation is related to the frame-like one by $\varepsilon(p):=\lambda\,\,\epsilon(X=0,P=\lambda^{-1}p)\,$.
The Lie bracket of the frame-like gauge algebra is nothing but the {\footnotesize$\bigstar$}-commutator of the gauge parameter coordinates $\epsilon(X,P)$ thus, at lowest order, the bracket of the metric-like gauge algebra is defined by 
\begin{equation}
\{\varepsilon_1(p),\varepsilon_2(p)\}_{(0)}\,:=\,\,i\,\lambda\,\left.[\,\epsilon_1(X,P)\,\stackrel{\bigstar}{,}\,\epsilon_2(X,P)\,]\right|_{X=0\,,\,P=\lambda^{-1}p}\,.
\label{lowestbracketmetric}
\end{equation}

The torsion-like constraint and the metric-like gauge impose that the components of the one-form $\Omega$ are given by (\ref{soltorsion}) in terms of the metric-like variables $\varphi\,$. This partial gauge fixing is preserved by gauge transformations where the parameters are related by (\ref{powerexpomegaeps}) to the metric-like variables $\varepsilon$ only in terms of the variables $D$ of Lemma \ref{Hgamma}. 
\begin{proposition}\label{generatingfunction}
The frame-like gauge parameter coordinates $\epsilon$ which preserve the metric-like gauge at lowest order are generated by the metric-like gauge parameter coordinates $\varepsilon\,$.

More explicitly,
$$\epsilon(X,P)\,=\,e^{-X^\mu\partial^T_\mu}\epsilon(0,P)\,+\,\stackrel{(0)}{\delta_\varepsilon}\alpha([\varphi])\,,$$
where $\partial^T$ is the total derivative on the infinite jet space of the paramer coordinates and $\alpha$ is a linear form on the infinite jet space of the field variables.
\end{proposition}
\proof{The power expansion 
\begin{eqnarray}
e^{-X^\mu\partial_\mu}\epsilon(P)
&=&\sum\limits_{t} \,\frac{(-1)^t}{t\,!}\,\,\partial_{\nu_t}\ldots\partial_{\nu_2}\partial_{\nu_1}\epsilon(P)\,
 X^{\nu_1}\,\ldots X^{\nu_t}
\nonumber\\
&=&\sum\limits_{t\,,\, r} \,\frac{(-1)^t}{t\,!\,r\,!}\,\,\partial_{\nu_t}\ldots\partial_{\nu_2}\partial_{\nu_1}\epsilon_{\mu_1\,\ldots\,\mu_r}\,
P^{\mu_1}\ldots P^{\mu_r} X^{\nu_1}\,\ldots X^{\nu_t}\nonumber
\end{eqnarray}
should be compared with (\ref{powerexpomegaeps}) where (\ref{trivialdefredef}) has been inserted.
}

\begin{theorem}

Modulo a trivial redefinition of the gauge parameters, the lowest order bracket on the metric-like gauge parameters arising from the unconstrained frame-like theory around Minkowski space-time ${\mathbb R}^{n-1,1}$ is the Moyal bracket on $C^\infty(T^*{\mathbb R}^n)\,$.
\end{theorem}
\proof{One should evaluate (\ref{lowestbracketmetric}). Due to Lemma \ref{trivdefgalg}, modulo a term coming from a trivial redefinition of the gauge parameter, one may assume that the frame-like gauge parameter is generated by the metric-like parameter following Proposition \ref{generatingfunction}. A straightforward computation shows that the lowest order bracket is precisely given by (\ref{explicitbracket}).
}
This ends the proof of the result stated in Section \ref{nonabsym} because, all collected results imply that, a first order deformation of the form (\ref{defogaugetransfo}) is indeed the most general one, modulo trivial deformations, which leads to the Moyal bracket at lowest order.

\subsection{Various remarks}\label{remarks}

The Lie algebra of vector fields on ${\mathbb R}^n$ already pops up at lowest order in a perturbative analysis of gravity \cite{Fang:1978rc}. This algebra is locally isomorphic to the Lie algebra of vector fields on any manifold ${\cal M}$ of dimension $n\,$. This property allows to shortcut the perturbative analysis order by order. Analogously, the algebras of differential operator are locally isomorphic, for any $\cal M\,$. This means that the first order deformation of the gauge algebra is presumably already the full gauge algebra of higher-spin transformations, which would suggest a shortcut of the perturbative analysis order by order in order to reconstruct the right gauge transformations, as in the case of gravity.

The first order deformation of the gauge algebra is nothing more than the good old algebra of quantum observables from our undergraduate studies. This implies a wide variety of possible origins of the present result from a (would-be) formulation at all orders. Subsection \ref{several} discusses the various brackets already encountered in the literature on higher-spin gauge theories. The unconstrained formulation and the Fedosov construction are briefly mentioned in Subsection \ref{Fed}.
A plausible generalisation of the present result around (anti) de Sitter background is presented in the subsection \ref{CCspacet}. 

\subsubsection{Several brackets}\label{several}

In the light of Lemma \ref{Liealgg}, the non-Abelian structure can be expected to be a Lie algebra on the space of gauge field parameters. This prompted the search of the right Lie algebra of gauge symmetries from the very beginning of the quest for non-Abelian higher-spin gauge theories. Since the gauge parameters may be encoded in a function $\varepsilon(x,p)$ on configuration or phase space, a series of natural brackets were suggested. If one looks for a Poisson bracket, then the Poincar\'e symmetries strongly restrict the possibilities.

\begin{proposition}\label{uniquenessP}
The canonical Poisson bracket is the unique Poincar\'e invariant Poisson bracket on the commutative algebra of functions on the (co)tangent bundle of the Minkowski space-time, up to trivial deformations of the gauge algebra (corresponding to redefinitions of the gauge parameters).
\end{proposition}

\proof{A Poisson bracket $\{\,\,,\,\}$ on $T{\mathbb R}^n$ is an antisymmetric biderivation on $C^\infty(T{\mathbb R}^n)\,$. (The proof also works for $T^*{\mathbb R}^n$ since they are related via the Minkowski metric). Hence, it takes the generic form:
\begin{eqnarray}
\{\,\,,\,\}\,=\,&&
A^{\mu\nu}(x,p)\,\frac{\overleftarrow{\partial}}{\partial x^{\mu}}\wedge\frac{\overrightarrow{\partial}}{\partial x^{\nu}}\,+\,
B^{\mu\nu}(x,p)\,\frac{\overleftarrow{\partial}}{\partial x^{\mu}}\wedge\frac{\overrightarrow{\partial}}{\partial p^{\nu}}\,+\,\nonumber\\
&&\quad\quad+\,C^{\mu\nu}(x,p)\,\frac{\overleftarrow{\partial}}{\partial p^{\mu}}\wedge\frac{\overrightarrow{\partial}}{\partial p^{\nu}}\,.
\label{ABC}
\end{eqnarray}
Translation invariance requires that the three functions $A$, $B$ and $C$ do not depend on the coordinates $x$ while Lorentz invariance requires that all the indices of the coordinates $(x,p)$ be contracted. The index $\mu$ of the partial derivative ${\partial}/{\partial x^{\mu}}$ can only be contracted with an index of another partial derivative, because the only alternative is to a conraction with the index of a coordinate $p^\mu$ which would produces the operator $p\cdot {\partial}/{\partial x}\,$, thus a trivial deformation of the gauge algebra due to Lemma \ref{trivdefgalg}. 

Let us examine all the possibilites:
The antisymmetric product implies that ${\partial}/{\partial x^{\mu}}\wedge{\partial}/{\partial x_{\mu}}=0$ so the first term in the right-hand-side of (\ref{ABC}) is zero. The third term is also zero because the two possibilities are either proportional to ${\partial}/{\partial p^{\mu}}\wedge{\partial}/{\partial p_{\mu}}=0$ 
or to $(p\cdot{\partial}/{\partial p})\wedge(p\cdot{\partial}/{\partial p})=0$. The only possiblity for the second term in the parenthesis is
$B(p^2)\,{\partial}/{\partial x^{\mu}}\wedge{\partial}/{\partial p_{\mu}}\,$. Finally one may check that the Jacobi identity for the corresponding ``bracket'' is satisfied only if the factor $B(p^2)$ is a constant.\footnote{Notice that, actually, for any function $B(p^2)\,$, the corresponding ``bracket'' satisfies a weaker version of the Jacobi identity, precisely as expressed by the equation (4.17) in \cite{Berends:1984rq}.} A constant factor may be removed by rescaling the gauge parameters.}

Read in terms of the covariant symmetric tensor coefficients of the power expansion in the momenta, the canonical Poisson bracket is the so-called Schouten bracket (reviewed in Appendix \ref{ApPoisson}). Consequently, in the low energy limit $\lambda\rightarrow\infty\,$, the Moyal bracket (\ref{explicitbracket}) is nothing but the Schouten bracket of the symmetric gauge parameters. Notice that the Schouten bracket (\ref{Poisson}) does not preserve the trace constraint. More precisely, the Poisson bracket $\{\varepsilon_1,\varepsilon_2\}_{_C}$ of two elements $\varepsilon_1\,,\varepsilon_2\in C^\infty(T{\mathbb R}^n)$ that are harmonic in the fibre is, in general, \textit{not} harmonic in the fibre because
$$\Big(\frac{\partial}{\partial p^\mu}\frac{\partial}{\partial p_\mu}\Big)\,\varepsilon_1\,=\,0\,,\quad \Big(\frac{\partial}{\partial p^\mu}\frac{\partial}{\partial p_\mu}\Big)\,\varepsilon_2\,=\,0\quad\Longrightarrow\quad \Big(\frac{\partial}{\partial p^\mu}\frac{\partial}{\partial p_\mu}\Big)\Big\{\varepsilon_1\,,\varepsilon_2\Big\}_{_C}\,=\,2\,\,\Big\{\frac{\partial\varepsilon_1}{\partial p^\mu},\frac{\partial\varepsilon_2}{\partial p_\mu}\Big\}_{_C}\,.$$
Thus the proposition \ref{uniquenessP} implies that deformations of the higher-spin gauge algebra cannot correspond to the Schouten bracket in the constrained case. Notice that the proposition does not take into account the possibility of modifying the correspondence between the tower of symmetric tensors and the functions on phase space. Therefore, the proposition \ref{uniquenessP} cannot be translated directly in a uniqueness statement about the Poisson brackets on the space of symmetric tensor fields, as will be seen explicitly below.
Nevertheless, another corollary of the proposition \ref{uniquenessP} states that that the deformation of the higher-spin gauge algebra which has been obtained from the unconstrained frame-like formulation is essentially the unique one coming from an associative structure on the commutative algebra of functions on phase space (with the same proviso just mentioned in the previous sentence).

\begin{corollary}
The Moyal bracket is the unique Poincar\'e invariant Lie bracket on the algebra of functions on the (co)tangent bundle of the Minkowski space-time arising as a star-commutator from an assocative deformation of the commutative algebra of functions, up to trivial deformations of the gauge algebra (corresponding to redefinitions of the gauge parameters).
\end{corollary}

\proof{This follows directly from three facts: (1) the property that any associative deformation of a commutative algebra defines a Poisson bracket, (2) the proposition \ref{uniquenessP} and (3) the uniqueness of star-products for a given Poisson manifold (modulo equivalence transformations).}

In any case, the fact that the seducing properties of the Schouten bracket could provide a promising departure for a higher-spin geometry is an old observation of Dubois-Violette \cite{D}. By the way, the $\cal W$ geometry of Hull is making use of symplectomorphisms in the context of higher-spin gauge symmetries \cite{Hull:1992vj}. The implementation of canonical transformations as part of conformal higher-spin gauge symmetries was discussed in a series of papers by Segal \cite{Segal:2000ke}. Actually, it seems that the Schouten bracket made several shadowed appearances in the early story of the subject. Surprisingly, the existence of the Schouten bracket was mentioned in \cite{Berends:1984rq} although no attempt was made to use it for higher-spins. Similarly, in \cite{Bengtsson:1986bz} it was observed that the symmetry properties of the tangent indices of the frame-like field $e^{a_1\ldots a_{s-1}}$ and the spin-like connection $\omega^{a_1\ldots a_{s-1},b}$ of \cite{Vframe80} fit elegantly the generators of the algebra mentioned in the remark \ref{remB2}. The metric-like unconstrained formulation of this frame-like theory should lead to the Schouten bracket, though this was not pointed in the original paper \cite{Bengtsson:1986bz}. Even earlier, Fr\o nsdal proposed to use the canonical Poisson bracket in higher-spin gauge theories, but with a different correspondence between tensors and functions on phase space from (\ref{varphi}). Instead, he proposed that gauge fields are given by \cite{F}
\begin{equation}
{\overline\phi}(x,p)\,\,=\,\frac12\,p^2\,+\,{\overline\varphi}(x,p)\,,
\end{equation}
with \begin{equation}
{\overline\varphi}(x,p)\,\,=\,\sum\limits_{s}\,\frac{1}{s\,!}\,(p^2)^{1-\,s/2}\,\varphi^{\mu_1\ldots\mu_s}(x)\,p_{\mu_1}\ldots p_{\mu_s}\,,
\end{equation}
and transform as
\begin{equation}
\delta_{\overline\varepsilon}{\overline\varphi}\,=\,\{{\,\overline\varepsilon}\,\,,{\overline\phi}\,\,\}_{_C}\,,
\label{nonabeliangaugetransfos2}
\end{equation}
under gauge transformations with parameters given by
\begin{equation}
{\overline\varepsilon}(x,p)\,\,=\,\sum\limits_{s}\,\frac{1}{s\,!}\,(p^2)^{(1-s)/2}\,\varepsilon^{\mu_1\ldots\mu_s}(x)\,p_{\mu_1}\ldots p_{\mu_s}\,.
\label{epsilonbar}
\end{equation}
The functions $\overline{\varphi}$ and ${\overline\varepsilon}$ are respectively homogeneous of degree two and one in $p\,$, but they are not analytic around $p=0\,$. This choice is particularly convenient in one respect: the fields and parameters can be assumed to be (single or double) traceless without loss of generality. In other words, the trace conditions may find a natural implementation, contrarily to the genuine Schouten bracket which does not preserve them. More precisely, the canonical Poisson bracket between functions ${\overline\varepsilon}$ of the form (\ref{epsilonbar}) induces a well-defined Lie bracket on the space of symmetric contravariant tensor fields quotiented by traceful tensor fields, which we might call ``Fr\o nsdal bracket.'' In other words, this bracket provides a candidate for the Lie algebra of gauge symmetries since the constrained parameters are traceless. The pointwise product between functions of the form (\ref{epsilonbar}) does not close, which prevents the space of traceless tensor fields endowed with the Fr\o nsdal bracket to inherit the (bi)derivation property for the symmetric product from the canonical Poisson bracket.
Maybe the difficulty, immediately recognised by Fr\o nsdal \cite{F}, of incorporating the trace constraints when using a genuine Poisson bracket prevented further exploration in such directions.

\subsubsection{Comparison with Fedosov's quantisation}\label{Fed}

The paper \cite{Bering} provides a concise review of the Fedosov procedure, of which the terminology is closely followed here. To start with a tentative dictionary, the analogue of the Fedosov connection is the Cartan-like connection $\cal A$ itself. In this regard, the background Cartan-like connection (\ref{A0}) is exactly the Hamiltonian for the Koszul-Tate differential in the case where the symplectic manifold is $T^*{\mathbb R}^n\,$. 
The main idea behind the Fedosov construction is that the star product {\footnotesize$\bigstar$} on the fibre (the Weyl algebra) induces the star product $\star$ on the base (here, the cotangent bundle).
Retrospectively, the short proof of the theorem in Subsection \ref{defo} can be read as an application of Fedosov's second theorem in the case of the cotangent bundle on flat space where the function on phase space is identified with the metric-like gauge parameter and the zero-form horizontal section with the frame-like gauge parameter. The horizontality condition is provided indirectly by Proposition \ref{generatingfunction}.
One of the many departure from the Fedosov scheme is the fact that the differential forms considered in the Vasiliev construction take value only in subalgebras of the Weyl algebra. Furthermore, the torsion-like constraint is weaker than the strict nilpotency of the exterior covariant derivative $\cal D\,$. In this respect, the unfolded equations are much closer to the Fedosov prescriptions. As mentioned in the section 7.2 of \cite{BCIV}, somehow, the Fedosov construction may be recovered from the general unfolding approach when no dynamical equations are imposed. Many other insights along these lines may be found in \cite{Vasiliev:2005zu,Grigoriev:2006tt}.

\subsubsection{Constant curvature spacetimes}\label{CCspacet}

Due to the successes encountered in the construction of interactions in the presence of a non-vanishing cosmological constant, it would be more satisfactory to generalise entirely the results of Subsection \ref{nonabsym} to (anti) de Sitter space-time. Unfortunately, it does not seem to be so straightforward for technical reasons.
Nevertheless, there exists a star product on the cotangent bundle $T^*AdS_n$ such that the tentative non-Abelian gauge transformations of the form (\ref{nonabeliangaugetransfos}), where the Moyal bracket is replaced with the corresponding star commutator and the $p^2$ term in (\ref{varphi}) is interpreted as the (anti) de Sitter background, reproduces at lowest order in the weak field expansion $\varphi$ the celebrated symmetrised derivative of the gauge parameter \cite{Fronsdal:1978vb}, modulo a field redefnition $\varphi\mapsto\varphi^\prime$,
\begin{equation}
\delta_\varepsilon\varphi^\prime\,=\,(p^\mu\nabla_\mu)\,\varepsilon\,+\,{\cal O}(\varphi)\,,
\label{abeliangaugetransfods}
\end{equation}
with $\nabla$ the $(A)dS_n$ covariant derivative.

Indeed, there exists a natural star product on the cotangent bundle $T^*AdS_n\,$, which is defined via an ambient construction. It was found in the seminal papers of deformation quantisation where it was introduced with humor as ``a star product is born'' \cite{Bayen:1977ha}.
\begin{lemma}[Bayen, Flato, Fr\o nsdal, Lichnerowicz and Sternheimer]\label{starisborn}

The commutative algebra $C^\infty(T^*AdS_n)$ of smooth functions $\varphi(x,p)$ on the cotangent bundle of the $n$-dimensional anti de Sitter space-time of curvature radius $R$ is isomorphic to the commutative algebra of smooth functions $\Phi(X,P)$ on the cotangent bundle of the ambient domain $X^2<0$ that are invariant under the scale transformations 
\begin{equation}
X^A\mapsto e^\alpha \,X^A\,,\quad P_B\mapsto e^{-\alpha}\,P_B\,,\quad(\alpha\in\mathbb{R})
\label{ctgtscale}
\end{equation}
and the longitudinal translations 
\begin{equation}
X^A\mapsto X^A\,,\quad P_B\mapsto P_B+\beta\, X^B\,,\quad(\beta\in\mathbb{R})
\label{cotgttransl}
\end{equation}
These algebras are endowed with their respective pointwise products.
The concrete correspondence is that a function $\varphi(x,p)$ is the evaluation of the ambient function $\Phi(X,P)$ on the cotangent bundle of the hyperboloid $X^2=-R^2$ defined by the transversality condition $X^AP_A=0$.

The pullback of the Moyal product for the cotangent bundle of the ambient domain $X^2<0$ induces a star product for the cotangent bundle $T^*AdS_n$ such that the former isomorphism becomes an isomorphism of associative algebras.
\end{lemma}

\proof{The surjection
\begin{equation}
X^A\mapsto \frac{R}{\sqrt{-X^2}}\,X^A\,,\quad P_B\mapsto \frac{1}{R}\Big(\sqrt{-X^2}\,P_B\,+\,\frac{X^AP_A}{\sqrt{-X^2}}\,X_B\Big)\,,
\label{mapstocotgt}
\end{equation}
sends any point of the cotangent bundle of the ambient domain $X^2<0$ on the cotangent bundle of the hyperboloid $X^2=-R^2$ defined by the transversality condition $X^AP_A=0$.
This surjection (\ref{mapstocotgt}) preserves the space of invariant functions, since it
is the composition of a longitudinal translation with a scale transformation, and thereby provides the ambient construction of $C^\infty(T^*AdS_n)\,$. 

The cotangent bundle of the ambient domain $X^2<0$ is a symplectic manifold where the scale transformations (\ref{ctgtscale}) and the longitudinal translations (\ref{cotgttransl}) are, respectively, the flow of the Hamiltonians $X^AP_A$ and $X^2\,$. Therefore the ambient Poisson bracket (and Moyal commutator) between the invariant functions $\Phi(X,P)$ and these two Hamiltonians must vanish (because these Hamiltonians are quadratic). Therefore, the derivation property of the commutator implies that the ambient Moyal product is well-defined in the subalgebra of invariant functions.
}

Following Lemma \ref{Fr}, the ambient transverse inverse metric $G^{AB}:=\eta^{AB}+X^AX^B/X^2$ defines the inverse metric $g^{\mu\nu}$ on $AdS_n\,$. 
The ambient invariant function 
\begin{equation}
P^2:=X^2\,G^{AB}P_AP_B=X^2\,\eta^{AB}P_AP_B\,-\,(X^AP_A)^2
\label{AdSbckgd}
\end{equation}
corresponds to the $AdS_n$ background $p^2:=g^{\mu\nu}p_\mu p_\nu\,$. The ambient Moyal commutator between any element $\epsilon(X,P)$ of the algebra of Lemma \ref{starisborn} and (\ref{AdSbckgd}) is equal to
\begin{equation}
[\,\epsilon(X,P)\,\stackrel{\bigstar}{,}\,P^2\,\,]\,=\,\frac{2\,i}{\lambda}\,\eta_{AB}\,\left(X^A X^B-\frac{\lambda^2}{4}\frac{\partial}{\partial P_A}\frac{\partial}{\partial P_B}\right)\,(P^C\nabla_C)\,\epsilon(X,P)\,,
\end{equation}
where {\footnotesize$\bigstar$} denotes the ambient Moyal product.
Therefore the commutator between the $AdS_n$ background field $p^2$ and any function $\varepsilon(x,p)$ on the cotangent bundle $T^*AdS_n$ for the star product $\star$ of Lemma \ref{starisborn} is equal to 
\begin{equation}
[\,\,\varepsilon(x,p)\,\stackrel{\star}{,}\,p^2\,\,]\,=\,\frac{2\,i}{\lambda}\,\left(1+\left(\frac{\lambda}{2R}\right)^2\,g_{\mu\nu}\,\frac{\partial}{\partial p_\mu}\frac{\partial}{\partial p_\nu}\right)\,(p^{\,\rho}\nabla_\rho)\,\varepsilon(x,p)\,,
\end{equation}
which produces the result in (\ref{abeliangaugetransfods}) in terms of the redefined field
\begin{equation}
\varphi^\prime(x,p):= \left(1+\left(\frac{\lambda}{2R}\right)^2\,g_{\mu\nu}\,\frac{\partial}{\partial p_\mu}\frac{\partial}{\partial p_\nu}\right)^{-1}\,\varphi(x,p)\,.
\end{equation}

\section{Conclusion}\label{conclusion}

Obviously, the present work is at most a very preliminary step towards the goal presented in the introduction: a metric-like version of interacting higher-spin gauge theories. Indeed, severe restrictions have been made: (i) no trace constraint, (ii) analysis at first order in a weak field expansion, and (iii) the dynamical equations have not been discussed. Nevertheless, it is plausible that the results given here may already deliver some impressionist view of the general picture. The analogy with gravity suggests that the non-Abelian algebra of gauge symmetries which has been exhibited might already capture the information at all orders. In such case, the group of unitary differential operators on the space-time $\cal M$ might extend the diffeomorphism group of gravity theory in some high energy/symmetry regime where the higher-spin gauge fields are included. The corresponding extension of the Lie bracket between vector fields would be the star-commutator between symbols for some (Hermitian) star-product on the cotangent bundle $T^*{\cal M}\,$, such as the one of \cite{Bordemann:1997ep}. Unfortunately, the adjoint action of the vector field Lie algebra $\Gamma(T{\cal M})$ through such star-commutators only reproduces the Lie derivative of symmetric tensor fields in the ``(semi)classical limit'' (\textit{i.e.} at lower orders in the dimensionful deformation parameter). Physically, this fact is unsatisfactory because one would like to identify the action of vector fields  with the infinitesimal diffeomorphisms. This annoying property might be related to the famous problem \cite{Aragone:1979hx} of the minimal\footnote{Recent (either algebraic or $S$ matrix) analyses \cite{Boulanger:2006gr,Boulanger:2008} even dropped any (respectively, either geometrical or variational) prejudice on this no-go result.} coupling between gravitons and higher-spin particles around flat space-time. 
Accordingly, it must be stressed that it is unclear whether the deformation of the gauge transformations written here defines a consistent deformation of the free higher-spin gauge theory because it may \textit{not} correspond to consistent cubic vertices (consistency is only guaranteed at the level of the gauge algebra). Anyway, the idea that higher-spin symmetries might arise from some gauging of higher-derivative rigid smmetries of the free fields is floating in the air since the early days of higher-spin studies.\footnote{A list (presumably inexhaustive) of references where this idea was advocated from various viewpoints is (sorted in chronological order): \cite{Berends:1985xx,Vasiliev:1999ba,Eastwood,Segal:2000ke,Mik,X,Boulanger:2006gr,Bekaert:2007mi,Fotopoulos:2007yq}.}
As argued in \cite{Bekaert:2007mi}, the Noether method applied to the gauging of the ``higher-translations'' (discovered in \cite{Berends:1985xx}) of a free complex scalar field on flat space-time leads to the Lie algebra of Hermitian differential operators on $L^2({\mathbb R}^n)\,$. Moreover, another argument in favor of such structure comes from the fact that Hermitian operators are symmetries of its mass term \cite{Fotopoulos:2007yq} since it is proportional to the $L^2$-norm of the massive scalar field. The fact that, due to the higher-derivatives, the higher-spin symmetry transformations are not derivations lies presumably at the heart of the difficulties with interacting higher-spin gauge fields because it prevents using the conventional methods of differential geometry, as was pointed out already in \cite{Bengtsson:1986bz}.

\section*{Acknowledgements}

N. Boulanger, M. Dubois-Violette and J. Mourad are especially thanked for numerous useful exchanges. G. Barnich, A.K.H. Bengtsson, P. Bieliavsky, E. Joung, K. Noui and J.H. Park are also acknowledged for interesting discussions. M. Grigoriev, E. Meunier and S. Pekar are acknowledged for pointing out two minor errors in the previous versions of the document.


\pagebreak

\appendix

\section{Weyl algebra}\label{ApWeyl}

A (linear) \textit{differential operator} $D$ on ${\mathbb R}^n$ is a function depending smoothly on the coordinates $x^\mu$ and polynomially on the partial derivatives ${\partial}/{\partial x^{\nu}}$ (the indices $\mu,\nu$ take $n$ values):
\begin{equation}
D=\sum\limits_{r=0}^m\,D^{\,\nu_1\ldots\,\nu_r}(x^\mu)\,\frac{\partial}{\partial x^{\nu_1}}\,\ldots\,\frac{\partial}{\partial x^{\nu_r}}\,.
\label{Differentialop}
\end{equation}
The degree $m\in\mathbb N$ in the partial derivatives is called the \textit{order} of the differential operator $D\,$. The functions $D^{\,\nu_1\ldots\,\nu_r}(x^\mu)$ are the \textit{coefficients} of $D\,$. They transform as contravariant symmetric tensors of rank $r$ under affine transformations of ${\mathbb R}^n$ but, for $r< m$ they do \textit{not} transform as tensor fields under general coordinate transformations, whereas the leading coefficient $D^{\,\nu_1\ldots\,\nu_m}(x^\mu)$, sometimes (\textit{e.g.} in \cite{Eastwood}) called the \textit{symbol} of $D\,$, is a contravariant symmetric tensor field of rank $m\,$, \textit{i.e.} an element of $\Gamma(\,\bigvee^m(T{\mathbb R}^n)\,)\,$.\footnote{Actually, if there is a connection defined on ${\mathbb R}^n$ then it is possible to replace the partial derivatives by covariant ones so that the coefficients properly transform as symmetric tensor fields (see \textit{e.g.} \cite{Bordemann:1997ep}).}
In this setting a \textit{vector field} is merely the symbol of a differential operator of order one. In physics, a differential operator is sometimes said to be ``higher-derivative'' if it is of order strictly greater than one.\vspace{1mm}

Let $\mathbb K$ be a field (either $\mathbb R$ or $\mathbb C$ here). The \textit{Weyl algebra} $A_{n}$ over $\mathbb K$ is the (unital) associative algebra of differential operators on ${\mathbb R}^n$
with polynomial coefficients, endowed with the composition $\circ$ as product. The Heisenberg algebra $\mathfrak{h}_n$ is the Lie algebra spanned by $\textsc{X}^\mu$ and $\textsc{P}_\nu$ with bracket defined by the only non-trivial relations
\begin{equation}
[\,\textsc{X}^\mu\,,\,\textsc{P}_\nu\,]\,=\,i\,\hbar\,\delta^\mu_\nu\,.
\label{Heisenberg}
\end{equation}
Abstractly, the Weyl algebra may be presented by its generators $\textsc{X}^\mu$ and $\textsc{P}_\nu$ modulo the commutation relations (\ref{Heisenberg}). (This definition means that $A_n$ is a realisation of the universal enveloping algebra of the Heisenberg algebra $\mathfrak{h}_n\,$.) In order to make contact with the previous concrete definition of the Weyl algebra one should of course perform the identification $\textsc{X}^\mu\mapsto x^\mu$ and $\textsc{P}_\nu\mapsto -i\,\hbar\,{\partial}/{\partial x^{\nu}}\,$. The \textit{polynomial algebra} ${\mathbb K}[x^\mu,p_\nu]$ is the commutative algebra spanned by all linear combinations over $\mathbb K$ with finite products of the $2n$ generators $x^\mu$ and $p_\nu$ modulo the commutation relations $[\,x^\mu\,,\,p_\nu\,]\,=\,0\,$. Physically, the Weyl/polynomial algebra respectively corresponds to the (associative) algebra of quantum/classical observables. Indeed, the polynomial algebra ${\mathbb K}[x^\mu,p_\nu]$ is isomorphic to the commutative algebra of polynomial functions on phase space. In other words, there is an injective morphism ${\mathbb K}[x^\mu,p_\nu]\hookrightarrow C^\infty(T^*{\mathbb R}^n)$ of commutative algebras. Actually, most of the sequel will be true for the smooth functions, as long as the operations involved (multiplication, \textit{etc}) are well defined. The focus on the Weyl/polynomial algebras is for the sake of simplicity only (in order to avoid convergence subtleties).\vspace{1mm}

A unital algebra $\cal A$ with product $*$ is \textit{graded} by a (semi)group $G$ of $\mathbb Z$ if
\begin{description}
  \item[(i)] it splits as the direct sum ${\cal A}=\bigoplus\limits_{i\in G}{\cal A}_i\,$.
  \item[(ii)] the unity $1$ belongs to
${\cal A}_0\,$.\vspace{1mm}
  \item[(iii)] the multiplication is such that ${\cal A}_i *{\cal A}_j\subseteq
{\cal A}_{i+j}$.
\end{description}
In other words, the property (iii) requires that the product $*$ is homogeneous of grading zero.   
The polynomial algebra ${\mathbb K}[x^\mu,p_\nu]$ is, for instance, $\mathbb N$-graded by the homogeneity degree in the indeterminates $p_\nu\,$. There is also an injective morphism of commutative graded algebras,
\begin{equation}
\iota\,:\,{\mathbb K}[x^\mu,p_\nu]\hookrightarrow \Gamma(\,\vee(T{\mathbb R}^n)\,)\,:\,
D^{\,\nu_1\ldots\nu_m}(x)\,p_{\nu_1}\,\ldots\,p_{\nu_m}\mapsto D^{\,\nu_1\ldots\nu_m}(x)\,,
\label{injection}
\end{equation}
from the polynomial algebra into the algebra of symmetric contravariant tensor fields endowed with the symmetric product 
\begin{equation}
(D_1\vee D_2)^{\,\nu_1\ldots\nu_{m_1+m_2}}=D_1^{\,(\nu_1\ldots\nu_{m_1}}D_2^{\,\nu_{m_1+1}\ldots\nu_{m_1+m_2})}
\label{symmetric product}
\end{equation}
In this morphism, the homogeneity degree in the ``classical'' momenta is mapped to the rank.\vspace{1mm}

For the Weyl algebra, a (weaker, so more general) notion than a graduation is needed.
A \textit{filtration} over an algebra $\cal A$ is a sequence $({\cal A}_i)_{i\in G}$ of subspaces of $\cal A$ such that 
${\cal A}_i\subseteq{\cal A}_{i+1}$ for any non-negative integer $i$ and satisfying the properties (ii) and (iii). 
The Weyl algebra $A_n$ is filtered by the order of differential operators (equivalently, by the degree in the ``quantum'' momenta $\textsc{P}_\nu$). The
\textit{graded algebra associated to the filtered algebra} $\cal A$ is denoted by
$gr({\cal A})$ and is defined as
$$gr({\cal A})=\bigoplus\limits_{i\in G} gr_i({\cal A})\,,\quad\quad gr_i({\cal A}):={\cal A}_i/{\cal A}_{i-1}\,,\quad i\in G-\{0\}\,.$$
The polynomial algebra ${\mathbb K}[x^\mu,p_\nu]$ (graded by the homogeneity degree in the ``classical'' momenta $p_\nu$) is isomorphic to the graded algebra $gr(A_n)$ associated to the Weyl algebra (filtered by the degree in the ``quantum'' momenta $\textsc{P}_\nu$). The representatives of the commutative algebra $gr(A_n)$ are the symbols of the differential operators and their multiplication is through the symmetric product (\ref{symmetric product}) of contravariant symmetric tensor fields:
$$D_1\circ D_2=(D_1\vee D_2)^{\,\nu_1\ldots\nu_{m_1+m_2}}(x)\,\frac{\partial}{\partial x^{\nu_1}}\,\ldots\,\frac{\partial}{\partial x^{\nu_{m_1+m_2}}}\,+\,\mbox{lower}\,.$$
The isomorphism $\Sigma$ of these two graded algebras may be realised as follows
\begin{eqnarray}
\Sigma&:&gr(A_n)\rightarrow {\mathbb K}[x^\mu,p_\nu]\nonumber\\
&:&\left[D^{\,\nu_1\ldots\nu_m}(x)\,\frac{\partial}{\partial x^{\nu_1}}\,\ldots\,\frac{\partial}{\partial x^{\nu_m}}\right]\,\,\longmapsto\,\, 
\,D^{\,\nu_1\ldots\nu_m}(x)\,p_{\nu_1}\,\ldots\,p_{\nu_m}\,.
\label{isomo}
\end{eqnarray}

Let $\cal A$ be an algebra over $\mathbb C$ with product $\star\,$. Let $\sigma:{\cal A}\rightarrow{\cal A}:x\mapsto\sigma(x)$ be a map. This map is \textit{anti-linear} iff $\sigma(\lambda x)=\lambda^* x$ for any $x\in\cal A$ and any $\lambda\in\mathbb C\,$. The map $\sigma$ is an \textit{anti-automorphism} iff $\sigma(x\star y)=\sigma(y)\star \sigma(x)$ for any $x,y\in{\cal A}\,$.
An anti-linear anti-automorphism $^\ast:{\cal A}\rightarrow{\cal A}:x\mapsto x^\ast$ is said to be an \textit{involution} if it is its own inverse, \textit{i.e.} $(x^\ast)^\ast=x$ for any $x\in{\cal A}\,$. An algebra with involution is called a $^\ast$-\textit{algebra\,}. The Hermitian conjugation $^\dagger$ sending the generators $\textsc{X}^\mu$ and $\textsc{P}_\nu$ to themselves endows the Weyl algebra with a structure of $^*$-algebra.\vspace{1mm}

The Lie $^*$-algebra obtained by endowing the space $\cal A$ of an associative $^*$-algebra with $-i$ times the commutator, $-i\,[\,\,\,\stackrel{\star}{,}\,\,\,]\,$, as Lie bracket is called here the \textit{commutator algebra} and it is denoted by $[{\cal A}]$ in the present paper. The elements such that $x^*=x$ are called \textit{self-adjoint}. The Lie subalgebra $[{\cal A}]_{\mathbb R}$ of its self-adjoint elements is a real form of the complex Lie algebra $[{\cal A}]\,$. Mathematically, the real algebra $[A_n]_{\mathbb R}$ corresponds to the Lie algebra of (polynomial) Hermitian differential operators on the Hilbert space with norm $L^2({\mathbb R}^n)$. Physically, it corresponds to the Lie algebra of quantum observables.\vspace{1mm}

The \textit{linear} anti-automorphism $\rho:A_n\rightarrow A_n$ of the Weyl algebra $A_n$ that is induced by the following transformations of the generators: $\rho(\textsc{X}^\mu)=\textsc{X}^\mu\,$, $\rho(\textsc{P}_\nu)=-\textsc{P}_\nu$
is \textit{not} an involution, but it is still ``involutive'' in the sense that it squares to the identity: $\rho^2=\rho\,$.
The Weyl algebra splits as the ${\mathbb Z}_2$-graded algebra $A_n=A^+_n\oplus A^-_n$ where $A_n^\pm$ is the eigenspace of eigenvalue $\pm 1\,$. The grading is essentially the parity in the generators $\textsc{P}_\nu\,$. The commutator algebra $[A^-_n]$ of elements that are odd in the momenta is a Lie subalgebra of $[A_n]\,$. Moreover, the Lie subalgebra $[A^-_n]_{\mathbb R}$ of self-adjoint elements that are odd in the momenta is a real Lie subalgebra of $[A_n]_{\mathbb R}\,$. This subalgebra corresponds to the Lie algebra of \textit{symmetric} differential operators on the Hilbert space of \textit{real} square-integrable functions.\vspace{2mm}

\noindent\textbf{Side remark:} Let $V\subseteq\cal A$ be a vector subspace of an associative algebra $\cal A\,$. (This vector subspace may not contain the unit $1\in\cal A$ and may not be a subalgebra.) The image ${\cal A}V$ of the left regular action of $\cal A$ on $V$ (\textit{i.e.} via multiplication from the left) is a left ideal of $\cal A\,$. The \textit{centraliser} ${\cal C}_V({\cal A})$ of $V$ in $\cal A$ is the subalgebra of $\cal A$ of elements that commute with all the elements of $V\,$.
The centraliser ${\cal C}_V({\cal A})$ has a two-sided ideal: the intersection ${\cal A}V\cap{\cal C}_V({\cal A})$ between the image ${\cal A}V$ and the centraliser ${\cal C}_V({\cal A})$ itself. The quotient of the centraliser by this ideal is denoted by $\overline{\cal C}_V({\cal A})$ here.

\section{Poisson bracket}\label{ApPoisson}

Let $\cal A$ be an associative algebra with $\cdot$ as product. A \textit{derivation} $\cal D$ over $\cal A$ is a linear operator obeying to the ``Leibnitz rule,'' that is ${\cal D}(x\cdot y)\,=\,({\cal D}x)\cdot y\,+\,x\cdot({\cal D}y)$ for any $x,y\in{\cal A}\,$. The space $Der({\cal A})$ of derivations over $\cal A$ is endowed with a structure of Lie algebra via the commutator $[\,\,\,\stackrel{\cdot}{,}\,\,]$ as Lie bracket. A \textit{Poisson bracket} $\{\,\,,\,\}$ for
$\cal A$ is a Lie bracket which is also a (bi)derivation, \textit{i.e.} $\{x,y\cdot z\}=y\cdot \{x,z\}+\{x,y\}\cdot z$
for any $x,y,z\in{\cal A}\,$. A (graded) \textit{Poisson algebra} is both a (graded) associative and Lie algebra $\cal A$ endowed
with an associative product and a Poisson bracket. 
The usual Poisson bracket $\{\,\,,\,\}_{_C}$ of classical mechanics will be called here the \textit{canonical Poisson bracket}. It is defined as
\begin{equation}
\{\,\,,\,\}_{_C}\,:=\,\frac{\overleftarrow{\partial}}{\partial x^{\mu}}\,\frac{\overrightarrow{\partial}}{\partial p_{\mu}}-\frac{\overleftarrow{\partial}}{\partial p_{\mu}}\,\frac{\overrightarrow{\partial}}{\partial x^{\mu}}\,,
\label{Poisson}
\end{equation}
where the arrows indicate on which factor they act. The canonical Poisson bracket endows the algebra $C^\infty(T^*{\mathbb R}^n)$ of function on the phase space with a structure of Poisson algebra. Explicitly, it acts as follows
$$\{\,P(x,p)\,,Q(x,p)\,\}_{_C}=\frac{\partial P}{\partial x^{\mu}}\frac{\partial Q}{\partial p_{\mu}}-\frac{\partial P}{\partial p_{\mu}}\frac{\partial Q}{\partial x^{\mu}}\,.$$
The canonical Poisson bracket endows ${\mathbb K}[x^\mu,p_\nu]$ with a structure of graded Poisson algebra if the gradation is taken to be the homogeneity degree in the ``classical'' momenta $p_\nu$ minus one (such that the grading of the Poisson bracket vanishes).\vspace{1mm}

The morphism (\ref{injection}) defines an injective morphism $\iota$ of Poisson algebras from the algebra of polynomial functions over the phase space into the algebra of symmetric contravariant tensor fields. The induced Poisson bracket is the so-called \textit{Schouten bracket} (see \textit{e.g.} \cite{DuboisV} and refs therein)
\begin{eqnarray}
\{\,\,,\,\}_S&:&\,\Gamma(\,\vee^{m_1}(T{\mathbb R}^n)\,)\otimes \Gamma(\,\vee^{m_2}(T{\mathbb R}^n)\,)
\rightarrow \Gamma(\,\vee^{m_1+m_2-1}(T{\mathbb R}^n)\,)\nonumber\\
&:&{\cal T}_1^{\,\nu_1\ldots\nu_{m_1}}(x)\,\otimes \,{\cal T}_2^{\,\nu_1\ldots\nu_{m_2}}(x)\,\,\longmapsto 
\,\,\{\,{\cal T}_1\,,{\cal T}_2\,\}_S^{\nu_1\ldots\nu_{m_1+m_2-1}}(x)\,,
\end{eqnarray}
where 
\begin{eqnarray}
\{\,{\cal T}_1\,,{\cal T}_2\,\}_S^{\nu_1\ldots\nu_{m_1+m_2-1}}\,\,:=&m_2&\partial_\mu {\cal T}_1^{\,(\nu_1\ldots\nu_{m_1}} {\cal T}_2^{\,\nu_{m_1+1}\ldots\nu_{m_1+m_2-1})\mu}\,-\nonumber\\
&&-\,\, m_1\,\,{\cal T}_1^{\,\mu(\nu_1\ldots\nu_{m_1}}\partial_\mu {\cal T}_2^{\,\nu_{m_1+1}\ldots\nu_{m_1+m_2-1})}\,.
\label{Schouten}
\end{eqnarray}
As one can see, this Lie bracket endows the algebra $\Gamma(\,\vee(T{\mathbb R}^n)\,)$ of symmetric contravariant tensors with a structure of graded Poisson algebra if the gradation is taken to be the rank minus one.\vspace{1mm}

The degree $p:=m-1$ defined by substracting one to the order of the differential operator $D$ in (\ref{Differentialop}) filters the commutator algebra $[A_n]$ because the order of the commutator of two differential operator $D_1$ and $D_2$ of respective orders $m_1$ and $m_2$ is not greater than $m=m_1+m_2-1\,$. Moreover, the Lie algebra $[A_n]$ does not possess a unit element (with respect to the commutator), so one may forget about (ii). 
Let $gr(\,[A_n]\,)$ be the graded Lie algebra associated to the commutator algebra $[A_n]$ (filtered by the degree in the ``quantum'' momenta $\textsc{P}_\nu$ minus one).
The representatives of the Lie algebra $gr(\,[A_n]\,)$ are the symbols of the differential operators and their multiplication is through the Schouten bracket (\ref{Schouten}) since
$$[\,D_1\,\stackrel{\circ}{,}\,D_2\,]\,=\,\{\,D_1\,,D_2\,\}_S^{\nu_1\ldots\nu_{m_1+m_2-1}}(x)\,\frac{\partial}{\partial x^{\nu_1}}\,\ldots\,\frac{\partial}{\partial x^{\nu_{m_1+m_2-1}}}\,+\,\mbox{lower}\,.$$
Therefore, the map (\ref{isomo}) is an isomorphism of graded Poisson algebras between the algebra of (polynomial) symbols of differential operators on ${\mathbb R}^n$ and the algebra of (polynomial) functions over the phase space $T^*{\mathbb R}^n\,$, where the degree in the ``quantum'' momenta $\textsc{P}_\nu$ is mapped to the homogeneity degree in the ``classical'' momenta $p_\nu\,$. 

By construction, the Schouten bracket of two symmetric tensor fields still transforms as a symmetric contravariant tensor field under coordinate transformations because it may be defined via the symbol of the commutator. The Poisson algebra of vector fields $\Gamma(T{\mathbb R}^n)$ is a subalgebra of the Poisson algebra $\Gamma(\,\vee(T{\mathbb R}^n)\,)$ of symmetric contravariant tensor fields and the restriction of the Schouten bracket to this subalgebra is precisely the \textit{Lie bracket of vector fields}. Again, this property follows directly from the construction. Actually, there is much more than that for the Schouten bracket: The adjoint action of the Lie subalgebra $\Gamma(T{\mathbb R}^n)$ of vector fields on the commutative algebra $\Gamma(\,\vee(T{\mathbb R}^n)\,)$ of symmetric contravariant tensor fields via the Schouten bracket is precisely through the \textit{Lie derivative of symmetric contravariant tensor fields}. Indeed,
\begin{equation}
ad\,:\,\Gamma(\,T{\mathbb R}^n\,)\rightarrow Der\Big(\,\Gamma(\,\vee(T{\mathbb R}^n)\,)\,\Big)\,:\,\xi\mapsto \{\,\xi\,,\,\,\}_S=-{\cal L}_\xi
\label{ad=S}
\end{equation}
is acting on a symmetric contravariant tensor $\cal T$ of rank $m$ as
\begin{equation}
\{\,\xi\,,{\cal T}\,\}^{\nu_1\ldots\nu_m}_S
\,=\,m\,\partial_\mu \xi^{\,(\nu_1} {\cal T}^{\,\nu_2\ldots\nu_m)\mu}\,-\,\xi^\mu\partial_\mu {\cal T}^{\,\nu_1\ldots\nu_m}=-({\cal L}_\xi{\cal T})^{\nu_1\ldots\nu_m}\,. 
\end{equation}

\section{Moyal product}\label{ApMoyal}

There is an isomorphism of commutative graded algebras  
\begin{equation}
\Gamma(\,\vee(T{\mathbb R}^n)\,)\,\rightarrow \,C_{pp}^\infty(\,T^*{\mathbb R}^n)\,:\,
{\cal T}^{\,\nu_1\ldots\nu_m}(x)\mapsto {\cal T}^{\,\nu_1\ldots\nu_m}(x)\,p_{\nu_1}\,\ldots\,p_{\nu_m}\,,
\label{symtphsp}
\end{equation}
between the algebra of symmetric contravariant tensor fields on ${\mathbb R}^n$ and the algebra $C_{pp}^\infty(\,T^*{\mathbb R}^n)$ of functions over the phase space $T^*{\mathbb R}^n$ that are polynomial in the fibre (`pp' stands for polynomial in the momenta $p$).
There even exists an isomorphism of Poisson algebras between the algebra of symbols of differential operators on ${\mathbb R}^n$ and the algebra of functions over the phase space $T^*{\mathbb R}^n$ that are polynomial in the fibre.
This isomorphism is best understood from the underlying associative algebra of differential operators, as explained in the previous appendix.
Nowadays, one way to define quantisation\footnote{The papers \cite{Waldmann:2003ce} provide excellent introductions to deformation quantisation.} is as the inverse problem of reconstructing the associate algebra from the Poisson structure alone.\vspace{1mm}

A differentiable manifold $\cal M$ endowed with a Poisson bracket $\{\,\,,\,\}$ for the commutative algebra $C^\infty({\cal M})$ of functions on
$\cal M$ is called a \textit{Poisson manifold}. The cotangent bundle $T^*{\mathbb R}^n$ endowed with the Poisson bracket (\ref{Poisson}) is a Poisson manifold. Let $\cal A$ be an algebra with $\cdot$ as product. The space  ${\cal A}[[\hbar]]$ is spanned by the formal power series in $\hbar$ with coefficient in $\cal A\,$.
A (formal) \textit{associative deformation} of
$\cal A$ is an associative product $\star$ for the space ${\cal A}[[\hbar]]$ of the
form
$$x\star y\,=\,\sum_{r=0}^\infty \,\hbar^r \,C_r(x,y)$$where $C_r:{\cal
A}\times{\cal A }\rightarrow \cal A $ are bilinear maps with
$C_0(x,y)=x\cdot y$. It may be shown that for any associative deformation of a commutative product $\cdot\,$, the antisymmetric part of
its first-order component, $C_1(x,y)-C_1(y,x)$ ($\forall x,y\in\cal A$), defines a Poisson bracket on $\cal A\,$.
A (formal) \textit{deformation quantisation} of a
Poisson algebra $\cal A$ with bracket $\{\,\,,\,\}$ is an
associative deformation of $\cal A$ with
$C_1(x,y)-C_1(y,x)\,=\,i\,\{x,y\}\,$.
A \textit{Hermitian deformation} of a
$^*$-algebra $\cal A$ with involution $^*$ is a deformation of $\cal
A$ such that $(x\star y)^*=y^*\star x^*$.
A \textit{star product} for a Poisson manifold
$\cal M$ is a product $\star$ for a deformation quantisation of the Poisson algebra $C^\infty({\cal M})$
such that (i) $x\star 1=x=1\star x$ for any $x\in\,C^\infty({\cal M})$, and (ii) $C_r$ is a (bi)differential operator for any $r\in{\mathbb N}\,$.
Two star products $\star$ and $\star{}^{\,\prime}$ are equivalent if
there exists an \textit{equivalence
transformation}, {\it i.e.} a formal power series
$$S\,=\,id\,+\,\sum_{r=1}^\infty\, \hbar^r \,S_r$$
of differential operators $S_r$ such that
$x\star{}^{\,\prime}\, y\,=\,S^{-1}(Sx\star Sy)$ and $
S\,1=1\,.$ Given the Poisson algebra, the star product is unique, modulo equivalence transformations.\vspace{1mm}

Let $P(x^\mu,p_\nu)$ be a polynomial of ${\mathbb K}[x^\mu,p_\nu]\,$.
Its image under the \textit{Weyl map}
\begin{equation}
{\cal W}\,:\,{\mathbb K}[x^\mu,p_\nu]\rightarrow A_n\,:\,P(x^\mu,p_\nu)\mapsto P_W(\textsc{X}^\mu,\textsc{P}_\nu)
\label{Weyl}
\end{equation}
is the Weyl/symmetric ordered polynomial $P_W(\textsc{X}^\mu,\textsc{P}_\nu)$ associated to $P(x^\mu,p_\nu)\,$.
The Weyl map (\ref{Weyl}) is an isomorphism of vector spaces whose inverse ${\cal W}^{-1}$ is called the \textit{Wigner map}. Its nicest property is that it relates the Hermitian conjugation $^\dagger$ of $A_n$ with the complex conjugation $^*$ of ${\mathbb C}[x^\mu,p_\nu]\,$,
\begin{equation}
{}^\dagger\,\circ\,{\cal W}\,=\,{\cal W}\,\circ \,{}^*\,.
\label{Herm-compl}
\end{equation}
The \textit{Moyal product} $\star$ is the pullback of the product in $A_n$ by the Weyl map which then becomes an isomorphism of associative algebras. It reads explicitly
\begin{equation}
\star\,:=\,\exp\Big(\,\frac{i\,\hbar}{2}\,\,\frac{\overleftarrow{\partial}}{\partial x^{\mu}}\wedge\frac{\overrightarrow{\partial}}{\partial p_{\mu}}\Big)\,.
\label{Moyal}
\end{equation}
The Weyl algebra $A_n$ may thus be seen as the Hermitian deformation quantisation of the following Poisson algebras: the algebra ${\mathbb K}[x^\mu,p_\nu]$ of (polynomial) functions on phase space $T^*{\mathbb R}^n\,$, the algebra $grad(A_n)$ of symbols of (polynomial) differential operators on ${\mathbb R}^n\,$, or the algebra of (polynomial) symmetric contravariant tensor fields on ${\mathbb R}^n\,$.
The Moyal product (\ref{Moyal}) provides a star product for the cotangent bundle $T^*{\mathbb R}^n$ endowed with the canonical Poisson bracket (\ref{Poisson}).
The Moyal product is extremely convenient because it is Hermitian, due to (\ref{Herm-compl}).
Another convenient property of the Moyal product is that its commutator is given by the simple relation
\begin{equation}
\{\,\,,\,\}_M\,:=\,\frac{1}{i\,\hbar}\,[\,\,\,\stackrel{\star}{,}\,\,\,]\,=\,\frac{2}{\hbar}\,\sin\Big(\,\frac{\hbar}{2}\,\,\frac{\overleftarrow{\partial}}{\partial x^{\mu}}\wedge\frac{\overrightarrow{\partial}}{\partial p_{\mu}}\Big)\,,
\label{starcommutator}
\end{equation}
where $\{\,\,,\,\}_M$ denotes the \textit{Moyal bracket} on ${\mathbb R}[x^\mu,p_\nu]\,$.  It is a deformation of the canonical bracket: $\{\,\,,\,\}_M=\{\,\,,\,\}_C+\,{\cal O}(\hbar^2)\,$. The image ${\cal W}^{-1}[A_n]_{\mathbb R}$ of the real form spanned by the self-adjoint elements under the Wigner map is therefore simply the real space ${\mathbb R}[x^\mu,p_\nu]\,$
endowed with the Moyal bracket. Physically, it corresponds to the Lie algebra of quantum observables.
As follows from the explicit expression (\ref{starcommutator}), notice that the Moyal bracket with any polynomial ${\mathbb R}[x^\mu,p_\nu]$ of degree two reduces to the mere Poisson bracket (\ref{Poisson}).

\section{Jet bundle}\label{Jspace}

In order to reformulate a field theoretical problem (\textit{i.e.} a functional problem) into a finite-dimensional algebraic problem (much more easy to address) via the hypothesis of locality, one usually treats the fields and their partial derivatives as independent coordinates of a so-called ``jet space''. From a mathematical perspective, jet bundles provide the right tool to address differential equations and their symmetries in a coordinate-free manner. This abstract way of introducing these objects is not chosen here because such a level of generality is not necessary for the present purpose.\vspace{1mm}

Let $V$ be the fibre of a vector bundle over a smooth manifold $\cal M$ of dimension $n\,$. The bundle is taken to be trivial for simplicity, but this construction can be generalised. The letter $\chi$ will collectively denote the coordinates of the vector space $V$ (thereby leaving any index implicit). The sections of the vector bundle ${\cal M}\times V$ are fields $\chi(x)\,$.
At any point of $\cal M\,$, let the $r$th partial derivatives of the field
variables $\chi$ be denoted by $\partial^r \chi$, that is $\partial^r\chi\,\sim\,\partial_{\mu_1}\ldots\partial_{\mu_r}\chi\,.$ 
A \textit{local function} of the field variables $\chi$ is a function $f(\,x,[\chi])$ of the space-time
coordinates $x\,$, of the field variables $\chi$ and a finite number of their derivatives,
where the notation $[\chi]$ stands for the variables $\chi$, $\partial\chi$, $\partial^2\chi$, ..., $\partial^k\chi$ for some finite but otherwise arbitrary integer $k\in{\mathbb N}_0\,$.
The \textit{jet space} $J^k V$ of order $k$ is taken to be the vector space with coordinates given by $[\chi]$ where the all the derivatives are taken as independent coordinates. The limiting case $k=\infty$ is actually admitted in the definition and will be referred to as the \textit{infinite jet space}. Since the partial derivatives are commuting, the following isomorphism of vector spaces holds:
\begin{equation}J^\infty V\,\cong\, \vee({\mathbb R}^{n*})\otimes V\,,
\end{equation}
where ${\mathbb R}^{n*}$ is isomorphic to the cotangent space at any point of $\cal M\,$.
A \textit{pseudolocal function} of the field variables $\chi$ is a function $f(\,x,[\chi];\lambda)$ of the space-time
coordinates $x\,$, the field variables $\chi$ and all their derivatives, which is also a formal power series in the expansion parameter $\lambda$ such that each Taylor coefficient is a local function. (The expansion parameter will be implicit most of the time.) 

The \textit{trivial jet bundle} of order $k$ is defined as the direct product ${\cal J}^k({\cal M}\times V):={\cal M}\times J^k V\,$, in other words the fibre is the corresponding jet space.\footnote{In the litterature, ``infinite jet space'' is sometimes a synonym for the limit $k=\infty$ of the jet bundles of order $k\,$. This terminology is not followed in the present paper.} A local function is thus a smooth function on a jet bundle of some finite order, that is, an element of $C^\infty\big(\,{\cal J}^k({\cal M}\times V)\,\big)\,$. As an example, the original vector bundle is the jet bundle of order zero, ${\cal J}^0({\cal M}\times V)={\cal M}\times V\,$. Any field $\chi(x)$ induces a natural section of the jet bundle ${\cal J}^k({\cal M}\times V)$ via
$$(\partial_{\mu_1}\ldots\partial_{\mu_\ell}\chi)(x)\,:=\,\frac{\partial}{\partial x^{\mu_1}}\ldots\frac{\partial}{\partial x^{\mu_\ell}}\,\chi(x)\,.$$

A derivation of the commutative algebra $C^\infty\big(\,{\cal J}^\infty({\cal M}\times V)\,\big)$ of smooth functions on the infinite jet bundle reads
\begin{equation}
A\,=\,a^\mu(x,[\chi])\,\frac{\partial}{\partial x^{\mu}}\,+\,\sum\limits_{r}\,\alpha^{\,\mu_1\ldots\,\mu_r}(x,[\chi])\,\frac{\partial}{\partial (\partial_{\mu_1}\ldots\partial_{\mu_r}\chi)}\,.
\label{genvectf}
\end{equation}
Such a derivation is called a \textit{generalised vector field}
if $\alpha^{\,\mu_1\ldots\,\mu_r}=0$ for all integers $r\neq 0\,$. It may be written as
\begin{equation}
a\,=\,a^\mu(x,[\chi])\,\frac{\partial}{\partial x^{\mu}}\,+\,\alpha(x,[\chi])\,\frac{\partial}{\partial \chi}\,.
\label{genvectfield}
\end{equation}
An \textit{evolutionary vector field} is a vertical generalised vector field, \textit{i.e.} even $a^\mu(x,[\chi])=0$ in (\ref{genvectf}). The $V$-valued function $\alpha(x,[\chi])$ on the jet bundle is called the \textit{characteristic} of this evolutionary vector field.
The evolutionary vector field $\breve{a}$ associated to the generalised vector field $a$ written in (\ref{genvectfield}) is equal to
\begin{equation}
\breve{a}\,=\,\breve{\alpha}(x,[\chi])\,\frac{\partial}{\partial \chi}\,,\quad\quad
\breve{\alpha}\,:=\,\alpha-a^\mu\partial_\mu\chi\,.
\label{evolutionary}
\end{equation}
The \textit{total derivatives} are the $n$ derivations
\begin{equation}
\partial_\mu^T\,:=\,\frac{\partial}{\partial x^{\mu}}\,+\,\sum\limits_{r}\,\partial_\mu\partial_{\nu_1}\ldots\partial_{\nu_r}\chi
\,\,\frac{\partial}{\partial (\partial_{\nu_1}\ldots\partial_{\nu_r}\chi)}\,,
\label{totalder}
\end{equation}
The \textit{infinite prolongation of a generalised vector field} (\ref{genvectfield}) is defined as
\begin{equation}
A\,:=\,a^\mu\,\partial^T_\mu\,+\,\sum\limits_{r}\,\partial^T_{\mu_1}\ldots\partial^T_{\mu_r}\breve{\alpha}\,
\,\frac{\partial}{\partial (\partial_{\mu_1}\ldots\partial_{\mu_r}\chi)}\,,
\label{infprol}
\end{equation}
for consistency with the natural sections of the jet bundle. The infinite prolongation of an evolutionary vector field with characteristic $\alpha:{\cal J}^\infty({\cal M}\times V)\rightarrow V$ defines the \textit{infinite prolongation of a characteristic}, which is denoted here by $[\alpha]:{\cal J}^\infty({\cal M}\times V)\rightarrow J^\infty V\,$.


\pagebreak


\end{document}